\documentclass[12pt]{article}
\pdfoutput=1
\usepackage[utf8]{inputenc}
\usepackage{bm}
\usepackage{mathrsfs}
\usepackage{empheq}
\usepackage{verbatim}
\usepackage{bibentry}
\usepackage{jheppub}
\usepackage{hyperref}
\usepackage{slashed}

\usepackage{subcaption}
\usepackage{multirow}
\usepackage{color}
\usepackage{tikz}
\usepackage{array}
\usepackage{tablefootnote}
\usepackage{latexsym}
\usepackage{amssymb}
\usepackage{amsmath}
\usepackage{mathtools}
\usepackage{stackrel}
\usepackage{youngtab}
\usepackage{cancel}
\usepackage[nosort]{cite}
\usepackage{graphicx}
\usepackage{epsfig}
\usepackage{psfrag}
\usepackage{tensor}
\usepackage{ulem}
\usepackage{bm}
\usepackage{url}
\DeclareMathAlphabet{\pazocal}{OMS}{zplm}{m}{n}
\usepackage{braket}
\usepackage{amsmath}
\usepackage{amssymb}

\usepackage{cleveref}

\usepackage{tocloft} 
\usepackage{ulem}
\usepackage{soul,xcolor}
\setstcolor{red}

\newcommand{\be}{\begin{equation}}
\newcommand{\ee}{\end{equation}}

\newcommand{\bi}{\begin{itemize}}
\newcommand{\ei}{\end{itemize}}
\newcommand{\bea}{\begin{eqnarray}}
\newcommand{\eea}{\end{eqnarray}}
\newcommand{\dd}{\text{d}}

\numberwithin{equation}{section}

\title{Quite Discrete for a fermion}

\author[1, 2]{Vasileios A. Letsios,}
\author[1]{Ben Pethybridge,}
\author[1]{and Alan Rios Fukelman}

\affiliation[1]{Department of Mathematics, King's College London \\ The Strand, London WC2R 2LS, U.K.}
\affiliation[2]{Current address: Physique de l’Univers, Champs et Gravitation,\\ Université de Mons – UMONS \\
Place du Parc 20, 7000 Mons, Belgium}

\emailAdd{vasileios.letsios@umons.ac.be}
\emailAdd{ben.pethybridge@kcl.ac.uk}
\emailAdd{alan.rios\_fukelman@kcl.ac.uk}

\abstract{We study Discrete Series representations of $SL(2,\mathbb{R})$ with half-integer scaling dimension $\Delta$. At the classical level, we show that these UIRs are realised in the space of mode solutions of spinor fields with \emph{imaginary mass parameters} on a fixed two-dimensional de Sitter, dS$_{2}$, background. Upon such tuning of the mass, the field develops a fermionic shift symmetry that we characterise. We show that in the Euclidean section this manifests itself in the presence of zero-modes which preclude the definition of a Hadamard two-point function for these UIRs. We propose a Euclidean procedure to deal with the zero-modes, define a two-point function with the right singularity structure, and analyse its late-time behaviour. We end this note by proposing two interacting theories containing the fermionic discrete series in their spectrum. }

\begin{document}
\maketitle

\section{Introduction}

In recent years a growing body of research has provided tools and models to understand features of de Sitter (dS) spacetimes. These include efforts to define a microscopic theory that captures some of the features of expanding universes \citep{Anninos:2017hhn,Anninos:2018svg,Anninos:2020cwo,Anninos:2022qgy,Susskind:2021esx,Ecker:2022vkr,Anninos:2022hqo,Witten:2020ert,Gorbenko:2018oov,Lewkowycz:2019xse,Coleman:2021nor,Shyam:2021ciy,Dong:2018cuv} and the development of CFT techniques applied to cosmological correlation functions  \citep{Bzowski:2013sza,Baumann:2022jpr,DiPietro:2021sjt, Strominger:2001pn,Maldacena:2002vr, Anninos:2014lwa,Arkani-Hamed:2017fdk,Jazayeri:2021fvk,Sleight:2020obc,Hogervorst:2021uvp,DiPietro:2021sjt} or more general observables for quantum field theory (QFT) in de Sitter spacetimes \citep{Bzowski:2013sza,Anninos:2014lwa,Gorbenko:2018oov,DiPietro:2021sjt,Benincasa:2022gtd,RiosFukelman:2023mgq}. 

While these developments unfold, it is becoming more evident that representation theory of the de Sitter isometry group plays a prominent role in this discussion. Wigner's classification in Minkowski spacetime \citep{wigner}, which identifies unitary irreducible representations (UIRs) with elementary particles, serves as a powerful organising principle that has guided developments of QFT in flat space. These particles serve then as the building blocks to develop theories consistent with spacetime symmetries; it is essential to have at our disposal a similar toolkit for expanding universes. 

For de Sitter spacetime, the single particle Hilbert space of the elementary particles should be in one to one correspondence with the UIRs of $SO(1,d+1)$.
Within the study of non-compact Lie groups, $SO(1,d+1)$ plays a prominent role pioneered by Harish-Chandra \citep{harish1,harish2}. An early reference describing the application to physics is \citep{Dobrev:1977qv}, and a modern review with specific applications to field theory is given in \citep{Sun:2021thf, Sengor:2022kji}. 

While the exposition of \citep{Dobrev:1977qv,Sun:2021thf} is complete from a group-theoretic viewpoint, from the perspective of de Sitter physics there are two missing pieces. There are still some UIRs, such as the Exceptional Series type I, that have yet to be realised in the single-particle Hilbert space of a free field theory. Secondly, these works focus on bosonic representations given by totally symmetric traceless tensors,\footnote{Mixed-symmetry fields are discussed in \citep{Basile:2016aen}.} thus leaving the fermionic representations out of the discussion. 

From the group-theoretic viewpoint, fermionic representations are addressed in \citep{hiraiI,ottoson,Gavrilik:1975ae, schwarz},  where the powerful techniques of Gelfand-Tsetlin patterns are used. Albeit useful, the physical intuition behind these techniques is obscure. A self-contained group-theoretic treatment along the lines of \citep{Sun:2020sgn} for fermionic representations was recently given in \citep{Schaub:2024rnl}. On the physics side, previous work focused on fermions in general-dimensional de Sitter has appeared in various references such as \citep{Pethybridge:2021rwf,Schaub:2023scu,Letsios:partI,Letsios:partII,Letsios:2020twa, Letsios:2023awz, Cotaescu:1998ay, Epstein:spinors, Hinterbichler_fermions}. In this paper, we extend this study of fermions in dS by concentrating on the example of parity-odd discrete series representations of the Lie algebra $\mathfrak{so}(1,2)$.

Of special interest is the case of (partially) massless tensor-spinor fields de Sitter spacetimes. In four dimensions these fields are the fermionic analogues of the bosonic, (partially) massless gauge fields discussed in \citep{Higuchi:1986wu,Deser:1983mm,Deser:2001pe,Higuchi:1986py}. Our interest is motivated by the possibility of realising these fields as part of (supersymmetric) higher-spin gravity theories which are relevant to recent efforts at formulating dS holography \citep{Anninos:2011ui,Anninos:2013rza,Anninos:2017eib},\footnote{Higher-spin gravity theories with covariant actions in 4-dimensional flat and (anti-)de Sitter space are studied in \citep{Krasnov:2021nsq}.} as well as the existence of a spin-$\frac32$ UIR corresponding to a fermionic gauge field, the gravitino. As observed in \citep{Letsios:partI,Letsios:partII}, a free field theory realisation of these UIRs at the level of classical mode solutions involves making sense of a purely imaginary mass parameter in the Lagrangian. It is essential to understand whether this affects the unitarity of the field theory or if the imaginary-mass theory should be understood as a constrained sector of a bigger, unitary, theory.

In this paper, we argue that the features characterising the four-dimensional gravitino, namely the pure imaginary mass and the presence of a gauge symmetry are already present in a simpler scenario, namely fermionic UIRs of $\mathfrak{so}(1,2)$ with tuned masses. These are known as the discrete series and may be realised as the single-particle states of a QFT in two-dimensional de Sitter. A similar analysis discussing the bosonic version of these theories in two dimensions was considered in \citep{Anninos:2023lin}, here we extend this discussion to fermionic fields, make comments about their links to four-dimensional de Sitter and entertain the possibility of realising such representations in theories with dS supersymmetry. 

Models of dS$_2$ have previously been shown to capture features relevant to quantum fields in dS$_4$. In \citep{Anninos:2024fty}, a two-dimensional theory in a fixed dS background was shown to include the spontaneous generation of a mass parameter for a gauge field, the presence of topological gauge-theoretic sectors, and the appearance of a chiral condensate. Furthermore, it was possible to compute exact correlation functions that showcase the stability of dS in the deep IR region. A two-dimensional gravitational model was also considered in \citep{Anninos:2023exn,Anninos:2022ujl}, along with its non-supersymmetric version \citep{Muhlmann:2021clm,Muhlmann:2022duj,Anninos:2021ene}, where it was found that there is an unbounded conformal mode equivalent to the one appearing in four-dimensional gravity. Even more, the theory contains a gravitino that showcases the same imaginary mass problem as in four dimensions. 

In this paper, we continue to use two dimensions as an ideal playground to develop models which exhibit features endemic to four-dimensional theories. In section \ref{sec:groupthy} we review the group theory construction of the corresponding UIRs. We then introduce free field theories whose classical mode solutions carry two modules of discrete series UIRs for specific tuning of the masses. We analyse the emerging gauge symmetry of such theories,  derive a de Sitter invariant two-point function for these fields, and comment on the quantisation of these free field theories. Finally, in section \ref{GTFermions} we analyse the appearance of such discrete series UIRs as pieces of bigger gauge theories. We discuss a complex $q=2$ super-SYK theory containing a higher-spin tower of bosdonic and fermionic discrete UIRs and a supersymmetric JT-gravity that admits a de Sitter saddle point solution whose Gaussian fluctuations are captured by a bosonic and fermionic discrete multiplet. Conventions and technical details of certain computations are left for the appendices.

\section{Group theory \& unitary representations} 
\label{sec:groupthy}
The representation theory of the $(d+1)$-dimensional de Sitter isometry group, $G = SO(1,d+1)$, is well known in the mathematical literature \citep{Dobrev:1977qv,schwarz,ottoson}. Recently, it has also gained a prominent role in efforts to address quantum features of de Sitter spacetimes \citep{Anninos:2014lwa,Bzowski:2013sza,Baumann:2022jpr,Sengor:2023buj,Sengor:2021zlc}. 
From a field-theoretic perspective, UIRs of the isometry group should be identified with possible elementary particles propagating in a fixed de Sitter background. Yet, knowing the UIRs is just half of the solution to the problem; one also needs to have a sensible field theory that upon quantisation realises such UIRs in the corresponding single-particle Hilbert space. 

\noindent The classification of the UIRs of $SO(1,d+1)$ depends heavily on the dimensionality of spacetime one is interested in. For even $(d+1)$, the harmonic analysis of the group shows that there exist discrete series representations \citep{knapp}. While the properties of such UIRs for general even $(d+1)$ are somewhat mysterious and no explicit field theory constructions are known \citep{Basile:2016aen}, for $4$ spacetime dimensions they correspond to (partially) massless gauge fields \citep{Sun:2021thf,RiosFukelman:2023mgq, Basile:2016aen,Higuchi:1991tn, Higuchi:1986wu, Letsios:partI, Letsios:partII, Letsios:2023awz}. 

\noindent Although most of the field-theoretic discussions concern bosonic discrete series, these UIRs also have fermionic counterparts which correspond to symmetric tensor-spinor fields in four spacetime dimensions and to mixed-symmetry fermionic fields for $(d+1) > 4$. Contrary to the bosonic case, (partially) massless symmetric tensor-spinor gauge fields were recently shown to be unitary only for $4$ spacetime dimensions through an analysis of the classical mode solutions for spin $s =\frac{3}{2}, \frac{5}{2}$ \citep{Letsios:partI, Letsios:partII}, while it was also shown, based on group-theoretic arguments, that this extends to all symmetric tensor-spinor gauge fields with spin $s \geq \frac{3}{2}$. These results were later obtained through a group theoretic analysis in \citep{Schaub:2024rnl}. We thus actually have a tower of (strictly and partially massless) fermionic discrete series UIRs of $Spin(1,4)$  for all spins $s = \frac{1}{2} + n\, , n \in \mathbb{N}$, corresponding to symmetric tensor-spinors with certain imaginary mass parameters on dS$_{4}$.\footnote{This was further verified through a mode function analysis of the massless case in \citep{Letsios:2023awz}.} 

The classical field realisation studied in \citep{Letsios:partI, Letsios:partII} showed that the Lagrangian description requires a purely imaginary mass parameter for the fields.\footnote{The attentive reader might find the claim of \emph{massless} and pure imaginary mass contradictory. We recall that on a curved background, a massless field corresponds to a field propagating on the lightcone or equivalently in $4$-dimensions to a field with two propagating degrees of freedom.} This poses some tension with the unitarity of the quantum theory as the action is not Hermitian. Despite this, the classical analysis shows that the modes furnish a direct sum of two discrete series UIR \citep{Letsios:partI, Letsios:partII}.

The case of a two-dimensional de Sitter spacetime was not analysed in \citep{Letsios:partI, Letsios:partII,Schaub:2024rnl}. One could argue that in lower-dimensional theories many of the interesting features of higher dimensions get lost. Yet, the Euclidean analysis of \citep{Anninos:2023exn} suggests that there also exists a $2$-dimensional analogue of the spin$-\frac{3}{2}$ field discrete series UIR. Even more, it was shown to share many of the features appearing in the four-dimensional theory, such as the purely imaginary mass and the appearance of gauge symmetry. The observations of \citep{Anninos:2023exn} serve as the main motivation of our present paper.

\noindent In the next section we analyse the \emph{fermionic} discrete series representation of $SL(2,\mathbb{R})$, the double cover of $SO(1,2)$. We use the group matrix elements in an abstract representation space to compute the Harish-Chandra character of the $SL(2,\mathbb{R})$ discrete series UIR with $\Delta = \frac32$, reproducing known results \citep{sugiura}.

\subsection{Classification of \texorpdfstring{$SL(2,\mathbb{R})$}{SL(2, R)} UIRs} \label{Subsec_classification of UIRs}

Despite not having a standard notion of spin in 2 dimensions, we refer to fermionic UIRs as those that have half-integer quantum numbers with respect to the maximally compact subgroup $SO(2)$ of $SO(1,2)$. To be precise, since we are interested in fermionic fields propagating in a fixed dS$_2$ spacetime one should consider the double cover of $G$, namely $SL(2,\mathbb{R})$.\footnote{It should be noted that at first sight the UIRs of $SL(2,\mathbb{R})$ could well be adapted to both AdS or dS, as in two-dimensions they share the same isometry group. Yet this is not correct. When discussing UIRs of the AdS$_2$ isometry group one unwraps the time direction, which corresponds to the Universal cover, $\widetilde{SL}(2,\mathbb{R})$ \citep{Kitaev:2017hnr}.} To construct the representations, we will work at the Lie algebra level, where the two algebras are isomorphic, $\mathfrak{so}(1,2) \simeq \mathfrak{sl}(2,\mathbb{R})$. This will allow us to obtain all the possible representations, both bosonic and fermionic. It should be understood that upon exponentiation, half-integer spin representations of the Lie algebra exponentiate to UIRs of the double cover $SL(2,\mathbb{R})$. 

\noindent We choose anti-symmetric generators $\hat{J}_{AB} = - \hat{J}_{BA}, A,B = 0,1,\cdots,d+1$ for the $\mathfrak{so}(1,d+1)$ Lie algebra obeying 
\begin{equation} 
    [\hat{J}_{AB}, \hat{J}_{CD}] = \eta_{BC} \hat{J}_{AD}- \eta_{AC}\hat{J}_{BD} + \eta_{AD} \hat{J}_{BC} - \eta_{BD} \hat{J}_{AC} \, , \label{lie algebra commutators}
\end{equation}
and we use $\, {\hat{}} \,$ to denote the abstract operators.  The quadratic Casimir is defined as 
\begin{equation}\label{def:Casimir_emb}
    \mathcal{C}_{2} \equiv - \frac{1}{2}\hat{J}_{AB}\hat{J}^{AB}  \, .
\end{equation}
A unitary representation of $\mathfrak{so}(1,d+1)$ requires the Lie algebra generators to be realised as anti-hermitian operators on some Hilbert space equipped with a positive definite scalar product, thus
\begin{equation} \label{dS_realitycond}
    \hat{J}^\dagger_{AB} = - \hat{J}_{AB} \, .
\end{equation}

\noindent We induce representations on $G$, or its double cover, from representations of the maximal compact subgroup. Restricting our attention to $d=1$ and working at the level of the Lie algebra, the maximal compact subalgebra $\mathfrak{so}(2)$ is generated by the compact generator $\hat{J}_{21}$, and according to~(\ref{lie algebra commutators}) we have:
\begin{align} \label{so(1,2) commutators clearly written anti-herm}
    [\hat{J}_{21}, \hat{J}_{01}] = \hat{J}_{02},~ ~~[\hat{J}_{21}, \hat{J}_{02}] =- \hat{J}_{01},~~~[\hat{J}_{01}, \hat{J}_{02}] =- \hat{J}_{21}.
\end{align}
Introducing the raising and lowering operators
\begin{align}\label{raising/lowering ops J basis}
    \hat{J}^{(\pm)}\equiv \hat{J}_{02}\pm i \hat{J}_{01},
\end{align}
the commutation relations~(\ref{so(1,2) commutators clearly written anti-herm}) are re-expressed as
\begin{align}
    [\hat{J}_{21}, \hat{J}^{(\pm)}] = \pm i\, \hat{J}^{(\pm)},~~~~~~[\hat{J}^{(+)} ,\hat{J}^{(-)}] =-2i \,\hat{J}_{21}.
\end{align}
The anti-hermitian generators, $\hat{J}_{21}, \hat{J}_{01}$ and $\hat{J}_{02}$, are related to the familiar hermitian generators, $\hat{L}_{0}, \hat{L}_{\pm 1}$, of the $\mathfrak{so}(1,2)$ subalgebra of the Virasoro algebra as follows:
\begin{align}
    \hat{J}_{21} =- i \hat{L}_{0},~~~~~~~\hat{J}^{(\pm)}=\hat{J}_{02} \pm i \hat{J}_{01} = -i \,\hat{L}_{\pm 1}.
\end{align}
Then, the commutation relations take the form
\begin{equation}
    [\hat{L}_n,\hat{L}_m] = (n-m)\hat{L}_{m+n} ~, 
\end{equation}
with $n,m = -1,0,1$. We label states in our representation as $\ket{\Delta, n}$ such that:
\begin{align}\label{def:Cas_groupEigen}
    &\mathcal{C}_2 \ket{\Delta,n} =\Delta (\Delta-1) \ket{\Delta,n}\, , \quad \hat{L}_0 \ket{\Delta, n} = -n\ket{\Delta, n}\, , \nonumber \\ &\hat{L}_{\pm 1} \ket{\Delta,n} = \sqrt{n(n \pm 1) - \Delta(\Delta -1)}  \ket{\Delta,n\pm1} \, ,
\end{align}
thus, $\Delta$ labels the quadratic Casimir of the corresponding representation and $n$ is an $\mathfrak{so}(2)$ quantum number.\footnote{Note that the states defined in this note $\ket{\Delta,n}$ have a different normalisation to the ones presented in recent works \citep{Sun:2021thf,Anninos:2023lin,Anous:2020nxu}. This allow us to make a direct identification with the mode solutions constructed below and agree with the normalisation of the original group theory works \citep{ottoson,schwarz}.} Our states are normalised with respect to a positive-definite and $\mathfrak{so}(1,2)$ invariant inner product as 
$$  \braket{\Delta , n | \Delta, n'} = \delta_{n,n'}  .$$

 Although at the level of the algebras,  $\mathfrak{so}(1,2)$ and $\mathfrak{sl}(2,\mathbb{R})$ are isomorphic, when exponentiating the generator of $\mathfrak{so}(2)$ compactness requires
\begin{equation}
\begin{split}
    e^{2\pi i \hat{L}_0} \ket{\Delta, n} &= \ket{\Delta, n} \, \textnormal{for } SO(1,2) \, , \\ 
    e^{2\pi i \hat{L}_0} \ket{\Delta, n} &= \pm \ket{\Delta, n} \, \textnormal{for } SL(2,\mathbb{R}).
\end{split}
\end{equation}
For $SO(1,2)$ the element $e^{2\pi i \hat{L}_0}$ corresponds to the identity operator and thus the $SO(2)$ eigenvalues are integers, $n \in \mathbb{Z}$. In contrast, for $SL(2,\mathbb{R})$ we clearly have $n \in \mathbb{Z}$ or $n \in \mathbb{Z}+\frac12$ for bosonic and fermionic representations respectively.

\noindent Imposing the anti-hermiticity condition \eqref{dS_realitycond} on the operators applied to the action of the ladder operators and restricting to states such that $\braket{n,\Delta \lvert n,\Delta} \geq 0$ one can see that only specific values of \eqref{def:Cas_groupEigen} will yield an UIR of $SL(2,\mathbb{R})$. These are summarised as follows \citep{sugiura, schwarz, ottoson, Thieleker1973, Thieleker1974}:

\begin{itemize} \label{reps}
    \item{\textbf{Principal series} $\pi_\nu$: Where $\Delta = \frac{1}{2} + i \nu$ with $\nu \in \mathbb{R}_+$}.  The $\mathfrak{so}(2)$ label takes values in  $n \in \mathbb{Z}$ or $n \in \mathbb{Z}+\frac12$ for the bosonic or fermionic representations, respectively.

    \item{\textbf{Complementary series} $\gamma_\Delta$: Where $\Delta \in (0,1)$} and the $\mathfrak{so}(2)$ label  takes only integer values ($n \in \mathbb{Z}$). Complementary series representations do not exist for the fermionic case \citep{sugiura, schwarz, ottoson, Thieleker1973, Thieleker1974}. 
    \item{\textbf{Discrete series $D_\Delta^\pm$:}  Where $\Delta \in \mathbb{Z}_+/2$. Half-integer and integer values correspond to bosonic and fermionic representations respectively. The $\mathfrak{so}(2)$ label then takes values $n \in \mathbb{Z}/2$ which are also (half-) integer in the (fermionic) bosonic case. These are the UIRs with an annihilating highest or lowest weight state. The range of $n$ is 
    \begin{align}
       & n \geq \Delta \geq \frac12 \qquad ~\textnormal{for } D_\Delta^+ \, , \label{condition_for_discrete_series_+}\\
        \qquad & n \leq -\Delta \leq -\frac12 \qquad\textnormal{for } D_\Delta^- \label{condition_for_discrete_series_-}.
    \end{align}}
\end{itemize}

\subsection{Discrete series Harish-Chandra characters for $SL(2 ,\mathbb{R})$}

Given the UIRs of a compact Lie group $G$, we can encode the information of a representation $R$ associated to some element $g \in G$ through the group character $\chi_R(g)$. For finite dimensional representations, the character can be defined unambiguously as the trace of $R(g)$ over the representation space. For infinite dimensional representations, such as the ones discussed in \cref{reps}, the definition of a \emph{trace} is more subtle. The general theory addressing this issue was studied by Harish-Chandra \citep{harish1,harish2,atiyah}. In the case of UIRs of $SO(1,d+1)$ or $Spin(1,d+1)$, one can appropriately define a trace class operator encoding the action of the group on a given representation space, and associate to it the \emph{Harish-Chandra character} of $R$ over the group element $g$.

\noindent From a physical point of view, the Harish-Chandra character plays a prominent role in the study of the one-loop contributions to the de Sitter entropy \citep{Anninos:2017eib}. It was shown that, in a suitable regularisation scheme, the Euclidean sphere path integral encodes the degrees of freedom associated with the Lorentzian UIRs through the corresponding Harish-Chandra character. In addition, for UIRs associated to spinning massive fields and (partially) massless spinning gauge fields, the Euclidean computation also contains novel contributions stemming from a co-dimension two surface. These were interpreted as \emph{edge mode} degrees of freedom, localized at the cosmological horizon of the de Sitter static patch \citep{Law:2023ohq,Grewal:2024emf}.\footnote{The fact that massive spinning fields also showcase the appearance of edge modes might seem confusing at first, as they are normally associated to the existence of a gauge symmetry. This can be further understood by noting that in order to compute the Euclidean path integral for such fields, one needs to formulate them a la Stuckelberg which in turns introduces a gauge symmetry, see \citep{Anninos:2020hfj}.} While fermionic tensor-spinor fields where not analysed in \citep{Anninos:2017eib}, the case of a spin$-\frac32$ field in a dS$_2$ spacetime was studied in \citep{Anninos:2023exn}; it was shown that after a suitable gauge fixing procedure, the sphere path integral also encodes the Harish-Chandra character of the $\Delta = \frac32$ UIR of the 2D dS group double cover, $SL(2, \mathbb{R})$. 

In arbitrary dimensions, the characters for bosonic representations were first computed in the classical papers \citep{hiraiI,hiraiII,hiraiIII} and were recently re-derived in \citep{Basile:2016aen} using Bernstein-Gelfand-Gelfand resolutions. As noted in \citep{Basile:2016aen}, these results disagree for even $d$ (odd spacetime dimensions) with the classical results. Furthermore, the characters computed in \citep{Anninos:2020hfj} through the Euclidean one-loop sphere path integral were shown to agree with \citep{hiraiI,hiraiII,hiraiIII}, while it was also pointed out that there is a mismatch with \citep{Basile:2016aen} for odd $d$ (even spacetime dimensions). As far as fermionic characters for $d > 1$ are concerned, Hirai states in \citep{hiraiI} that his methods for bosonic characters can be used to derive fermionic characters of $Spin(1,d+1)$. Moreover, the characters presented in \citep{Basile:2016aen} should hold for both fermionic and bosonic representations although the unitarity of the fermionic representations was not explicitly studied.\footnote{We would like to thank Thomas Basile for clarifying this.} 
For $d=1$, characters have much simpler expressions; these  $Spin(1,2) \cong SL(2,\mathbb{R})$ characters are known both for bosonic and fermionic representations - see, e.g.~\citep{sugiura}.

For the discrete series UIRs $D^\pm_\Delta$ (\ref{condition_for_discrete_series_-}) of $ SL(2,\mathbb{R})$, the characters are known to be \citep{sugiura}: 
\begin{align}\label{discrete series characters any Delta}
    \chi_{D^\pm_\Delta}(t) = \frac{e^{-\Delta |t|}}{1-e^{-|t|}},
\end{align}
with $\Delta$ being a positive integer or half-integer for bosonic or fermionic UIRs respectively. Such characters correspond to group elements of the form $g = e^{-t \hat{J}_{01}}$ with $\hat{J}_{01}$ a boost generator associated with the $\mathfrak{so}(1,1) \subset \mathfrak{so}(1,2)$ generator which in turn can be associated with the timelike Killing vector of the static patch of de Sitter.

{Given a state in an UIR, group matrix elements encode the action of the group on the state.} Matrix elements have been calculated for $SL(2,\mathbb{R})$ in \citep{Wilson}, as well as in \citep{Marolf:2008it,Marolf:2008hg,Higuchi:1991tm} in the context of group averaging in de Sitter space. In what follows, we review the construction of boost group matrix elements and use them to compute the corresponding character of $SL(2,\mathbb{R})$. {While the result is known, the explicit and detailed calculation of the infinite sum of diagonal boost matrix elements in order to compute the $\Delta = \frac32$ character appears here for the first time {and might be useful for efforts in computing the fermionic Harish-Chandra character of higher dimensional UIRs of $Spin(1,d+1)$.} 

The generators $\hat{J}_{21}, \hat{J}_{01},\hat{J}_{02}$ act according to \eqref{def:Cas_groupEigen}, and let us recall that we consider the basis on which $\hat{J}_{21}$ acts diagonally 
\begin{equation}
    \hat{J}_{21} \ket{\Delta,n} = i n \ket{\Delta,n} \, ,
\end{equation}
where $\ket{\Delta,n}$ are the states in a given UIR, see \eqref{def:Cas_groupEigen}. We define the group matrix element as 
\begin{equation}\label{def:MelementsMAIN}
    M_{n',n}^\Delta (t) \equiv \bra{\Delta,n'} e^{-t \hat{J}_{01}} \ket{\Delta,n} \, .
\end{equation}
Knowing the action of the generators on arbitrary states, allows us to obtain relations between different group matrix elements, which in turn gives a differential equation for $M_{n',n}^\Delta(t)$ allowing to obtain an explicit expression for the group matrix elements. See Appendix \ref{Appendix_matrix elements} for details on the derivation. 

Having obtained an explicit form for the group matrix elements, the character is then defined by 
\begin{equation}\label{def:Char_gen}
    \chi_R(t) = \sum_{n} \bra{\Delta,n} e^{-t \hat{J}_{01}} \ket{\Delta,n} \, = \sum_{n} M^\Delta_{n,n}(t) ,
\end{equation}
while the allowed values of $n$ depend on the UIR under consideration - for the discrete series $D_\Delta^\pm$ see \eqref{condition_for_discrete_series_+} and \eqref{condition_for_discrete_series_-}. Each matrix element $M^{\Delta}_{n,n'} (t)$ obeys a hypergeometric-type differential equation~(\ref{diff eqn matrix elements}), which takes the following form for the diagonal elements 
\begin{align}\label{diff eqn matrix elements n'=n}
  \left(  \frac{d^{2}}{  d   t^{2}}   + \coth{t}  \frac{d}{d  t}    - \frac{2n^{2}}{\sinh^{2}{t}}  (1- \cosh{t})  - \Delta (\Delta-1) \right)   M^\Delta_{n,n} (t) = 0\, .
\end{align}
Demanding regularity for $\cosh t = 1$ we obtain the solution 
\begin{equation}\label{solution for SL(2,R) matrix element}
    M^\Delta_{n,n}(t) = \frac{(1+\cosh t)^n}{2^n} {_2}F_1 \left(-\Delta + n + 1, \Delta + n ; 1 ; \frac{1-\cosh t}{2} \right) \, .
\end{equation}
Note that the solution~(\ref{solution for SL(2,R) matrix element}) is expressed using an integral representation for the Gauss hypergeometric function in \citep{Wilson}. Furthermore, for both fermionic and bosonic discrete series UIRs the $D_\Delta^+$ matrix elements \eqref{solution for SL(2,R) matrix element} are equal to the $D_\Delta^-$ matrix elements, this explains why both representations have the same character - see also appendix \ref{Appendix_matrix elements}. To compute the character one just needs to substitute~(\ref{solution for SL(2,R) matrix element}) into \eqref{def:Char_gen} and compute the sum. In particular, for the case of $\Delta = \frac32$ we have explicitly computed the sum in appendix \ref{Appendix_matrix elements}, as
\begin{equation}\label{Delta=3/2 character}
    \chi_{D_{\frac{3}{2}}^\pm}(t) = \sum_{n=3/2}^\infty M^{\frac{3}{2}}_{n,n}(t) = \frac{e^{-\frac{3}{2}\lvert t \lvert}}{1-e^{-\lvert t \lvert}} \,. 
\end{equation}
We readily see that (\ref{Delta=3/2 character}) coincides with \eqref{discrete series characters any Delta} as expected. We note that this character has appeared in the study of a supergravity theory defined on a two-dimensional de Sitter spacetime \citep{Anninos:2023exn}. For the case of $\mathcal{N} = 1$, upon computing the sphere path integral around the $S^2$ saddle, the fluctuations of the complexified gravitino field (integrated along a half-dimensional contour) yield two times \eqref{Delta=3/2 character} - see eq. 3.25 of \citep{Anninos:2023exn} in the limit  
$\beta \to 0$ which means $\nu \to i$. 

\section{A free field theory for discrete fermions}
\label{sec:classFT}

In this section we first briefly review the geometry of dS$_2$, and discuss the action of the isometries on scalar and spinor fields. Moreover, for later convenience, we discuss the Euclidean section of the dS$_{2}$ geometry. We then review the field theory realisation of the bosonic discrete series and fermionic principal series in 2D de Sitter QFT, referring the reader to \citep{Anninos:2023lin} for further (bosonic) details. Then in \cref{subsec:discreteferm} we put forward a free spinor field theory on a fixed dS$_2$ background whose classical mode solution space (of positive frequency) furnishes a direct sum of two fermionic discrete series UIRs. In particular, we first show that upon tuning the mass of the Dirac fermion to a specific purely imaginary value, the mode functions of the theory furnish two discrete series modules of $SL(2, \mathbb{R})$ the scaling dimension of which is related to the imaginary \emph{mass}. We note that in this case, the theory develops a set of gauge symmetries that we characterise, which are the 2-dimensional analogues of the shift symmetries discussed in \citep{Hinterbichler_fermions}.  

Upon quantisation such theories should furnish discrete UIRs in their single particle Hilbert space. However, the purely imaginary mass makes quantisation not trivial since now the hermiticity of the action is lost. Instead of moving forward with a non-standard quantisation procedure, we will propose, in section \ref{GTFermions}, that such discrete series UIRs should be understood as pieces of an underlying gauge theory. In this context, the discrete series UIRs manifest themselves after gauge-fixing, as in the scalar case discussed in \citep{Anninos:2023lin}.

\subsection{Geometry and isometries of \texorpdfstring{dS$_{2}$}{dS}}

Two-dimensional de Sitter spacetime can be represented as a Lorentzian hypersurface embedded in a three-dimensional ambient Minkowski spacetime, described by 
\begin{equation}\label{embedding3d}
    -\left( X^0 \right)^2 + \left( X^1 \right)^2 + \left( X^2 \right)^2 = \ell^2 \, , \qquad X^A = (X^0, X^1, X^2) \in \mathbb{R}^{1,2} \, ,
\end{equation}
where $\ell$ is the de Sitter radius. The metric of dS$_2$ is induced from the ambient space one 
\begin{equation}
    \textnormal{d}s^2 = \eta_{AB} \textnormal{d}X^A \textnormal{d}X^B \, ,
\end{equation}
where $\eta_{AB} = \textnormal{diag}(-,+,+)$ is the standard Minkowski metric.
From this point of view, the isometry group of dS$_2$, $SO(1,2)$, is manifest. In the embedding space \eqref{embedding3d} the generators of the group $\hat{J}_{AB}$ \eqref{lie algebra commutators}, when acting on fields, can be decomposed as 
\begin{equation} \label{def: so(1,2) gen embedding}
    J_{AB} = L_{AB} + \Sigma_{AB} \, ,
\end{equation}
where 
\begin{equation}
    L_{AB} = X_A \partial_B - X_B \partial_A \, , 
\end{equation}
with $\partial_A = \frac{\partial}{\partial X^A}$, is the orbital part and $S_{AB}$, which acts on the field indices, is called the spin part \citep{Dirac:1935zz}.  

Working on a given dS$_2$ slice with a choice of chart, the generators $\hat{J}_{AB}$ \eqref{lie algebra commutators} act on the classical fields as differential operators, namely the (spinor) Lie derivatives \citep{Ortin:2002qb,derivees} with respect to Killing vectors $\xi_{(AB)}$ of dS$_{2}$. Concretely, if $\xi$ is a Killing vector, then the generators are
\begin{equation}
    \pounds_{\xi} \phi = \xi^a \partial_a \phi \, ,  
\end{equation}
\begin{equation} \label{eq:Liederivs}
    \mathbb{L}_{{\xi}}{\Psi}  =  \xi^{\nu} \nabla_{\nu} {\Psi} + \frac{1}{4}  \nabla_{\kappa} \xi_{\lambda}  \gamma^{\kappa   \lambda}   {\Psi},
\end{equation}
where $\phi$ and $\Psi$ are scalar and spinor fields, respectively, on dS$_{2}$. The covariant derivative, $\nabla_\mu$, when acting on a fermionic field is understood to be defined in terms of the spin connection, see \eqref{def:covDspin}. This identification allows us to realise the Casimir \eqref{def:Casimir_emb} as a second-order differential operator acting on fields. For a scalar field one can readily see that  
\begin{equation}
    \mathcal{C} \mathcal{\phi}(\mathbf{x}) = \left(- \mathcal{\pounds}_{\xi_{(21)}} \mathcal{\pounds}_{\xi_{(21)}} + \mathcal{\pounds}_{\xi_{(01)}} \mathcal{\pounds}_{\xi_{(01)}} +  \mathcal{\pounds}_{\xi_{(02)}} \mathcal{\pounds}_{\xi_{(02)}}  \right)\phi(\mathbf{x}) = -\ell^2 \Box_{\textnormal{dS}} \phi(\mathbf{x}) \, ,
    \label{eq:Cas_scalar}
\end{equation}
whereas for a spinor we have 
\begin{equation}\label{eq: clas casimir spinors}
    \mathcal{C} \Psi(\mathbf{x}) = \left(- \mathbb{L}_{\xi_{(21)}} \mathbb{L}_{\xi_{(21)}} + \mathbb{L}_{\xi_{(01)}} \mathbb{L}_{\xi_{(01)}} +  \mathbb{L}_{\xi_{(02)}} \mathbb{L}_{\xi_{(02)}}  \right)\Psi(\mathbf{x}) =-\ell^2 \left(  \slashed{\nabla}^2 +\frac{1}{4 \ell^{2}} \right) \Psi(\mathbf{x}) \, , 
\end{equation}
where $\Box_{\textnormal{dS}}$ is the scalar Laplacian on de Sitter and $\slashed{\nabla}^2 = (\gamma^\mu \nabla_\mu)^2 = \nabla^{\mu}   \nabla_{\mu} - \frac{R}{4}$, and $R$ is the Ricci scalar $R = \frac{2}{\ell^{2}}$.

\noindent Correlation functions computed on the Bunch-Davies vaccuum state will be written in terms of de Sitter invariant quantities. For example, a two-point function between $X^A$ and $Y^B$ will be a function of the invariant distance given by
\begin{equation}\label{ds_inv_Gen}
    u_{xy} \equiv \frac{(X^A-Y^A)^2}{2\,\ell^2}  = 1 - \frac{1}{\ell^2} \eta_{AB} X^A Y^B \, ,
\end{equation}
where contractions are made with the ambient space metric $\eta_{AB}$.

\subsection*{Global coordinates}

This coordinate chart covers the full spacetime and corresponds to the choice:
\begin{equation} \label{glob_Emb}
    X^A = \ell \left( \sinh \tau \, , \cos\vartheta \cosh \tau\, , \sin\vartheta \cosh\tau \right) \, .
\end{equation}
The induced metric on dS$_2$ is then given by
\begin{equation} \label{global}
    \frac{\textnormal{d}s^2}{\ell^2} = -\textnormal{d}\tau^2 + \cosh^2 \tau \textnormal{d} \vartheta^2 \, , \qquad \tau \in \mathbb{R}\,, \qquad \vartheta \in [0,2 \pi ) \, .
\end{equation}
The invariant distance (\ref{ds_inv_Gen}) can be readily expressed as 
\begin{equation}\label{ds_inv_Glob}
u_{xy} = 1 + \sinh\tau \sinh\tau' - \cos\left( \vartheta - \vartheta' \right) \cosh\tau \cosh\tau' \, ,     
\end{equation}
where the points $\mathbf{x}$ and $\mathbf{y}$ in dS$_2$ are given by \eqref{glob_Emb} at $(\tau,\vartheta)$ and $(\tau',\vartheta')$ respectively.  In these coordinates, the three Killing vectors $\xi^\mu$ are given by
\begin{equation}\label{Killing vectors global coordinates}
\begin{split}
    \xi_{(21)} &= \frac{\partial}{\partial \vartheta} \, , \\ 
    \xi_{(02)} &= \cos\vartheta \frac{\partial}{\partial \tau} - \sin \vartheta \tanh \tau \frac{\partial}{\partial \vartheta} \, , \\ 
    \xi_{(01)} &= \sin \vartheta \frac{\partial }{\partial \tau} + \cos \vartheta \tanh \tau \frac{\partial}{\partial \vartheta} \, ,
\end{split}
\end{equation}
where the first one corresponds to the generator of the maximally compact subalgebra $\mathfrak{so}(2)$ and the remaining ones are boost generators. 

\subsection*{Conformal Compactification} 
When trying to understand the late-time behaviour of physical quantities, it is often convenient to work on a global patch with a compactified time coordinate 
\begin{equation} \label{eq:compactif}
    \frac{\dd s^2}{\ell^2} = \frac{-\dd T^2 + \dd \vartheta^2}{\sin T^2} \, , \qquad T \in (-\pi,0) \, ,
\end{equation}
which can be obtained from \eqref{glob_Emb} by the following identification 
\begin{equation}
    \cosh \tau = -\frac{1}{\sin T} \, . 
\end{equation}
It is straightforward to obtain the invariant distance, which is given by 
\begin{equation} \label{eq:ucompactified}
    u_{xy} = \frac{\cos\left( T - T' \right) - \cos \left( \vartheta - \vartheta' \right)}{\sin T \sin T'} \, ,
\end{equation}
and the Killing vectors are given by 
\begin{equation}
    \begin{split}
        \xi_{(21)} &= \partial_\vartheta \, , \\ 
        \xi_{(01)} &= -\left( \cos \vartheta \sin T \partial_T + \sin \vartheta \cos T \partial_\vartheta  \right) \, , \\ 
        \xi_{(02)} &= -\left( \sin \vartheta \sin T \partial_T - \cos \vartheta \cos T \partial_\vartheta \right) \, .
    \end{split}
\end{equation}

\subsection*{Euclidean section} 

The global patch \eqref{global} can be analytically continued to Euclidean de Sitter spacetime by the identification 
\begin{align} \label{eq: anal cont}
    \tau \to i \left(\varphi - \frac{\pi}{2}\right),
\end{align}
from which we obtain the $S^{2}$ line element 
\begin{equation}
    \frac{\dd s^2}{\ell^2} = \dd \varphi^2 + \sin^2 \varphi \dd \vartheta^2 \, , \qquad \varphi \in [0,\pi] \, .
    \label{eq:metricS2}
\end{equation}
Similarly, we obtain the invariant distance between two points on $S^{2}$
\begin{equation} \label{EdS_invDist}
    u^E_{xy} = 1 - \left( \cos(\vartheta_x - \vartheta_y) \sin\varphi_x \sin\varphi_x + \cos\varphi_x \cos \varphi_y \right) \, .
\end{equation}
Another useful invariant distance is given by the chordal distance between two points, or the $SO(3)$ invariant distance, which can be defined via 
\begin{equation} \label{def: geodesic distance S2}
    \cos \Theta_{xy} \equiv \frac{X \cdot Y}{\ell^2} = \cos(\vartheta_x - \vartheta_y) \sin\varphi_x \sin\varphi_x + \cos\varphi_x \cos \varphi_y \,  .
\end{equation}
This is closely related to \eqref{EdS_invDist} as can be seen from 
\begin{equation} \label{EdS_invDist altern}
    u^E_{xy} = 2 \sin^2 \frac{\Theta_{xy}}{2} \, . 
\end{equation}

\subsection{Discrete Series field theories in dS$_2$: A review}
It is well known that the single-particle Hilbert space of a free massive scalar field $\phi(\mathbf{x})$ (\ref{eq:scalar_box}) on dS$_2$ with squared mass $m^2 \ell^2 > \frac{1}{4}$ furnishes a principal series UIR.\footnote{For interesting discussions on field theories on dS$_2$ see also Refs. \citep{Higuchi:automorphic, Epstein:spinors}.} Similarly, the single-particle Hilbert space of a free massive spinor field carries the fermionic principal series UIR. The realisation of the discrete series in a quantum field theory is a more subtle question. In the bosonic case, the equations of motion with discrete series solutions were initially shown to appear as the classical modes in tachyonic scalar field theories \citep{Higuchi:1986wu} and upon quantisation in \citep{Bros:2010wa,Epstein:2014jaa}. An analysis of the Euclidean partition function shows that the presence of zero modes renders it ill defined, having to define a suitable gauge fixing procedure to remove them \citep{Anninos:2023exn}. On the other hand, the fermionic discrete series on dS$_2$ have not been discussed. There is still no proposal for a field theory whose Hilbert space would carry such UIRs upon quantisation.

\subsection*{Review: bosonic discrete series} 
\label{subsec:discbo}

Let us briefly review the case of the bosonic discrete series representations using the formalism of \citep{Anninos:2023lin}, to set the stage for the fermionic discussion. 

Consider a massive scalar field $\phi(\mathbf{x})$ whose equation of motion is given by 
\begin{equation} \label{eq:scalar_box}
    \Box_{\textnormal{dS}} \phi(\mathbf{x}) = m^2 \phi(\mathbf{x}) \, .
\end{equation}
Comparing \eqref{eq:Cas_scalar} and \eqref{def:Cas_groupEigen} we identify 
\begin{equation}
    \Delta(\Delta-1) = - m^2 \ell^2 \quad \Longrightarrow \quad \Delta = \frac{1}{2} \left( 1 \pm \sqrt{1-4m^2 \ell^2} \right) \, , 
\end{equation}
from where we see that a heavy scalar field $m^2 \ell^2 > \frac{1}{4}$ realises a principal series representation, while a light scalar field $0 < m^2 \ell^2 < \frac{1}{4}$ a complementary series one - see \ref{reps}. In order to obtain a discrete series representation with $\Delta = 1 + t$ and $t \in \mathbb{N}_0$ we immediately see that this requires a tachyonic mass given by
\begin{equation} \label{def:tachyon_mass}
   \ell^{2} \, m^2 = -t(t+1) \, .
\end{equation}
This creates some tension with the Wigner paradigm, as an UIR of the isometry group enforces a tachyonic field theory which normally implies  an instability of the theory. The mode decomposition and quantisation schemes for the fields satisfying the tachyonic wave equations were first considered in \citep{Bros:2010wa,Epstein:2014jaa}. In order to alleviate the tension with Wigner's paradigm, one would hope that it is possible to define a bona fide field theory whose equations of motion reduce to the tachyonic ones through some physically sensible procedure. 

The simplest case corresponds to a massless scalar field or a $\Delta = 1$ discrete UIR. In this case, the theory is  invariant under constant shifts of the field $\phi(\mathbf{x}) \to \phi(\mathbf{x}) + c$. It was shown in \citep{Anninos:2023lin} that gauging this symmetry provides a well defined theory for the $\Delta = 1$ field after completely fixing the gauge. From an Euclidean analysis, if the mass is given by \eqref{def:tachyon_mass} the path integral has zero-modes. Integration over these modes leads to divergences that need to be taken care off, to see this consider the Euclidean field theory whose action is 
\begin{equation} \label{eq:EuclS_scalar}
    S_E[\phi] = \frac{1}{2} \int_{S^2} \dd^2 x \sqrt{g} \phi(\Omega) \left[ -\Box_{S^2}+m^2 \right] \phi(\Omega) \, ,
\end{equation}
where $\Omega$ is a point on $S^2$ and $\Box_{S^2}$ is the standard Laplacian on the round sphere. As usual, to evaluate the corresponding path integral one needs to expand the fields in a complete basis of eigen-functions of the two-sphere Laplacian which correspond to the spherical harmonics, see \cref{appenidx:expansions} for details. 

For the purpose of our discussion, we can just analyse the two-point function computed via the Euclidean path integral 
\begin{equation}
    G_E^\Delta(\Omega, \Omega') = \frac{\int \mathcal{D}\phi\, \phi(\Omega) \phi(\Omega') e^{-S_E[\phi]}}{\int \mathcal{D}\phi \, e^{-S_E[\phi]}} \, .
\end{equation}
The Euclidean correlation function defined by this procedure is computed in the Euclidean vacuum and, upon analytic continuation to Lorentzian signature, gives the Hadamard and dS invariant Wightman function on dS$_2$. Exploiting the complete set of scalar spherical harmonics on $S^{2}$ one obtains 
\begin{equation}
    G_E^\Delta(\Omega,\Omega') = \sum_{L=0}^\infty \sum_{M=-L}^L \frac{Y_{LM}(\Omega) Y_{LM}(\Omega')}{L(L+1) + m^2 \ell^2} \, .
\end{equation}
In the case of discrete series \eqref{def:tachyon_mass} it is clear that we need to be careful with the modes satisfying 
\begin{equation}
    L(L+1)-t(t+1) = 0 \, ,
\end{equation}
namely the zero-modes of the differential operator appearing in the action \eqref{eq:EuclS_scalar}. In order to make sense of the discrete series two-point function, one can instead consider the following expression:
\begin{equation}
    G_E^{\Delta=1+t}(\Omega,\Omega') = \sum_{L\neq t} \sum_{M=-L}^L \frac{Y_{LM}(\Omega) Y_{LM}(\Omega')}{L(L+1) -t(t+1) } \, ,
\end{equation}
where we have removed the problematic zero-modes. In general, this procedure is a non-local deformation of our theory that needs a guiding principle. Ultimately, as advocated in \citep{Anninos:2023lin}, this involves defining a gauge theory with a suitable gauge fixing procedure that both removes the zero-modes and whose gauge fixed equations of motions correspond to the discrete series UIR. We note in passing that, as was recently shown, the discrete series scalars have a hidden global conformal symmetry \citep{Farnsworth:2024yeh}.

As a further example, the discrete series field equations on dS$_{2}$ with $\Delta = 1 + n, \, n \in \mathbb{N}$ were embedded in a topological $SL(N,\mathbb{R})$ BF theory in  \citep{Anninos:2023lin}. This case is an analogue of higher spin gravity in two dimensions discussed in \citep{Alkalaev:2013fsa,Alkalaev:2019xuv}. Here the discrete series appear at the level of the operators, while the Hilbert space can be shown to collapse onto a finite-dimensional space of states when the gauge constraints are taken into account. These models include the topological two-dimensional graviton for the case $t\geq 2$ in JT gravity, as well as the gauged massless scalar and the analogy to higher spin gravity in 2D of \citep{Alkalaev:2013fsa}.

\subsection*{Review: quantisation of real-mass fermions} 

Here we review the quantisation of a free Dirac fermion with real mass in dS$_2$ \citep{Schaub:2023scu,Pethybridge:2021rwf, Letsios:2020twa}. 
We show that upon quantisation the Hilbert space of single particle states transforms in a principal series UIR of $SL(2, \mathbb{R})$. We begin by considering a Dirac fermion $\Psi(\mathbf{x})$ defined on Lorentzian dS$_2$ with action given by
\begin{equation} \label{def:MassiveFermAction}
    S[\bar{\Psi}, \Psi] = -\int \dd^2 x \sqrt{-g} \, \bar{\Psi}(\mathbf{x}) \left( \slashed{\nabla} + m \right) \Psi(\mathbf{x}) \, , 
\end{equation}
where $m \in \mathbb{R}$, $\slashed{\nabla} = \gamma^\mu \nabla_\mu$, and Dirac conjugation is defined as: 
\begin{equation}
    \bar{\Psi}(\mathbf{x}) \equiv i \Psi^\dagger(\mathbf{x}) \gamma^0 \, .
\end{equation}
Our conventions for the gamma matrices can be found in \cref{App:fermions}. The equation of motion is readily found to be the Dirac equation
\begin{equation} \label{eq:PSeriesEom}
    \left( \slashed{\nabla} + m \right) \Psi(\mathbf{x}) = 0 \, .
\end{equation}
Using the field-theoretic expression for the Casimir \eqref{eq: clas casimir spinors} acting as a differential operator on spinors, yields
\begin{equation}
    \mathcal{C}_2 \Psi(\mathbf{x}) = \left(i \, \ell m + \frac12 \right)\left(i \, \ell  m - \frac{1}{2} \right) \Psi(\mathbf{x})  \, .
\end{equation}
Comparing with the representation-theoretic expression \eqref{def:Cas_groupEigen}, we find 
\begin{equation} \label{res:Casimir_PSeries}
    \Delta = \frac{1}{2} + i  \, \ell m \, ,
\end{equation}
where we see that the massive Dirac fermion furnishes a principal series representation. Furthermore, one could also obtain the shadow representation that corresponds to 
\begin{equation}
    \bar{\Delta} = 1-\Delta \, ,
\end{equation}
which is isomorphic to \eqref{res:Casimir_PSeries}. In other words, the theory defined by the action \eqref{def:MassiveFermAction} and the theory with the opposite sign for the mass are equivalent to each other.

In global coordinates, the Dirac equation is expressed as
\begin{align} \label{eq:KosherEOM}
  & \gamma^{0}\left( \frac{\partial}{\partial{\tau}}+\frac{1}{2}\tanh{\tau} \right) \Psi(\tau, \vartheta)+\gamma^1\frac{\partial_\vartheta\Psi(\tau , \vartheta)}{\cosh{\tau}}  = -\ell \, m \Psi(\tau, \vartheta) \, ,
 \end{align}
where we have used the conventions of Appendix \ref{App:fermions} for the zweibein, spin connection and gamma matrices.
It is known that the mode solutions can be obtained by a suitable analytic continuation of the Euclidean modes on $S^2$ \citep{Letsios:2020twa}. Instead, we work directly on dS$_{2}$ and consider the following Ansatz \citep{Camporesi:1995fb}
\begin{equation} \label{eq: Ansatz principal}
    \Psi_L^\pm(\tau,\vartheta) = \tilde{\Psi}_{L}^\pm(\tau)\frac{e^{ \pm i(L+\frac12) \vartheta}}{\sqrt{2\pi}} \, ,
\end{equation}
where we have expressed $\Psi_{L}^\pm(\tau,\vartheta)$ as a product of a time-dependent 2-component spinor and the spinor harmonics $e^{\pm i(L+\frac12) \vartheta}$ on $S^1$, where $L \in \mathbb{Z}_{\geq 0}$.\footnote{For details on spinor spherical harmonics see Appendix \ref{appenidx:expansions}.} Substituting this into \eqref{eq:KosherEOM}, one finds four families of mode functions; the first two are given by:
\begin{equation} \label{eq: pos freq modes principal}
    \Psi^{+}_L(\tau, \vartheta) = \frac{c_{L}}{\sqrt{2}} \begin{pmatrix}
    i\beta_{L}(\tau) \\
    - \alpha_{L}(\tau)
\end{pmatrix} \frac{e^{i(L+1/2) \vartheta}}{\sqrt{2\pi}} \, , \qquad \Psi^{-}_L(\tau,\vartheta) =\frac{c_{L}}{\sqrt{2}} \begin{pmatrix}
    \alpha_{L}(\tau) \\
    - i \beta_{L}(\tau)
\end{pmatrix}\frac{e^{-i(L+1/2) \vartheta}}{\sqrt{2\pi}} \, ,
\end{equation}
while the other two correspond to their charge conjugates
\begin{equation} \label{fermionCC}
   \left(\Psi_L^{\pm}(\tau, \vartheta)\right)^\mathcal{C}  = \gamma_{*}  \left(  \Psi^{\pm}_L(\tau,\vartheta) \right)^{*} ~.
\end{equation}
 The charge conjugation matrix $\gamma_{*} = (\gamma_{*})^{-1}$ satisfies
\begin{align}
    \gamma_{*}  \gamma^{a}  \left(\gamma_{*} \right)^{-1} = \left(  \gamma^{a}   \right)^{*}.
\end{align}
The explicit form of the functions $\alpha_L(\tau),\beta_L(\tau)$ is \citep{Camporesi:1995fb, Letsios:2020twa} 
\begin{equation}\label{eq: a(t) principal series function}
\begin{split}
    \alpha_L(\tau) = \Big( \cos{\left( \frac{\pi/2 - i \tau }{2}\right)}\Big) 
 ^{L+1} &   \Big(\sin{\left( \frac{\pi/2 - i \tau }{2}\right)} \Big)^L \\ 
    &\times {_2}F_1 \left(L+1+i\, \ell \,m, L+1-i \,\ell \,m; L+1 ;\zeta(\tau) \right) \, ,
\end{split}
\end{equation}
\begin{equation}
\begin{split}\label{eq: b(t) principal series function}
    \beta_L(\tau) = -\frac{i \, \ell \,m}{L+1} 
 \Big( \cos{\left( \frac{\pi/2 - i \tau }{2}\right)} \Big)^L &\Big( \sin{\left( \frac{\pi/2 - i \tau }{2}\right)} \Big)^{L+1} \\ 
    \times &{_2}F_1\left(L+1-i\,\ell \, m, L+1+i \, \ell \, m;L+2;\zeta(\tau) \right) \, , 
\end{split}
\end{equation}
where 
\begin{equation}\label{def:variable zeta(t)}
    \zeta(\tau) = \frac{1-i \sinh \tau}{2}= \frac{1- \cos{\left( \frac{\pi}{2} - i \tau  \right)}}{2}.
\end{equation}

The constant $c_{L}$ is a normalisation factor depending on the mass $m$ and angular momentum quantum number $L$. It is given by 
\begin{equation}\label{normln fact principal}
     c_{L}  = \frac{\left( \Gamma(L+1+ i \ell m) \,\Gamma(L+1- i \ell m) \right)^{1/2}}{ \Gamma(L+1)} \, ,
\end{equation}
and it ensures that the mode functions are normalised as~\citep{Letsios:2020twa}
\begin{equation}
    \left( \Psi_L^\sigma, \Psi_{L'}^{\sigma'} \right) = \left( \left(\Psi_L^\sigma \right)^{\mathcal{C}}, \left(\Psi_{L'}^{\sigma'} \right)^{\mathcal{C}} \right)   =\delta_{L L'} \delta_{\sigma \sigma'} \, , \quad \left( \Psi_L^\sigma, \left(\Psi_{L'}^{\sigma'} \right)^{\mathcal{C}}  \right) = 0 ,
\end{equation}
where $\sigma,\sigma' \in \{+,-\}$, $L,L' \in \mathbb{Z}_{\geq0}$, while we have denoted the standard Dirac inner product between two solutions $\Psi, \Psi^{'}$ as 
\begin{align} \label{def: Dirac inner prod}
    \left( \Psi , \Psi'  \right) = \int_{0}^{2 \pi} \,\dd \vartheta \sqrt{-g}  \, \left(\Psi(\tau , \vartheta)\right)^{\dagger}  \Psi' (\tau , \vartheta),
\end{align}
where $\sqrt{-g}= \cosh{t}$.
For any two solutions $\Psi, \Psi'$, the Dirac inner product is time-independent, dS invariant and positive definite.\footnote{See \citep{Letsios:2020twa} for a proof and further details.} Namely, the $SL(2, \mathbb{R})$ infinitesimal generators \eqref{eq:Liederivs} are realised as anti-hermitian operators 
\begin{align}
     \left( \mathbb{L}_{\xi} \Psi , \Psi'  \right) +\left( \Psi , \mathbb{L}_{\xi} \Psi'  \right)=0,
\end{align}
for any dS Killing vector $\xi$. According to the standard representation theoretic definition of unitarity \citep{BARUT}, the anti-hermiticity of the generators and the positive-definiteness of the Dirac inner product ensure that the  $SL(2, \mathbb{R})$ representation realised on the space of mode solutions is unitary.

The modes $\Psi_L^{\sigma}(\tau, \vartheta)$ exhibit Minkowskian behaviour in the large angular momentum $(L \gg 1)$ limit. This can be seen by expressing the functions (\ref{eq: a(t) principal series function}) and (\ref{eq: b(t) principal series function}) as
\begin{equation}\label{eq: a(t) principal series function altern}
\begin{split}
    \alpha_L(\tau) = \Big( \cos\left( \frac{\pi/2 - i \tau}{2} \Big) 
 \right)^{-L-1} &   \Big(\sin \left( \frac{\pi/2 - i \tau}{2}  \right)  \Big)^L \\ 
    &\times {_2}F_1 \left(i\, \ell \,m,-i \,\ell \,m; L+1 ;\zeta(\tau) \right) \, ,
\end{split}
\end{equation}
\begin{equation}
\begin{split}\label{eq: b(t) principal series function altern}
    \beta_L(\tau) = -\frac{i \, \ell \,m}{L+1} 
 \Big( \cos\left( \frac{\pi / 2 - i  \tau}{2} \right) \Big)^{-L} &\Big( \sin\left( \frac{\pi / 2 - i \tau}{2} \right) \Big)^{L+1} \\ 
    \times &{_2}F_1\left(1-i\,\ell \, m, 1+i \, \ell \, m;L+2;\zeta(\tau) \right) \, , 
\end{split}
\end{equation}
respectively, where we have made use of the following transformation formula \citep{NIST:DLMF}:
\begin{align}
    _{2}F_{1}\left( A,B;C;x  \right) =  (1-x)^{A-B-C}~ _{2}F_{1}\left( C-A,C-B;C;x  \right).
\end{align}
Then, in the $L \gg 1$ limit we find
\begin{equation} \label{pos freq behaviour}
    \frac{\partial}{\partial \tau} \Psi_L^{\sigma}(\tau, \vartheta) \sim - \frac{i L}{\cosh \tau} \Psi_L^{\sigma}(\tau, \vartheta) \, . 
\end{equation}
This corresponds to the generalised positive frequency condition used to define the Bunch-Davies vacuum on global de Sitter \citep{Birrell:1982ix,Letsios:2020twa}. Similarly, the charge conjugate modes \eqref{fermionCC} correspond to generalised negative frequency modes.

{Let us also demonstrate the dS invariance of the space of positive frequency modes. In other words, we will show that the infinitesimal dS transformations of positive frequency modes, $\mathbb{L}_{\xi_{(02)}} \Psi_L^{\sigma}$, $\mathbb{L}_{\xi_{(01)}} \Psi_L^{\sigma}$ and $\mathbb{L}_{\xi_{(21)}} \Psi_L^{\sigma}$, are expressed as linear combinations of only positive frequency modes. We find}\footnote{See \citep{Letsios:2020twa} for details of the computation for arbitrary spacetime dimensions. Note that there is a misprint in the coefficients in equation (C11) of \citep{Letsios:2020twa} - the correct expression is given by our eqs. (\ref{eq: spinor Lie (0,2) principal}) and (\ref{eq:coef_spinor_Lie02principal}). Note that our $L$ and $\ell m$ are denoted as $\ell$ and $M$, respectively, in \citep{Letsios:2020twa}.}
{\begin{align}
\mathbb{L}_{\xi_{(21)}}\Psi_L^{\sigma}  (\tau, \vartheta)=& \, \partial_{\vartheta} \Psi_L^{\sigma}(\tau, \vartheta)= {i}\,\sigma \,(L+1/2)\, \Psi_{L}^{\sigma}(\tau , \vartheta) ,\label{eq: spinor Lie (2,1) principal}
\end{align}
and
\begin{align*}
\mathbb{L}_{\xi_{(02)}}\Psi_L^{\sigma}(\tau, \vartheta)=&
\left(\xi^{\mu}_{(02)}\, \partial_{\mu} + \frac{\sin{\vartheta}}{2 \cosh{\tau}} \gamma^{1} \gamma^{0}\, \right)\Psi_{L}^{\sigma}(\tau, \vartheta),
\end{align*}
which gives
\begin{align}\label{eq: spinor Lie (0,2) principal}
\mathbb{L}_{\xi_{(02)}}\Psi_L^{\sigma}(\tau, \vartheta)= ~ -\frac{i}{2}A_{m,L} \, \Psi_{L+1}^{\sigma}(\tau, \vartheta) -\frac{i}{2}A_{m,L-1} (1 - \delta_{0L} )\, \Psi_{L-1}^{\sigma}(\tau, \vartheta) -\frac{i\,\ell |m|}{2} \delta_{0L}\, \Psi_{0}^{-\sigma}(\tau, \vartheta),
\end{align}
where the coefficient $A_{m,L}$ is given by
\begin{align}\label{eq:coef_spinor_Lie02principal}
    A_{m,L} = \sqrt{\left(L+\frac{1}{2} \right)  \left(L+\frac{3}{2} 
 \right)+\ell^{2}\,m^{2}+\frac{1}{4}  } = \sqrt{(L+1)^{2}  + \ell^{2} m^{2}}.
\end{align}
The spinor Lie derivative $\mathbb{L}_{\xi_{(01)}} \Psi_{L}^{\sigma}$ is simply determined as $ \mathbb{L}_{\xi_{(01)}} \Psi_{L}^{\sigma} = [\mathbb{L}_{\xi_{(02)}},   \mathbb{L}_{\xi_{(21)}}] \Psi_{L}^{\sigma} $:
\begin{align}\label{eq: spinor Lie (0,1) principal}
\mathbb{L}_{\xi_{(01)}}\Psi_L^{\sigma}(\tau, \vartheta)= -\frac{\sigma}{2}A_{m,L} \, \Psi_{L+1}^{\sigma}(\tau, \vartheta) +\frac{\sigma}{2}A_{m,L-1} (1 - \delta_{0L} )\, \Psi_{L-1}^{\sigma}(\tau, \vartheta) + \frac{\sigma\,\ell |m|}{2} \delta_{0L}\, \Psi_{0}^{-\sigma}(\tau, \vartheta).
\end{align}
Note that, in the last terms of eqs. (\ref{eq: spinor Lie (0,2) principal}) and (\ref{eq: spinor Lie (0,1) principal}), the sign of the $\sigma$ label in the mode functions is flipped - however, this term is non-vanishing only for $L=0$.
From equations (\ref{eq: spinor Lie (2,1) principal})-(\ref{eq: spinor Lie (0,1) principal}), it is clear that positive frequency modes transform among themselves under $sl(2, \mathbb{R}).$\footnote{A similar computation shows that negative frequency modes also transform among themselves under $sl(2, \mathbb{R})$.} Recalling that the spinor Lie derivatives $\mathbb{L}_{\xi_{(AB)}}$ correspond to the $sl(2,\mathbb{R})$ generators $\hat{J}_{AB}$ ($A,B \in \{0,1,2 \}$) acting on mode solutions, we identify the positive frequency modes $\Psi_{L}^{+}$ and $\Psi_{L}^{-}$ with the states $\Big|{\frac{1}{2} + i \, \ell m,  L+1/2 \Big> }$ and $\Big|{\frac{1}{2} + i \, \ell  m, -(L+1/2) \Big> }$, respectively. Then, comparing equations (\ref{eq: spinor Lie (2,1) principal})-(\ref{eq: spinor Lie (0,1) principal}) with (\ref{J21 on states abstract})-(\ref{J01 on states abstract}) it is clear that the positive frequency modes form the representation space for the principal series UIR of $sl(2,\mathbb{R})$ with $\Delta= 1/2 +i \, \ell m $, where the positive-definite inner product is given by the Dirac inner product (\ref{def: Dirac inner prod}).}

We quantise the theory using the mode functions presented above. The mode expansion for the Dirac quantum field $\Psi$ on global dS$_{2}$ is
\begin{equation}
\begin{split}
    \hat{\Psi}(\tau, \vartheta) =  \frac{1}{\sqrt{\ell}}\sum_{\sigma = \pm} \, \sum_{L=0}^\infty   &\left(\hat{a}^{\sigma}_{L}  \Psi^{\sigma}_{L}(\tau, \vartheta)  + \hat{b}^{\sigma \dagger  }_{L}    \left(\Psi^{\sigma}_{L}(\tau , \vartheta)\right)^{C} \right) \, .
\end{split}
\end{equation}
Following the standard canonical quantisation procedure, we find
\begin{equation}
\begin{split}
     \left \{ \hat{\Psi}(\tau,\vartheta)_\alpha, \hat{\Psi}^\dagger(\tau,\vartheta')^\beta  \right \} &= \frac{1}{\sqrt{-g}} \delta(\vartheta-\vartheta') \delta_\alpha{\hspace{0.1mm}}^\beta \, , \\ 
     \left \{ \hat{\Psi}(\tau,\vartheta)_\alpha, \hat{\Psi}(\tau,\vartheta')_\beta \right \} &= \left \{ \hat{\Psi}^\dagger(\tau,\vartheta)^\alpha, \hat{\Psi}^\dagger(\tau,\vartheta')^\beta \right \} = 0 \, ,
\end{split}   
\end{equation}
where $\alpha , \beta \in \{ 1,2 \}$ are spinor indices. Using the Dirac inner product (\ref{def: Dirac inner prod}) we find
\begin{align}
    \hat{a}^{\sigma}_{L} = \left(\Psi^{\sigma}_{L}, \hat{\Psi}   \right), ~~~~  \hat{b}^{\sigma  \dagger}_{L} = \left(\left(\Psi^{\sigma}_{L}\right)^{C}, \hat{\Psi}   \right)   ,
\end{align}
which leads to the standard anti-commutation relations for the annihilation and creation operators
\begin{equation}
    \left \{ \hat{a}^\sigma_L, \hat{a}^{\sigma' \dagger}_{L'} \right \} = \delta_{L L'} \delta_{\sigma \sigma'} \, , \qquad  
    \left \{ \hat{b}^\sigma_L, \hat{b}^{\sigma' \dagger }_{L'} \right \} = \delta_{L L'} \delta_{\sigma \sigma'} \, ,
\end{equation}
while all other anti-commutators are zero. The Bunch-Davies vacuum is defined by 
\begin{equation}
    \hat{a}_L^\sigma \ket{\Omega} = \hat{b}_L^\sigma \ket{\Omega} = 0 \, , 
\end{equation}
for all $\sigma$ and $L$. The dS invariance of this vacuum is ensured by the dS invariance of the space of positive frequency modes.

{The $sl(2 , \mathbb{R})$ generators are expressed in terms of creation and annihilation operators as
\begin{align}
  \hat{L}_{+1}= i \left(\hat{J}_{02} + i\hat{J}_{01}   \right)  =&  ~{\ell|m|} \,  \hat{a}^{+ \dagger}_{0}  \hat{a}^{-}_{0} +\sum_{L=0}^{\infty}   {A_{m,L}}\, \hat{a}^{+ \dagger}_{L+1} \hat{a}^{+}_{L} +\sum_{L=1}^{\infty}   {A_{m,L-1}}\, \hat{a}^{- \dagger}_{L-1} \hat{a}^{-}_{L} \nonumber \\
+& {\ell|m|} \,  \hat{b}^{+ \dagger}_{0}  \hat{b}^{-}_{0} +\sum_{L=0}^{\infty}   {A_{m,L}}\, \hat{b}^{+ \dagger}_{L+1} \hat{b}^{+}_{L} +\sum_{L=1}^{\infty}   {A_{m,L-1}}\, \hat{b}^{- \dagger}_{L-1} \hat{b}^{-}_{L}  
\end{align}
\begin{align}
  \hat{L}_{-1}  ={i}  \left(  \hat{J}_{02}  
 - i \hat{J}_{01} \right) =&  ~{\ell|m|} \,  \hat{a}^{- \dagger}_{0}  \hat{a}^{+}_{0} +\sum_{L=0}^{\infty}   {A_{m,L}}\, \hat{a}^{- \dagger}_{L+1} \hat{a}^{-}_{L} +\sum_{L=1}^{\infty}   {A_{m,L-1}}\, \hat{a}^{+ \dagger}_{L-1} \hat{a}^{+}_{L} \nonumber \\
+& {\ell|m|} \,  \hat{b}^{- \dagger}_{0}  \hat{b}^{+}_{0} +\sum_{L=0}^{\infty}   {A_{m,L}}\, \hat{b}^{- \dagger}_{L+1} \hat{b}^{-}_{L} +\sum_{L=1}^{\infty}  {A_{m,L-1}}\, \hat{b}^{+ \dagger}_{L-1} \hat{b}^{+}_{L}  
\end{align}
\begin{align}
 \hat{L}_{0}  =i \hat{J}_{21}= -\sum_{L=0}^{\infty}  \sum_{\sigma = \pm}  {\sigma} (L+1/2)\, \left( \hat{a}^{\sigma \dagger}_{L} \hat{a}^{\sigma}_{L}  +  \hat{b}^{\sigma \dagger}_{L} \hat{b}^{\sigma}_{L} \right). 
\end{align}
}
Finally, one can obtain  for the $sl(2,\mathbb{R})$ transformations of the creation operators:
\begin{equation}
    [\hat{L}_0,\hat{a}^{\sigma \, \dagger}_L ] =-  \sigma \left(L+\frac12  \right) \hat{a}^{\sigma \, \dagger}_L ,
\end{equation}
\begin{equation}
    [ \hat{L}_\pm ,  \hat{a}^{\sigma \, \dagger}_{L}  ] = \ell \lvert m \lvert \, \delta_{0 L} \delta_{ \mp, \sigma}~  \hat{a}^{\pm \, \dagger }_0 + A_{m,L} \delta_{\pm, \sigma} ~ \hat{a}^{\pm \, \dagger}_{L+1} + A_{m,L-1} \delta_{\mp, \sigma}~(1 - \delta_{L0}) ~\hat{a}^{\mp \, \dagger}_{L-1}    \, ,
\end{equation}
while similar expressions can be obtained for $  [\hat{L}_0,\hat{b}^{\sigma \, \dagger}_L ] $ and $ [ \hat{L}_\pm ,  \hat{b}^{\sigma \, \dagger}_{L}  ]$. Using these expressions, it is easy to check that the single-particle states $\hat{a}^{\sigma 
 \dagger}_{L}  \ket{\Omega}$ transform in the same way as the mode functions under $sl(2,\mathbb{R})$, and thus, they furnish the Principal Series UIRs, i.e. $\hat{a}^{\sigma 
 \dagger}_{L}  \ket{\Omega} \equiv \Big|\frac{1}{2} + i \, \ell  m, \sigma(L+1/2) \Big>  $.

\subsection{Discrete Series UIRs in the space of classical modes}
\label{subsec:discreteferm}

As discussed in the previous section, a massive Dirac fermion on dS$_2$ furnishes principal series UIRs of $SL(2, \mathbb{R})$. We would now like to obtain a classical field-theoretic realisation for the fermionic discrete series representations - see \cref{reps}. Let us start by making the following observation. The field-theoretic Casimir is given as a differential operator by \eqref{eq: clas casimir spinors}, while the representation-theoretic Casimir (\ref{def:Cas_groupEigen}) for the fermionic discrete series is parameterised by 
\begin{equation}
    \Delta = \frac{1}{2} + r \, , \quad r \in \mathbb{N} \, .
\end{equation}
Comparing with the result of the real mass Dirac field in \cref{res:Casimir_PSeries}, we identify 
\begin{equation} \label{res:fermion_imass}
   \ell \, m = \pm i r \, ,
\end{equation}
which corresponds to a non-standard \emph{purely imaginary} `mass' fermion. 

Before moving forward let us make a few comments regarding the appearance of an imaginary mass parameter. In the context of four-dimensional theories, similar values for the mass of a spin-$\frac32$ field were found in \citep{Pilch:1984aw} when trying to develop a supergravity theory for de Sitter spacetime. While at that point the group theory structure was, mostly unknown, the existence of a spin-$\frac32$ discrete series UIR of $Spin(1,4)$ was established in \citep{ottoson, schwarz}. Furthermore, building on this literature, recent attempts to realise this UIR in terms of a Rarita-Schwinger field in dS$_4$ also suggest an imaginary `mass' for the spin-3/2 gauge potential \citep{Letsios:partI,Letsios:partII}. In the context of two-dimensional theories, a $\Delta = \frac{3}{2}$ imaginary mass field was found in the fluctuations around a $S^2$ saddle point of a $\mathcal{N}=1$ super-Liouville theory \citep{Anninos:2023exn}. Interestingly, in this case, the discrete series field represents pieces of an underlying gauge theory that is subject to gauge fixing procedures and constraints that remove all propagating degrees of freedom. As was shown in \citep{Anninos:2023exn}, the gauge fixed Euclidean path integral of the theory is well define and it is possible to recover the $\Delta=\frac32$ Harish-Chandra character (\ref{Delta=3/2 character}) of the corresponding UIR.

Comparing with the bosonic case \eqref{def:tachyon_mass}, we see that the imaginary mass \eqref{res:fermion_imass} is just the fermionic counterpart of the tachyonic mass encountered before. Even more, in the bosonic case it is known that the fields enjoy a gauge (shift) symmetry. Fermions with the imaginary mass \eqref{res:fermion_imass} enjoy a similar gauge symmetry of the form 
\begin{equation} \label{eq:shift_symFerm}
    \Psi(\tau,\vartheta) \to \Psi(\tau,\vartheta) + \epsilon_{(r)}(\tau,\vartheta)  . 
\end{equation}
For $r=1$, the spinor parameters correspond to Killing spinors on dS$_2$, i.e they obey the standard Killing spinor equation 
\begin{equation}\label{eq:shift_symFerm r=1}
    \left( \nabla_\mu + \frac{i}{2\ell} \gamma_\mu \right)\epsilon_{(1)} = 0 \, .
\end{equation}
Similarly, for $r=2$, 
\begin{equation}\label{eq:shift_symFerm r=2}
    \left( \nabla_{(\mu} \nabla_{\nu)} + \frac{i}{\ell} \gamma_{{(\mu}}\nabla_{\nu)} + \frac{3}{4\ell^2} g_{\mu \nu} \right) \epsilon_{(2)} = 0 \, ,
\end{equation}
where round brackets denote symmetrisation as $a_{(\mu}  b_{\nu)} = \frac{1}{2}  (a_{\mu}  b_{\nu}   + a_{\nu}  b_{\mu})$. For any integer $r\geq 3$ the spinor parameter $\epsilon_{(r)}(\tau,\vartheta)$ satisfies higher-derivative generalisations of the previous equations. Although we will not give the explicit form of these equations, we will provide the explicit form of the spinors $\epsilon_{(r)}(\tau,\vartheta)$, $r \in \mathbb{N}$ - see also \citep{Hinterbichler_fermions}.

Shift-symmetric fermions with imaginary massess on dS$_{d+1}$, with $d+1 \geq 3$, have been studied in~\citep{Hinterbichler_fermions}, while the corresponding representations of $so(1,d+1)$ are non-unitary.\footnote{Interestingly, totally symmetric tensor spinors on $dS_{d+1}$ ($d+1 \geq 3$) with generic imaginary mass are unitary only for $d=3$ and only if the mass is tuned to the strictly or partially massless values \citep{Letsios:partI, Letsios:partII}.} One of the interesting features of the two-dimensional shift-symmetric fermionic theories under consideration is that they are unitary in dS$_{2}$. 

In what follows we construct the mode solutions for the fermionic discrete series representations of $SL(2,\mathbb{R})$ corresponding to \emph{purely imaginary} `mass' fermions \eqref{res:fermion_imass} obeying 
\begin{equation} \label{eq: imaginary mass Dirac eqn}
    \left( \slashed{\nabla} + i \frac{r}{\ell} \right) \Psi_{(r)}(\tau,\vartheta) = 0.
\end{equation}

\noindent We note that the physical modes of the theory with the opposite sign for the mass parameter, $ \left( \slashed{\nabla} - i \frac{r}{\ell} \right) \Psi_{(-r)} = 0$, furnish equivalent UIRs with the modes $\Psi_{(r)} $. This happens because the two sets of modes are related as $ \Psi_{(-r)} = \gamma_{*} \Psi_{(r)}$, while $ \gamma_{*}$ commutes with the spinor Lie derivative. Thus, the $SL(2 , \mathbb{R})$ generators act on the same way on both  $\Psi_{(r)} $ and $\Psi_{(-r)}$.

We show that the physical, \emph{i.e.} non-zero-norm, mode solutions  furnish the direct sum of two discrete series modules,
\begin{equation}\label{def:Deltadisc}
    D^+_\Delta \bigoplus D_\Delta^- \, , \qquad \Delta = r + \frac12 ,~~ r \in \mathbb{N}\, ,
\end{equation}
and we finally give explicit expressions for the spinors $\epsilon_{(r)}(\tau,\vartheta)$ which correspond to pure-gauge modes, zero-modes, of the theory. These pure-gauge modes can be considered as the `real part' of the full Lorentzian zero-modes, while the `imaginary part' of the full zero-modes will not be discussed here as it is not relevant for our representation-theoretic analysis. In fact,  the physical modes of the fermionic discrete series theories will be found to transform among themselves up to pure-gauge contributions of the form $\epsilon_{(r)}(\tau,\vartheta)$. However, for a full Lorentzian field-theoretic treatment, both the real and imaginary parts of the zero-modes must be taken into account. This situation is similar to the case of the shift-symmetric scalars - see, e.g.  \citep{Allen:vacuua, Guillito+Vas, Bonifacio:2018zex}.\footnote{The analyis of the massless (minimally coupled) scalar field in \citep{Allen:vacuua} concerns 4-dimensional de Sitter, but it shares many similarities with the 2-dimensional case. Also, details for the case of mode solutions of shift-symmetric scalars of any mass on dS$_{d}$ can be found in \citep{Guillito+Vas, Higuchi:1986wu}. In particular, in Appendix C of \citep{Guillito+Vas}, details can be found concerning the dS transformations of the physical modes, as well as the zero-modes - both their real and imaginary parts, called \emph{pure-gauge modes} and \emph{seed modes} respectively in \citep{Guillito+Vas}.}

To construct the mode solutions we use the same Ansatz as in the real-mass case \eqref{eq: Ansatz principal}, \emph{i.e.}:
\begin{equation}
    \Breve{\Psi}_{r, L}^{\pm}(\tau,\vartheta) =  \chi^{\pm}_r(\tau) \frac{e^{\pm i (L+\frac12)  \vartheta}}{\sqrt{2\pi}},
\end{equation}
where $L=0,1, \dots$, while the breve denotes that the modes are not normalised, the normalised modes will be denoted without a breve, \emph{i.e.} ${\Psi}_{r, L}^{\pm} $.
 
Working as in the real-mass case, we obtain the following two families of positive frequency mode solutions of \eqref{eq: imaginary mass Dirac eqn}: 
\begin{equation}\label{eq: pos freq discrete}
   \Breve{\Psi}^{+}_{r,L}(\tau, \vartheta)   =\begin{pmatrix}
    i\beta_{r,L}(\tau) \\
    - \alpha_{r,L}(\tau)
\end{pmatrix} \frac{e^{i(L+1/2) \vartheta}}{\sqrt{2\pi}} \, , \qquad  \Breve{\Psi}^{-}_{r,L}(\tau, \vartheta) =\begin{pmatrix}
    \alpha_{r,L}(\tau) \\
    - i \beta_{r, L}(\tau)
\end{pmatrix}\frac{e^{-i(L+1/2) \vartheta}}{\sqrt{2\pi}} \, ,
\end{equation}
where  $\alpha_{r,L}(\tau)$ and $ \beta_{r, L}(\tau) $ are given by eqs. (\ref{eq: a(t) principal series function}) and (\ref{eq: b(t) principal series function}), respectively, with $ \ell \, m = i r$, while $L= 0,1, \dots$. 

\noindent Moreover, since $r$ is a positive integer the hypergeometric series terminates, allowing us to rewrite the corresponding functions \eqref{eq: a(t) principal series function}, \eqref{eq: b(t) principal series function} as 
\begin{equation}
    \alpha_{r,L}(\tau) = \Big(\cos \left( \frac{\pi/2 -i \tau}{2}\right)\Big )^{-L-1}  ~ \Big(\sin\left( \frac{\pi/2   - i   \tau}{2} \right)  \Big)^L  \sum_{n=0}^r(-1)^n {r \choose n} \frac{(r)_n}{(L+1)_n} \left(\zeta(\tau)\right)^n \, ,
\end{equation}
\begin{align}
    \beta_{r,L}(\tau) =& \frac{r}{L+1} \Big(\cos\left( \frac{\pi / 2 - i  \tau}{2}\right)\Big)^{-L}~~\Big( \sin\left( \frac{\pi /2   -   i   \tau}{2} \right) \Big)^{L+1} \nonumber \\
    \times & \sum_{n=0}^{r-1}(-1)^n {r-1 \choose n} \frac{(r+1)_n}{(L+2)_n} \left(\zeta(\tau)\right)^n \, ,
\end{align}
where $(a)_{n} = \Gamma(a + n) / \Gamma(a)$ is the Pochhammer symbol, while $\zeta(\tau)$ is given by \cref{def:variable zeta(t)}. In the short wavelength limit ($L \gg 1$), the modes (\ref{eq: pos freq discrete}) satisfy the generalised positive frequency condition (\ref{pos freq behaviour}). 

\noindent  \textbf{Negative frequency.} Another set of solutions can be obtained by applying charge conjugation to the modes (\ref{eq: pos freq discrete}).  However, the charge conjugation matrix that must be used now is different from the one used in the real-mass or principal series case discussed earlier. The charge conjugation matrix $B_{-}$ that we have to use satisfies
\begin{align}
    B_{-} \gamma^{\mu}  B_{-}^{-1} = - \left(\gamma^{\mu}\right)^{*}
\end{align}
and the charge conjugate of the spinor $\Psi$ is $\Psi^{C} = B^{-1}_{-} \Psi^{*}$. In particular, if $\Psi$ is a solution of (\ref{eq: imaginary mass Dirac eqn}), then so is $\Psi^{C}$.\footnote{In 2 dimensions there are two charge conjugation matrices $B_{\pm }$ satisfying $B_{\pm} \gamma^{\mu} B^{-1}_{\pm} = \pm (\gamma^{\mu})^{*}$ \citep{Freedman:book}. The charge conjugation matrix $B_{+}$ has been used in \eqref{fermionCC} for the principal series case. In particular, if $\Psi$ is a solution of the Dirac equation (\ref{eq:PSeriesEom}), then so is $B_{+}^{-1} \Psi$. However, if $\Psi_{(r)}$ is a solution of \eqref{eq: imaginary mass Dirac eqn} with an imaginary mass parameter, it is easy to check that $B_{+}^{-1} \Psi_{(r)}$ satisfies the same equation with the opposite mass parameter, i.e. with $- i \frac{r}{\ell}$. Thus, charge conjugation defined in terms of $B_{+}$ does not preserve the solution space of the Dirac equation in this case. However, if one defines charge conjugation of $\Psi_{(r)}$ as $B_{-}^{-1} \Psi_{(r)}^{*}$, then this charge conjugate is a solution of (\ref{eq: imaginary mass Dirac eqn}) without flipping the sign of the imaginary mass parameter. }
Since our gamma matrices are imaginary, see (\ref{even_gammas}), we conclude that we can take $B_{-} = \mathbf{1}$, and thus, charge conjugation coincides with usual complex conjugation. Now, applying charge conjugation to the modes (\ref{eq: pos freq discrete}), we find another set of modes, as
\begin{equation}\label{eq: neg freq discrete}
   \left(\Breve{\Psi}^{+}_{r,L}(\tau, \vartheta)\right)^{*}   =-\begin{pmatrix}
   i \left(\beta_{r,L}(\tau)\right)^{*} \\
     \left(\alpha_{r,L}(\tau) \right)^{*}
\end{pmatrix} \frac{e^{-i(L+1/2) \vartheta}}{\sqrt{2\pi}} \, , \qquad  \left(\Breve{\Psi}^{-}_{r,L}(\tau, \vartheta)\right)^{*} =\begin{pmatrix}
    \left(\alpha_{r,L}(\tau) \right)^{*} \\
     i \left(\beta_{r, L}(\tau)\right)^{*}
\end{pmatrix}\frac{e^{i(L+1/2) \vartheta}}{\sqrt{2\pi}} \, .
\end{equation}
For $L \geq r$ these modes are independent of the positive frequency ones (\ref{eq: pos freq discrete}), while for short wavelengths $L \gg 1$ they satisfy the negative frequency analogue of (\ref{pos freq behaviour}). However, for $0 \leq L \leq r-1$ the modes (\ref{eq: neg freq discrete}) are not independent of (\ref{eq: pos freq discrete}). This can be easily checked for the $r=1$ case, where the functions describing the time-dependence are:
\begin{equation}
    \alpha_{r=1,L}(\tau) = \Big(\cos \left( \frac{\pi/2 -i \tau}{2}\right)\Big )^{-L-1}  ~ \Big(\sin\left( \frac{\pi/2   - i   \tau}{2} \right)  \Big)^L  \left(  1-\frac{\sin^{2}\left(\frac{\pi/2 - i \tau}{2} \right)}{L+1}    \right) \, ,
\end{equation}
\begin{align}
    \beta_{r=1,L}(\tau) =& \frac{1}{L+1} \Big(\cos\left( \frac{\pi / 2 - i  \tau}{2}\right)\Big)^{-L}~~\Big( \sin\left( \frac{\pi /2   -   i   \tau}{2} \right) \Big)^{L+1} ,
\end{align}
while, in particular for $L=r-1=0$, we have
$$  \beta_{r=1,0}(\tau)  =  \sin\left( \frac{\pi /2   -   i   \tau}{2} \right) = \left( \cos\left( \frac{\pi /2   -   i   \tau}{2} \right)  \right)^{*} = \left( \alpha_{r=1,0}(\tau)  \right)^{*},$$
and thus
\begin{equation}\label{neg freq depend on pos}
   \left(\Breve{\Psi}^{+}_{r=1,0}(\tau, \vartheta)\right)^{*}   =-i \, \Breve{\Psi}^{-}_{r=1,0}(\tau, \vartheta) \, , \qquad  \left(\Breve{\Psi}^{-}_{r=1,0}(\tau, \vartheta)\right)^{*} = -i\,\Breve{\Psi}^{+}_{r=1,0}(\tau, \vartheta) .
\end{equation}
These modes, as well as modes in the range $r \leq  L$ for general $r$, will be shown to be pure-gauge modes, \emph{i.e.} modes with zero norm.

\noindent \textbf{Physical and pure-gauge modes.} To understand which modes are physical and which pure-gauge we will calculate their norm using a dS invariant and time-independent  scalar product. Let $\Psi$ and $\Psi'$ be any two solutions of (\ref{eq: imaginary mass Dirac eqn}) with fixed imaginary mass $i \, r/ \ell$. With this specific mass parameter, the usual Dirac inner product (\ref{def: Dirac inner prod}) is neither dS invariant nor time-independent because, unlike the real-mass case, the Dirac current $J^{\mu}(\Psi, \Psi')= i \, \bar{\Psi} \gamma^{\mu} \Psi'$ is not conserved,
$$ \nabla_{\mu} \left( i \, \bar{\Psi} \gamma^{\mu} \Psi'  \right)  \neq 0 . $$
 However, now, the axial current is conserved,
$$J_{ax}^{\mu}(\Psi, \Psi')= i \, \bar{\Psi} \gamma^{\mu} \gamma_{*}\Psi',~~~\nabla_{\mu}J_{ax}^{\mu}(\Psi, \Psi')  = 0. $$ 
Thus, the axial scalar product 
\begin{align} \label{def: axial inner prod}
    \left( \Psi , \Psi'  \right)_{ax} &=  \int_{0}^{2 \pi} \,\dd \vartheta \sqrt{-g} ~J_{ax}^{0}(\Psi, \Psi') \nonumber \\
    &= \int_{0}^{2 \pi} \,\dd \vartheta \sqrt{-g}  \, \left(\Psi(\tau , \vartheta)\right)^{\dagger}  \, 
  \gamma_{*}\,\Psi' (\tau , \vartheta),
\end{align}
is both time-independent, $\partial_{t}\left( \Psi , \Psi'  \right)_{ax} = 0$, and dS invariant which means that the $sl(2, \mathbb{R})$ generators (Lie derivatives) are anti-hermitian\footnote{See \citep{Letsios:partI, Letsios:partII} for more details and the proof of dS invariance of the axial scalar product.}
\begin{align}
     \left( \mathbb{L}_{\xi} \Psi , \Psi'  \right)_{ax} +\left( \Psi , \mathbb{L}_{\xi} \Psi'  \right)_{ax}=0 \, .
\end{align}

Let us now compute the norm of the modes (\ref{eq: pos freq discrete}) and (\ref{eq: neg freq discrete}). A straightforward computation gives
\begin{align}
   & \left( \Breve{\Psi}^{\sigma}_{r,L} ,  \Breve{\Psi}^{\sigma'}_{r,L'}\right)_{ax} = (-\sigma) \left( |\alpha_{r,L}(\tau)|^{2}   -  |\beta_{r,L}(\tau)|^{2}    \right) \delta_{L L'}  \delta_{\sigma \sigma'}, \nonumber \\
   &\left( \left(\Breve{\Psi}^{\sigma}_{r,L} \right)^{*},  \left(\Breve{\Psi}^{\sigma'}_{r,L'}\right)^{*}\right)_{ax} = (-\sigma) \left( |\alpha_{r,L}(\tau)|^{2}   -  |\beta_{r,L}(\tau)|^{2}    \right) \delta_{L L'}  \delta_{\sigma \sigma'},   \\
   & \left( \Breve{\Psi}^{\sigma}_{r,L},  \left(\Breve{\Psi}^{\sigma'}_{r,L'}\right)^{*}\right)_{ax} = 0.
\end{align}
As the axial scalar product is $\tau$-independent, we can just let $\tau = 0$ to simplify the computation, and we find
\begin{align} \label{norms discrete series}
   & \left( \Breve{\Psi}^{\sigma}_{r,L} ,  \Breve{\Psi}^{\sigma'}_{r,L'}\right)_{ax} = (-\sigma) \frac{2\, |\Gamma(L+1)|^{2}}{\Gamma(L+1+r) \Gamma(L+1-r)} \delta_{L L'}  \delta_{\sigma \sigma'}, \nonumber \\
   &\left( \left(\Breve{\Psi}^{\sigma}_{r,L} \right)^{*},  \left(\Breve{\Psi}^{\sigma'}_{r,L'}\right)^{*}\right)_{ax} = (-\sigma) \frac{2\, |\Gamma(L+1)|^{2}}{\Gamma(L+1+r) \Gamma(L+1-r)}\delta_{L L'}  \delta_{\sigma \sigma'},  \nonumber  \\
   & \left( \Breve{\Psi}^{\sigma}_{r,L},  \left(\Breve{\Psi}^{\sigma'}_{r,L'}\right)^{*}\right)_{ax} = 0.
\end{align}
It is clear that the modes with $L \geq r$ have non-zero norm and we call them physical modes, while the modes with $0 \leq L \leq r-1$ have zero norm, and we call them pure-gauge modes. For $r=1 $ the pure-gauge modes are Killing spinors satisfying (\ref{eq:shift_symFerm r=1}), while for $r=2$ they satisfy (\ref{eq:shift_symFerm r=2}) and so forth. As the pure-gauge modes are orthogonal to themselves and to all physical modes we identify them with zero in the solution space. Thus, the physical modes form $sl(2, \mathbb{R})$ representations with the following equivalence relation: if for any two physical modes, $\Psi^{(1)}$ and $\Psi^{(2)}$, the difference $\Psi^{(1)} -\Psi^{(2)}$ is a linear combination of pure-gauge modes, then $\Psi^{(1)}$ and $\Psi^{(2)}$ belong to the same equivalence class.

Recall that the upper label `$\pm$' in the modes $\Breve{\Psi}^{\pm}_{r,L}(t, \vartheta)$ refers to the sign of the $\vartheta$-dependence of the spinor harmonics on $S^1$ which is associated to the $\frak{so}(2)$ generator $\partial_\vartheta$. We will then use the term `chirality', in analogy with the $4-$dimensional case, to distinguish between the two signs. 
 
From (\ref{norms discrete series}) we see that the physical positive  frequency modes $\Breve{\Psi}^{-}_{r,L}$  have positive norm, while the modes $\Breve{\Psi}^{+}_{r,L}$ have negative norm. The same happens in the case of the negative frequency modes
$\left(\Breve{\Psi}^{-}_{r,L} \right)^{*}$ and $\left( 
 \Breve{\Psi}^{+}_{r,L} \right)^{*}$. 
We see then that the sign of the norm depends on the `chirality', stemming from the appearance of $\gamma_{*}$ in the axial scalar product (\ref{def: axial inner prod}) as in the case of the 4-dimensional fermionic discrete series  modes \citep{Letsios:partI, Letsios:partII, Letsios:2023awz}. We now show that modes of definite `chirality' do not mix with each other, and thus, unitarity is ensured by a redefinition of the scalar product in the case of negative-norm modes. More specifically, we show that the positive frequency modes $\Breve{\Psi}^{-}_{r,L}$ and  $\Breve{\Psi}^{+}_{r,L}$ form the discrete series UIRs $D_{\Delta=r + \frac{1}{2}}^-$ and $D_{\Delta=r + \frac{1}{2}}^+$, respectively. Similarly, the negative frequency modes $\left(\Breve{\Psi}^{-}_{r,L}\right)^{*}$ and  $\left(\Breve{\Psi}^{+}_{r,L}  \right)^{*}$ form the discrete series UIRs $D_{\Delta=r + \frac{1}{2}}^+$ and $D_{\Delta=r + \frac{1}{2}}^-$, respectively.\footnote{Complex conjugation, which is equivalent to charge conjugation in the imaginary-mass case, as explained before, flips the `chirality'.} 

\noindent   \textbf{dS transformations of mode solutions.} Let us now examine the action of the $sl(2 , \mathbb{R})$ generators (spinor Lie derivatives) on the discrete series mode solutions. As in the principal series case (\ref{eq: spinor Lie (0,2) principal}), (\ref{eq: spinor Lie (2,1) principal}),\footnote{Note that our expressions for the principal series dS transformations (\ref{eq: spinor Lie (2,1) principal}), (\ref{eq: spinor Lie (0,1) principal}) refer to normalised mode solutions, while in (\ref{eq: spinor Lie (2,1) discrete}), (\ref{eq: spinor Lie (0,2) discrete L=0}) and (\ref{eq: spinor Lie (0,2) discrete L>=1}) we have un-normalised modes as indicated by the breve symbol - see eqs. (\ref{eq: pos freq discrete}) and (\ref{eq: neg freq discrete}).} we find
\begin{align}
\mathbb{L}_{\xi_{(21)}} \Breve{\Psi}_{r,L}^{\sigma}  (\tau, \vartheta)= \partial_{\vartheta}\Breve{\Psi}_{r,L}^{\sigma}  (\tau, \vartheta)  = {i}\,\sigma \,(L+1/2)\, \Breve{\Psi}_{r,L}^{\sigma}(\tau , \vartheta) ,~~L \geq 0, \label{eq: spinor Lie (2,1) discrete}
\end{align}

\begin{align}\label{eq: spinor Lie (0,2) discrete L=0}
\mathbb{L}_{\xi_{(02)}}\Breve{\Psi}_{r,L=0}^{\sigma}(\tau, \vartheta)= ~-\frac{i}{2}\frac{1-r^{2}}{L+1} \, \Breve{\Psi}_{r,L=1}^{\sigma}(\tau, \vartheta) + \frac{r}{2} \Breve{\Psi}_{r,L=0}^{-\sigma}(\tau, \vartheta),
\end{align}

\begin{align}\label{eq: spinor Lie (0,2) discrete L>=1}
\mathbb{L}_{\xi_{(02)}} \Breve{\Psi}_{r,L}^{\sigma}(\tau, \vartheta)=& ~ -\frac{i}{2}\frac{(L+1)^{2}-r^{2}}{L+1} \, \Breve{\Psi}_{r,L+1}^{\sigma}(\tau, \vartheta)-\frac{i}{2}L  \, \Breve{\Psi}_{r,L-1}^{\sigma}(\tau, \vartheta),~~L \geq 1,
\end{align}
where $\sigma = \pm$, while $\mathbb{L}_{\xi_{(01)}}$ follows directly from the Lie bracket between $\mathbb{L}_{\xi_{(02)}}$ and $\mathbb{L}_{\xi_{(21)}}$.
It is easy to see from these equations that pure-gauge modes ($0 \leq L \leq r-1$) transform only into pure-gauge modes, as they should. Now, let us normalise the physical modes ($L\geq r$) using (\ref{norms discrete series}), as
\begin{align}
    \Psi^{\text{phys}; \, \sigma}_{r,L} = \left( \frac{2\, |\Gamma(L+1)|^{2}}{\Gamma(L+1+r) \Gamma(L+1-r)} \right)^{-1/2} \Breve{\Psi}^{\sigma}_{r,L}, ~~~~~\text{for}~~~~~~L \geq r,
\end{align}
such that
\begin{align} \label{norms discrete series norm to 1}
   & \left( {\Psi}^{\text{phys}; \, \sigma}_{r,L} ,  {\Psi}^{\text{phys} ;\,\sigma'}_{r,L'}\right)_{ax} = (-\sigma) \delta_{L L'}  \delta_{\sigma \sigma'}, \nonumber \\
   &\left( \left({\Psi}^{\text{phys};\,\sigma}_{r,L} \right)^{*},  \left({\Psi}^{\text{phys};\sigma'}_{r,L'}\right)^{*}\right)_{ax} = (-\sigma) \delta_{L L'}  \delta_{\sigma \sigma'},  \nonumber  \\
   & \left( {\Psi}^{\text{phys}; \, \sigma}_{r,L},  \left({\Psi}^{\text{phys};\,\sigma'}_{r,L'}\right)^{*}\right)_{ax} = 0,~~~~~~~~~~~~~~~~~~~~~~~~~~L \geq r.
\end{align}
Then, using the transformations of the un-normalised physical modes (\ref{eq: spinor Lie (2,1) discrete}), (\ref{eq: spinor Lie (0,2) discrete L>=1}), we find

\begin{align}
\mathbb{L}_{\xi_{(21)}} {\Psi}_{r,L}^{\text{phys};\sigma}  (\tau, \vartheta)= {i}\,\sigma \,(L+1/2)\, {\Psi}_{r,L}^{\text{phys};\sigma}(\tau , \vartheta) ,~~L \geq r, 
\end{align}
\begin{align}
\mathbb{L}_{\xi_{(02)}} {\Psi}_{r,L}^{\text{phys}; \sigma}(\tau, \vartheta)=& ~ -\frac{i}{2}  \sqrt{(L+1)^{2}-r^{2}} \, {\Psi}_{r,L+1}^{\text{phys};\,\sigma}(\tau, \vartheta) \nonumber \\
&-\frac{i}{2} \sqrt{L^{2}-r^2}  \, {\Psi}_{r,L-1}^{\text{phys};\,\sigma}(\tau, \vartheta),~~\text{for}~~L \geq r+1,
\end{align}
\begin{align}
\mathbb{L}_{\xi_{(02)}} {\Psi}_{r,r}^{\text{phys}; \sigma}(\tau, \vartheta)=& ~ -\frac{i}{2}  \sqrt{(r+1)^{2}-r^{2}} \, {\Psi}_{r,r+1}^{\text{phys};\,\sigma}(\tau, \vartheta)+ (\text{pure-gauge}),~~\text{for}~~L=r
\end{align}
where $(\text{pure-gauge}) \propto \Breve{\Psi}_{r,r-1}^{\sigma}$ is a pure-gauge mode since its angular momentum quantum number is $L=r-1 \leq r$. As we mentioned earlier we can identify the pure-gauge modes with zero, and by doing so, we may write the transformation of all physical modes in a compact way as follows
\begin{align}
\mathbb{L}_{\xi_{(02)}} {\Psi}_{r,L}^{\text{phys}; \sigma}(\tau, \vartheta)=& ~ -\frac{i}{2}  \sqrt{(L+1)^{2}-r^{2}} \, {\Psi}_{r,L+1}^{\text{phys};\,\sigma}(\tau, \vartheta)-\frac{i}{2} \sqrt{L^{2}-r^2}  \, {\Psi}_{r,L-1}^{\text{phys};\,\sigma}(\tau, \vartheta),~~L \geq r.
\end{align}
It is clear that physical modes of definite `chirality' do not mix with each other under dS transformations; the modes $\{ {\Psi}_{r,L}^{\text{phys}; -}\}$  and $\{ {\Psi}_{r,L}^{\text{phys}; +}\}$ separately form irreducible representations of $sl(2, \mathbb{R})$. The representation furnished by  $\{ {\Psi}_{r,L}^{\text{phys}; -}\}$  is unitary  with the axial scalar product (\ref{def: axial inner prod}) being the positive-definite and dS invariant inner product. Similarly, the representation furnished by  $\{ {\Psi}_{r,L}^{\text{phys}; +}\}$  is unitary  with the \textbf{negative} of the axial scalar product (\ref{def: axial inner prod}) being the positive-definite and dS invariant inner product. Finally, introducing the scaling dimension $\Delta = r+1/2$ and defining $n_{+} \equiv L +1/2 \geq \Delta$ in the case of  $\{ {\Psi}_{r,L}^{\text{phys}; +}\}$ (see (\ref{condition_for_discrete_series_+})), and  $n_{-} \equiv -(L +1/2) \leq -\Delta$ in the case of $\{ {\Psi}_{r,L}^{\text{phys}; -}\}$ (see (\ref{condition_for_discrete_series_-})), we have
\begin{align}
\mathbb{L}_{\xi_{(21)}} {\Psi}_{r,L}^{\text{phys};\pm}  (\tau, \vartheta)= {i}\, n_{\pm}  \, {\Psi}_{r,L}^{\text{phys};\pm}(\tau , \vartheta) , 
\end{align}
\begin{align}
\mathbb{L}_{\xi_{(02)}} {\Psi}_{r,L}^{\text{phys}; \pm}(\tau, \vartheta)=& ~ -\frac{i}{2}  \sqrt{n_{\pm} (n_{\pm} + 1) - \Delta(\Delta-1)} \, {\Psi}_{r,L+1}^{\text{phys};\,\pm}(\tau, \vartheta)\nonumber\\
&  -\frac{i}{2} \sqrt{n_{\pm} (n_{\pm} - 1)-\Delta(\Delta-1)}  \, {\Psi}_{r,L-1}^{\text{phys};\,\pm}(\tau, \vartheta).
\end{align}
Comparing these equations with (\ref{J21 on states abstract}) and (\ref{J02 on states abstract}), and recalling that the discrete series conditions (\ref{condition_for_discrete_series_+}) and (\ref{condition_for_discrete_series_-}) are satisfied, we conclude that the physical modes 
$\{ {\Psi}_{r,L}^{\text{phys}; +}\}$  and $\{ {\Psi}_{r,L}^{\text{phys}; -}\}$ furnish a direct sum of discrete series UIRs, as shown in (\ref{def:Deltadisc}).


\section{Fermionic two-point functions}

In the previous section, we have seen that upon tuning the mass parameter to specific imaginary values, the classical mode solutions of the Dirac equation \eqref{eq:PSeriesEom} in a fixed dS$_2$ background furnish a direct sum of discrete series UIRs $D_\Delta^+   \bigoplus D_\Delta^+ $. Relative to the principal series case, these UIRs present several peculiar traits such as the need for a purely imaginary mass and the emergence of gauge symmetry. In this section, we focus on the structure of the free fermionic two-point function of such UIRs. We show that it is not possible to obtain a dS invariant two-point function obeying the Hadamard condition. We then analyse the problem in the Euclidean section and show that this is a consequence of the presence of zero-modes in the theory, we then propose a modification of the two-point function (and of the corresponding differential equation) that, upon analytic continuation to Lorentzian de Sitter, is both Hadamard and invariant under the de Sitter isometries. We propose then that sensible theory containing the discrete series fermions after gauge-fixing should reproduce the results of this section.

\subsection{Correlation function for principal series fermions}

Let us consider the free fermionic propagator for spinor fields with real mass parameters on dS$_{2}$, 
\begin{equation}
    S_{\textnormal{f}}(\mathbf{x},\mathbf{y}) \equiv \langle \Omega \lvert {\Psi}(\mathbf{x}) \, {\overline{\Psi}}(\mathbf{y}) \lvert \Omega \rangle \, ,
\end{equation}
where the index `f' stands for `free' , while $\ket{\Omega}$ is the  Bunch-Davies vacuum. The two-point function satisfies
\begin{equation} \label{2ptPseries}
    \left( \slashed{\nabla} + m \right) S_{\textnormal{f}}(\mathbf{x},\mathbf{y}) = 0 \,.
\end{equation}
 This is the Wightman two-point function, the corresponding advanced/retarded or Feynman ones can be obtained by a suitable modification of the right-hand side. The spinor two-point function can be computed using several standard methods \citep{Schaub:2023scu,Letsios:2020twa,Mueck:1999efk,Moschella:2018mxp}. Here, we will first review the embedding space formalism for principle series fermions, following the conventions of \citep{Pethybridge:2021rwf}. We will then re-apply this method to derive the discrete series two-point function, using the euclidean two point function to verify 

\noindent \textbf{Embedding space approach.} The differential equation \eqref{2ptPseries} defining $S_{\textnormal{f}}(\mathbf{x},\mathbf{y})$ can be written in terms of the de Sitter invariant distance $u_{xy}$ (\ref{ds_inv_Gen}),  while it is convenient to introduce the variable
\begin{equation}\label{def: define z=1-u/2}
    z =\frac{1}{2}\left(1+ \frac{X\cdot Y}{\ell^2}\right) = 1-\frac{u_{xy}}{2}~.
\end{equation}
The embedding space approach allows us to map a two-point function in the ambient space onto one on a dS$_2$ slice. Let {$\Psi^A(X)$ denote a spinor living on the 3-dimensional Minkowski embedding space, where the label $A$ has been used to denote that the fermion lives on ambient space. In order to define} a spinor on the $2-$dimensional dS slice we need to consider instead constrained spinors given by
{
\begin{equation} \label{constraintmain}
    X^M\Gamma_M \Psi^A= \Psi^A~.
\end{equation}
}
{The $\Gamma_A$'s are the Dirac matrices of the $3$-dimensional ambient space, and we define the uplift of $\gamma_*$ in 2-dimensions as $\Gamma_*$ see \citep{Pethybridge:2021rwf}.   
The constraint effectively maps this object to a single, transverse spinor in the de Sitter slice.}
Instead of working with the corresponding spinorial structures, we can introduce a dummy spinor $S$ in embedding space and construct a scalar object {$\tilde{\Psi}^A(X,\bar{S})$ defined as 
\begin{equation}
    \tilde{\Psi}^A(X,\bar{S}) \equiv \bar{S} \Psi^A(X) \, , 
\end{equation}}
now the constraint \eqref{constraintmain} translates to the dummy spinor $S$, see \cref{App:fermions} or \citep{Pethybridge:2021rwf} for details on the construction.

Then, the ambient space fermion two-point function can be obtained through the two-point function of these ambient space scalar objects. 
As we are only currently interested in the singular behaviour of the two-point function in terms of $z$, we do not need to project down to a particular coordinate parameterisation of the slice, however these projections into planar and global coordinates are available in \citep{Pethybridge:2021rwf, Schaub:2023scu}.  We now note that the quadratic Casimir \eqref{def:Casimir_emb} in the embedding space can be written as 
\begin{equation} \label{Cas_Emb}
    \mathcal{C} = -\frac{1}{2} \left( X_A \frac{\partial}{\partial X^B} - X_B \frac{\partial}{\partial X^A} + \bar{S} \Sigma_{AB}  \frac{\partial}{\partial\bar{S}}\right) \left( X^A\frac{\partial}{\partial X_B} - X^B \frac{\partial}{\partial X_A} + \bar{S} \Sigma^{AB}  \frac{\partial}{\partial\bar{S}} \right) \, ,
\end{equation}
where $\Sigma_{AB} = \frac{1}{4} [\Gamma_A,\Gamma_B]$. From this expression we recognise the $\mathfrak{so}(1,2)$ generators $J_{AB}$ (\ref{def: so(1,2) gen embedding}), while we also have a spinorial component given by $\Sigma_{AB}$ that acts on the corresponding spinor indices \citep{Dirac:1935zz}. We also stress that this form of the generators is due to the fact that we are working in the embedding space. If we were to work instead in de Sitter space we would have to utilise the spinor-Lie derivative with respect to the Killing vectors of dS$_{2}$ \eqref{eq:Liederivs}. 

The two-point function $S_{\textnormal{f}}(\mathbf{x},\mathbf{y})$ may be mapped directly onto the two-point function of these scalar objects as
{
\begin{equation}
    \braket{{\tilde{\Psi}}^A(X,\overline{S}_1)\,  {\overline{\tilde{\Psi}^A}}(Y,S_2)} = \bar{s}_1 S_f(x,y) s_2.
\end{equation}}
Here the $s_i$'s are dummy spinors on the dS$_{2}$ slice defined such that {$\bar{S_i} {{\Psi}^{A}}(X) = \bar{s_i} {\Psi}(x)$}. 
The most general two-point function of scalar objects in the embedding space, compatible with the symmetries of the problem, has a pair of structures
{
\begin{equation}\label{def:Structures}
    \braket{{\tilde{\Psi}}^A(X,\overline{S}_1)\,  {\overline{\tilde{\Psi}^A}}(Y,S_2)} = \overline{S}_1 S_2 \,g^\Delta_+(z) + \overline{S}_1\Gamma_*S_2 \,g^\Delta_-(z)~.
\end{equation}}
Here $\Delta$ parameterises the UIR of the fermion field as in \cref{reps}.   The first-order coupled equations of motion on the spinor components derived from the equation of motion may be expressed as a pair of decoupled second-order equations. These equations are precisely the same as the action of the Casimir \eqref{Cas_Emb} on the two-point function of \eqref{def:Structures} 
{
\begin{equation} \label{FermEOMDef}
    \mathcal{C}_X\braket{{\tilde{\Psi}}^A(X,\overline{S}_1)\,  {\overline{\tilde{\Psi}^A}}(Y,S_2)} = \Delta(\Delta-1)\braket{{\tilde{\Psi}}^A(X,\overline{S}_1)\,  {\overline{\tilde{\Psi}^A}}(Y,S_2)}~.
\end{equation}}
For the principal series the result is known to be, setting $d = 1$, 
\begin{equation}\label{eq: principal hyperg 2pt functn mass}
\begin{split}
    g_+\left( z \right) &= -\frac{\left| \Gamma\left(1 + i \ell m  \right) \right|^{2}}{16\pi} {_2}F_1\left(1 + i \ell m, 1 - i \ell m, 1, z\right) \, ,\\ 
    g_-\left( z \right) &= i \ell m \,\frac{\left| \Gamma\left(1 + i \ell m  \right) \right|^{2}}{16\pi} {_2}F_1\left(1 + i \ell m, 1 - i\ell m, 2, z\right) \, ,
\end{split}
\end{equation}
where to recover the result in terms of $\Delta$ and $\bar{\Delta} = 1-\Delta$ one just needs use \eqref{res:Casimir_PSeries}. The normalisation is chosen in such a way that we recover the standard flat space normalisation in the $z \to 1$ limit (i.e. $u_{xy} \to 0$). Furthermore, the solutions are chosen to only have singularities at the coincident point limit, \emph{i.e} we discard the solution with antipodal singularities as it is screened behind the cosmological horizon. These conditions define a Hadamard two-point function for the Principal Series fermion. 

The two-point function \eqref{def:Structures} is given in terms of geometric structures defined on the embedding space. One could have instead worked directly on the dS slice, as in \citep{Mueck:1999efk, Letsios:2020twa}. As we discuss below, the geometric structure appearing in this case is known as the spinor parallel propagator, which is the spinor analogue of the vector parallel propagator used in \citep{Allen:1985wd}.

\noindent \textbf{Euclidean approach.} Another way of computing the fermionic two-point function, which will be helpful when studying discrete series two-point functions, is through the Euclidean formulation of the theory, see \cref{appenidx:expansions} for technical details. The Euclidean action is
\begin{equation} \label{def:MassiveFermActionEuclidean}
    S_{E}[\bar{\Psi},\Psi] =  \int_{S^2} \dd^2 x \sqrt{g} \bar{\Psi}\left( \mathbf{x} \right) \left( \slashed{\nabla}+m \right)\Psi\left( \mathbf{x} \right) \, ,
\end{equation}
where $\Psi$ is a (Euclidean) Dirac spinor field on $S^{2}$ and $\bar{\Psi}$ is defined to be
\begin{equation}
    \bar{\Psi} = \Psi^\dagger \, ,
    \label{EuclideanDirac}
\end{equation}
and its the Euclidean Dirac conjugate of $\Psi$. One can then proceed to define the Euclidean two-point function 
\begin{equation} \label{def: Euclidean two-point princ Dirac}
    S_{\textnormal{f}}\left(\Omega,\Omega'\right) = \frac{\int \mathcal{D}\bar{\Psi}\mathcal{D}\Psi\, {\Psi}(\Omega) \bar{\Psi}(\Omega') e^{- S[\bar{\Psi},\Psi]}}{\int \mathcal{D}\bar{\Psi} \mathcal{D}\Psi \, e^{-S[\bar{\Psi},\Psi]}} \, ,
\end{equation}
where $\Omega,\Omega'$ denote two points on $S^2$.
The standard approach is to expand the fields in a complete basis of spinor spherical harmonics on $S^2$ given by $\psi^{(\pm)}_{\pm, N L}$, with Grassmann-valued coefficients, see \cref{appenidx:expansions}.  Evaluation of the Grassmann integrals in the free theory is straightforward and one obtains  
\begin{equation} \label{Sf_pSeries_modes}
    S_{\textnormal{f}}\left(\Omega, \Omega'\right) = \frac{1}{\ell} \sum_{N=0}^\infty \sum_{L=0}^N \sum_{\sigma = \pm} \left( \frac{ \psi_{+,NL}^{(\sigma)}\left(\Omega\right)\otimes\left( \psi_{+,NL}^{(\sigma)}\left(\Omega'\right) \right)^\dagger}{m \ell + i (N+1)} + \frac{ \psi_{-,NL}^{(\sigma)}\left(\Omega\right)\otimes\left( \psi_{-,NL}^{(\sigma)}\left(\Omega'\right) \right)^\dagger}{m \ell - i (N+1)}\right) \, ,
\end{equation}
where a point on $S^2$ is parameterized by two angles $\Omega = \left( \varphi, \vartheta \right)$, see \eqref{eq:metricS2}, while the spinor spherical harmonics $\psi_{\pm,NL}^{(\sigma)}$ are given by \eqref{eq: spinor eigenmodes S2}. Note that \eqref{Sf_pSeries_modes} is well-defined for all real values of the mass; including the massless fermion  which lies at the boundary between principal and discrete series UIRs, with $m = 0$, i.e. $\Delta = \frac12$.

In order to compute the two-point function \eqref{Sf_pSeries_modes}, one needs to perform the sum over the eigen-spinors on $S^2$. Contrary to the bosonic case, the spinor correlation function carries more structure that is encoded in the spinorial nature of \eqref{Sf_pSeries_modes}. This can be simplified by working in terms of purely geometric objects. As discussed before, on $S^{d+1}$ the geometric structure of the two-point function is encoded in the spinor parallel propagator. This object, $\Lambda(\Omega , \Omega ' )$, parallel transports a spinor from $\Omega$ to $\Omega'$ along the shortest geodesic connecting the two points \citep{Mueck:1999efk, Camporesi:1995fb,Letsios:2020twa}.  While we will not delve into the details of the geometric construction here, let us recall some useful definitions that we will need below. One can define at the point $\Omega$, a tangent vector $n^\mu$ to the shortest geodesic connecting the points $\Omega, \Omega'$ as 
\begin{equation}
    n_\mu =  \partial_\mu \Theta_{xy} \, , \qquad \mu \in \{\varphi,\vartheta\}\, .
\end{equation}
where $\Theta_{xy}$ is defined in \eqref{def: geodesic distance S2}. 
Note that the way we have defined $n_{\mu}$ is somehow unconventional as it has dimensions of $\ell^{-1}$ (with $n^{\mu} n_{\mu} = \ell^{-2}$), while it is defined to be dimensionless in \citep{Mueck:1999efk}. 
Furthermore, on $S^2$, the spinor two-point function is $\mathfrak{so}(3)$ invariant, and as such it will be written as a function of the $SO(3)$ invariant distance. We can thus, without loss of generality, let $\vartheta = \vartheta'$ and recover the general dependence on $\vartheta$, once the dust has settled, through \eqref{def: geodesic distance S2}. 

With this choice, the two-point function is independent of $\vartheta$ because of the explicit form of the spinor spherical harmonics (\ref{eq: spinor eigenmodes S2}). Furthermore, for $\vartheta = \vartheta'$, we have \citep{Camporesi:1995fb, Letsios:2020twa}
\begin{equation}
    \Lambda(\Omega , \Omega ' ) = \bm{1} \, , \qquad \Theta_{xy}= \varphi - \varphi' \, , \qquad n_{\mu} = (n_{\varphi} , n_{\vartheta}) = \ell^{-1}(1,0) \, .
\end{equation}
Making use of the explicit form of the spinor spherical harmonics ~(\ref{eq: spinor eigenmodes S2}), we can express the two-point function (\ref{Sf_pSeries_modes}) as 
\begin{equation}
    S_{\textnormal{f}}\left(\Omega, \Omega'\right) = \frac{1}{\ell} \sum_{N=0}^\infty \sum_{L=0}^N \frac{c_{NL}^2}{\ell^2 m^2 + \left(N+1 \right)^2} \left( \mathcal{A}_{NL} \left( \varphi,\varphi' \right) - \mathcal{B}_{NL}(\varphi,\varphi') {\ell \slashed{n}} \right) \bm{1} \, ,
\end{equation}
where {$\ell \slashed{n} = \ell \gamma^{\mu} n_{\mu} = \gamma^{\varphi} = \gamma^{2}$},  $c_{NL}$ are the normalisation factors of the spinor spherical harmonics (\ref{eq: normalistn const S2}), while
\begin{equation}
    \begin{split}
        \mathcal{A}_{NL}\left( \varphi, \varphi' \right) &= \ell m \left( \Phi_{NL}\left(\varphi\right)\Phi_{NL}\left( \varphi' \right) + \Psi_{NL}\left( \varphi \right) \Psi_{NL}\left(\varphi'\right) \right) \, , \\ 
        \mathcal{B}_{NL}\left( \varphi, \varphi' \right) &= (N+1) \left( \Phi_{NL}(\varphi) \Psi_{NL}(\varphi') - \Psi_{NL}(\varphi) \Phi(\varphi') \right) \, ,
    \end{split}
\end{equation}
and the explicit form of the functions $\Phi_{NL}(\varphi)$ and $\Psi_{NL}(\varphi)$ is given by (\ref{eq: spinor eigenmodes S2 Phi}) and (\ref{eq: spinor eigenmodes S2 Psi}), respectively.

We can further exploit the rotational symmetry of the problem by fixing one of the points to lie on the North pole of $S^{2}$, that is $\Omega' = \left( 0 , \vartheta \right)$. This will simplify the computation as only the $L=0$ terms will survive in the sum because of eqs. (\ref{eq: spinor eigenmodes S2 Phi varphi=0}) and (\ref{eq: spinor eigenmodes S2 Psi varphi=0}). We thus find
\begin{equation}
    \begin{split}
      S_{\textnormal{f}} & \left((\varphi, \vartheta), (0, \vartheta)\right)  \\
        &= \frac{1}{2\pi  \ell} \sum_{N=0}^\infty  \frac{N+1}{\ell^2 m^2 + \left(N+1 \right)^2} \left( \ell m \cos \frac{\varphi}{2} P_N^{(0,1)} \left( \cos \varphi \right) +  (N+1) \, \sin \frac{\varphi}{2}   \, P_N^{(1,0)}(\cos \varphi) ~{\ell \slashed{n}} \right) \bm{1} \, ,
    \end{split}
\end{equation}
where $P_{n}^{(a,b)}(x)$ are the Jacobi polynomials (\ref{def: Jacobi in terms of 2F1}). We recall now that we have fixed $\Omega = \left( \varphi, \vartheta \right)$ and $\Omega' = (0,\vartheta)$. Thus, the $SO(3)$ invariant distance \eqref{def: geodesic distance S2} can be read off, as well as the spinor parallel propagator, which for our current choice of points is $\Lambda(\Omega,\Omega') = \bm{1}$. We thus obtain
\begin{align}\label{eq:pre-final 2pt principal}
   S_{\textnormal{f}}\left(  \Omega, \Omega'\right) = &~ \frac{1}{2\pi \ell } \ell m\, \cos \frac{\Theta_{xy}}{2} ~\Bigg(\sum_{N=0}^\infty \frac{N+1}{\ell^2 m^2 + \left(N+1 \right)^2}  ~ {_2}F_1 \left(-N,N+2;1; \frac{1- \cos{\Theta_{xy}}}{2} \right) \Bigg) \, \Lambda(\Omega,\Omega') \nonumber  \\ 
    &+ \frac{1}{2\pi \ell}  \sin \frac{\Theta_{xy}}{2} \Bigg(\sum_{N=0}^{\infty} \frac{(N+1)^3}{\ell^2 m^2 + \left(N+1 \right)^2}\, {_2}F_1\left(-N,N+2;2;\frac{1- \cos{\Theta_{xy}}}{2}  \right)\Bigg) ~{ \ell \slashed{n}}  \Lambda(\Omega,\Omega') \, ,
\end{align}
for any two points $\Omega, \Omega'$, and\footnote{This is the Euclidean version of \eqref{def: define z=1-u/2}.}
\begin{equation}
    \frac{1-\cos \Theta_{xy}}{2}=1-z = \frac{u^{E}_{xy}}{2} \, ,
\end{equation}
while we have expressed the Jacobi polynomials in terms of the hypergeometric function \eqref{def: Jacobi in terms of 2F1}. One can identify the appearance of the two geometrical structures, $\slashed{n} \Lambda$ and $\Lambda$, in (\ref{eq:pre-final 2pt principal}) in agreement with the construction of \citep{Mueck:1999efk}. Evaluating the sums in (\ref{eq:pre-final 2pt principal}) numerically, we finally find   
\begin{equation}\label{eq:final 2pt principal}
\begin{split}
     S_{\textnormal{f}}\left(\Omega, \Omega'\right) = \frac{\lvert \Gamma(1+i\ell m) \lvert^2}{4\pi   {\ell}}  &\left[ \ell m \,\left( 1 - \frac{u^E_{xy}}{2} \right)^{1/2} {_2}F_1\left( 1+i\ell m, 1- i \ell m; 2; 1-\frac{u^E_{xy}}{2} \right) \Lambda\left( \Omega, \Omega' \right) \right. \\ 
     &\left. + \, \left( \frac{u^E_{xy}}{2} \right)^{1/2} {_2}F_1 \left(1+i\ell m, 1-i \ell m; 1; 1-\frac{u^E_{xy}}{2}\right) {\ell \slashed{n}} \Lambda(\Omega, \Omega') \right] \, ,
\end{split}
\end{equation}
where we have written the spinor two-point function in terms of the invariant distance $u^{E}_{xy}$ \eqref{EdS_invDist altern}. This is exactly the result in terms of intrinsic geometric objects given in \citep{Mueck:1999efk, Letsios:2020twa}.\footnote{Note that there the result presented in \citep{Mueck:1999efk} has a misprint corresponding to the form of the hypergeometric function multiplying $\slashed{n} \Lambda$ that has been corrected in \citep{Letsios:2020twa}.} Interestingly, comparing (\ref{eq:pre-final 2pt principal}) and (\ref{eq:final 2pt principal}) we obtain two conjectures, (\ref{Conjecture 1}) and (\ref{Conjecture 2}), for closed-form expressions of series involving hypergeometric functions - see \cref{App:2ptdetails} for details.

We have demonstrated that the Euclidean path integral computation of the spinor principal series two-point function agrees with the construction in terms of intrinsic geometric objects. The expression for the  principal series 2-point function in (\ref{eq:final 2pt principal}) coincides with the projection of the embedding space 2-point function \eqref{def:Structures} on the $dS_{2}$ slice. One can re-introduce the scaling dimension of the representation in (\ref{eq:final 2pt principal}) via \eqref{res:Casimir_PSeries} which also makes the $\Delta \to 1-\Delta$ invariance explicit. Furthermore, the Lorentzian two-point function at space-like separated points can be obtained by making the analytic continuation of the invariant distance, namely $u^E_{xy} \to u^L_{xy}$, which in turn allows us to discuss the late-time behaviour of such correlation function. 

\subsection{Correlation function for discrete series fermions}

We now turn our attention to the discrete series UIRs \eqref{def:Deltadisc}, where $\Delta = r + \frac12$. The second-order differential equations obtained from \eqref{FermEOMDef} are 
\begin{equation}
  z (z-1) 
   \partial_z^2 g^r_\pm(z)+ \left(3 z-\frac{3 {\mp 1}}{2}\right) \partial_z g^r_\pm(z)-\left(    r^2{-}1\right) g^r_\pm(z) = 0~ ,
\end{equation}
where we now denote the solutions as $g^r_\pm$. Again, the general solution to this differential equation is given in terms of a hypergeometric polynomial and the Meijer-G function
\begin{equation} \label{singstructureDS}
  g^r_\pm (z) =  \alpha \, {_2}F_1\left({1-r,1+r;1;z}\right) +\beta ~G_{2,2}^{2,0}\left(z\left|
\begin{array}{c}
 {-r,r} \\
 \tfrac{-1{\mp}1}{2},0 \\
\end{array}
\right.\right)~.
\end{equation}
The physiscal two-point function will be given by a specific combination of $\alpha,\beta$ in such a way that the resulting correlation function is Hadamard. But, just as in the bosonic case \citep{Anninos:2023lin}, it is impossible to find $\alpha, \beta$ for which the correlation function includes the coincident point singularity of a local quantum field theory, and evades the antipodal singularity; the Meijer-G function has a logarithmic singularity at both $z=0,1$, while the hypergeometric function, being a polynomial on the arguments, has none. This is a sign the equation of motion has missed an important contribution, the analogue of the contribution given by removing the Euclidean zero-modes in the bosonic case.

\noindent \textbf{Euclidean approach.} To define a suitable discrete series two-point function with the right singularity structure, we will exploit the Euclidean formulation of the theory. As discussed in detail in the previous section, the discrete series UIRs arise at special imaginary tunings of the mass parameter \eqref{res:fermion_imass}. One can define the fermionic discrete series two-point function as in the case of the principal series Euclidean two-point function (\ref{Sf_pSeries_modes}), while now we choose to omit the zero-modes from the mode-sum, as:
\begin{equation} \label{Sf_DSeries_diracTRUE}
\begin{split}
     H^{\Delta = r+\frac12}_{\textnormal{f}}\left(\Omega, \Omega'\right) &= \frac{1}{\ell} \sum_{N\neq r-1} \sum_{L=0}^N \sum_{\sigma = \pm}  \frac{ \psi_{+,NL}^{(\sigma)}\left(\Omega\right)\otimes\left( \psi_{+,NL}^{(\sigma)}\left(\Omega'\right) \right)^\dagger}{i r + i (N+1)} \\ 
     &+\frac{1}{\ell} \sum_{N\neq r-1} \sum_{L=0}^N \sum_{\sigma = \pm} \frac{ \psi_{-,NL}^{(\sigma)}\left(\Omega\right)\otimes\left( \psi_{-,NL}^{(\sigma)}\left(\Omega'\right) \right)^\dagger}{ ir - i (N+1)} \, .
\end{split}
\end{equation}
As in the case of the Principal Series UIRs, the two-point function is the same for both Dirac and Majorana spinors.\footnote{Instead of considering a Dirac spinor we could define a (Euclidean) Majorana spinor and compute the corresponding correlation function, but it produces the same result for the 2-point function as the Euclidean Dirac spinor.} To proceed we act with $\left(\slashed{\nabla} +  i \frac{r}{\ell}\right) $ and we find 
\begin{equation}
\begin{split}
  {\left( \slashed{\nabla} + i \frac{r}{\ell} \right)H^{\Delta = r+\frac12}_{\textnormal{f}}\left(\Omega, \Omega'\right) = \frac{1}{\ell^2} \frac{\delta\left( \Omega, \Omega' \right)}{\sqrt{g}}} 
  &- {\frac{1}{\ell^2}\sum_{L=0}^{r-1}\sum_{\sigma = \pm} \sum_{P = \pm}\psi^{(\sigma)}_{P, (r-1) L}(\Omega) \, \otimes \left(\psi^{(\sigma)}_{P, (r-1) L}(\Omega')\right)^\dagger  } \, ,
\end{split}
\end{equation}
where we have used the completeness relation (\ref{completeness spinors}) of the spinor harmonics on $S^2$. Then, we can perform the sum of the spinor harmonics \eqref{app:spinor_addition_pm} obtaining  
\begin{equation}
\begin{split}
     \left( \ell \slashed{\nabla} + i r \right)H_{\textnormal{f}}^{\Delta = r + \frac12} \left( \Omega, \Omega' \right) = \frac{1}{\ell} \frac{\delta\left( \Omega, \Omega' \right)}{\sqrt{g}} 
     - \frac{r}{2\pi \ell}  \cos \frac{\Theta_{xy}}{2} P_{r-1}^{(0,1)}\left( \cos \Theta_{xy}\right)   \Lambda(\Omega, \Omega').
\end{split}
\label{discrete2ptdif}
\end{equation}
As in the bosonic case, we see that now the two-point function for the fermionic discrete series obeys an inhomogeneous differential equation. To construct the solution, let us make the following Ansatz 
\begin{equation}
    H_{\textnormal{f}}^{\Delta = r+\frac12}(\Omega,\Omega') = \left[ f_+^r(\Theta_{xy}) \bm{1} + f^r_-(\Theta_{xy})\, {\ell \slashed{n}} \right] \Lambda(\Omega, \Omega') \, ,
\end{equation}
and directly substitute in \eqref{discrete2ptdif}. Then, using the following relations \citep{Mueck:1999efk}:
\begin{equation}
    \slashed{\nabla} \Lambda(\Omega,\Omega') = -\frac12 \tan \frac{\Theta_{xy}}{2} \slashed{n} \Lambda(\Omega,\Omega')\, , \quad  \slashed{\nabla} \slashed{n} = {\frac{1}{\ell^{2}}} \cot \Theta_{xy}  \bm{1} \, ,
\end{equation}
we obtain a system of coupled differential equations 
\begin{align}
      \frac{d}{d\Theta_{xy}}f^{r}_{-}\left( \Theta_{xy} \right) + \frac12 \cot{\frac{\Theta_{xy}}{2}}  f^{r}_{-}\left( \Theta_{xy} \right) + i r  f^{r}_{+}\left( \Theta_{xy} \right) = \frac{\delta(\Omega - \Omega')}{\ell \sqrt{g}}-\frac{r}{ 2 \pi \ell}\cos{\frac{\Theta_{xy}}{2}}  P_{r-1}^{(0,1)}(\cos{\Theta_{xy}}) 
\end{align}
\begin{align}
       \frac{d}{d\Theta_{xy}}f^{r}_{+}\left( \Theta_{xy} \right) - \frac12 \tan{\frac{\Theta_{xy}}{2}}f^{r}_{+}\left( \Theta_{xy} \right) + i r  f^{r}_{-}\left( \Theta_{xy} \right) =0 \, .
\end{align}
Introducing the variable $z$ (\ref{def: define z=1-u/2}), which write here again for convenience,
\begin{equation}
    z = \cos^2\left( \frac{\Theta_{xy}}{2} \right) = \frac{1}{2} \left(1+ \cos\left( \Theta_{xy} \right) \right) \, ,
\end{equation}
we re-write the system as
\begin{equation}
    \begin{split}
        \sqrt{z(1-z)}\frac{\dd}{\dd z} f^r_+(z) + \frac{\sqrt{1-z}}{2\sqrt{z}} f^r_+(z) - i r f^r_-(z) &= 0 \\ 
        \sqrt{z(1-z)} \frac{\dd}{\dd z} f^r_-(z) - \frac{\sqrt{z}}{2\sqrt{1-z}} f^r_-(z) - i r f^r_+(z) &=- \frac{\delta(\Omega - \Omega')}{\ell \sqrt{g}} + \frac{r}{2\pi \ell} \sqrt{z} P_{r-1}^{(0,1)} (2z-1) \, ,        
    \end{split}
\end{equation}
from which we  obtain 
\begin{align}
    \frac{i}{ r } \Big[ z(1-z) \frac{\dd^2}{\dd z^2} + (1-2z) \frac{\dd}{\dd z} +& \frac{1}{4z}(-1+(4r^2-1) z) \Big]  f^r_+(z)\nonumber \\
    &= \frac{\delta(\Omega - \Omega')}{\ell \sqrt{g}}- \frac{r}{2\pi \ell} \sqrt{z} P_{r-1}^{(0,1)}(2z-1) \, ,
\end{align}
and we will proceed by considering $\Omega \neq \Omega'$.
Here we will present the solution for $r=1$, but it is straightforward to find the solutions for other values of $r$. In particular, for  $r=1$ ($\Delta = \frac32$), the second-order equation becomes 
\begin{align}
   \left( z(1-z) \frac{\dd^{2}}{\dd z^{2}}   + (1-2z) \frac{\dd}{\dd z} + \frac{3z-1}{4z} \right) f^{r=1}_{+}(z) = i \sqrt{z}   \frac{1}{2 \pi \ell},
\end{align} 
and the solution can be found to be 
\begin{equation}
    f^{r=1}_+\left(z; \alpha  \right) = \frac{\alpha}{\ell} \sqrt{z} - \frac{i}{4\pi \ell} \sqrt{z}\log(1-z) \, .
\end{equation}
{The solution has only the coincident point singularity for $z \to 1$, to achieve this we have set the coefficient of the term with the antipodal singularity to $0$.} We note that $\alpha$ is an arbitrary parameter proportional to a homogeneous solution which can be traced back to the ambiguity in the process of removing the corresponding zero-mode. We can now obtain  $f^{r=1}_-(z)$ as
\begin{equation}
    f^{r=1}_-(z;\alpha) = \frac{z}{4\pi\ell \sqrt{1-z}} - \frac{\sqrt{1-z}\log(1-z)}{4\pi \ell} - i \frac{\alpha}{\ell} \sqrt{1-z} \, .
\end{equation}
Finally, the  $\Delta = \frac{3}{2}$ discrete series two-point function is given by
\begin{equation}
\begin{split}
    H_{\textnormal{f}}^{\Delta = \frac32}(z;\alpha) =\frac{1}{\ell} &\left[ \left( \alpha \sqrt{z} - \frac{i}{4\pi} \sqrt{z}\log(1-z) \right) \bm{1}  \right. \\ 
    &+ \left. \left( \frac{z}{4\pi \sqrt{1-z}} - \frac{\sqrt{1-z}\log(1-z)}{4\pi} - i \alpha \sqrt{1-z} \right)\, {\ell \slashed{n}} \right] \Lambda(\Omega, \Omega') \, .
\end{split}
\label{DS2ptcorrect}
\end{equation}


\subsection*{Comment on the late-time behaviour}

We want now to analyse the late-time behaviour of the correlation functions described above. To do so, we first re-introduce the invariant distance on $S^2$ through the relation \eqref{def: define z=1-u/2}. This allows us to write \eqref{DS2ptcorrect} as a function of the Euclidean invariant distance $u^E_{xy}$ and then perform the analytic continuation to the Lorentzian section by replacing $u^E_{xy} \to u^L_{xy}$, as long as we are avoiding the singularities present in \eqref{DS2ptcorrect}. For spacelike separated points we have in the patch \eqref{eq:compactif} 
\begin{equation}
    u^E_{xy} \to u^L_{xy} = \frac{\cos\left( T_x - T_y \right)-\cos \left( \varphi_x - \varphi_y\right)}{\sin T_x \sin T_y} \, . 
\end{equation}
The equal-time late-time limit, for this choice, corresponds to $T_x = T_y \equiv T$ with $T \to 0^-$. We then obtain 
\begin{equation}
\begin{split}
    \lim_{T \to 0^-} H_{\textnormal{f}}^{\Delta = \frac32} (\mathbf{x}, \mathbf{y}) &= \frac{1}{2\pi \ell}\left( \frac{T^2}{\sin^2\left( \frac{\varphi_x-\varphi_y}{2} \right)}\right)^{\bar{\Delta}} \left[ 2\pi i \alpha - \log\left( \frac{T^2}{\sin^2\left( \frac{\varphi_x-\varphi_y}{2} \right) } \right)^{\bar{\Delta}}\right] \Lambda(\Omega,\Omega') \\ 
    &-\frac{1}{4\pi \ell} \left( \frac{T^2}{\sin^2\left( \frac{\varphi_x-\varphi_y}{2} \right)}\right)^{\bar{\Delta}} \left[ 1 + 4\pi i  - 2 \log\left( \frac{T^2}{\sin^2\left( \frac{\varphi_x-\varphi_y}{2} \right) } \right)^{\bar{\Delta}} \right] \ell \slashed{n} \Lambda(\Omega,\Omega') \, ,
\end{split}
\end{equation}
where we have made explicit the dependence on $\bar{\Delta} = -\frac12$. Similar to the case of the bosonic discrete series \citep{Anninos:2023exn}, we see that the two-point function grows in the deep IR of the theory. It is instructive to compare this with the late-time limit of a principal series two-point function 
\begin{equation}
\begin{split}
     \lim_{T\to 0^-} S_{\textnormal{f}}(\mathbf{x},\mathbf{y}) = \frac{i}{4\pi \ell}& \left[ \left( \frac{\Gamma(\bar{\Delta}-\Delta)\Gamma(\frac12+\Delta)}{\Gamma\left(\frac12+\bar{\Delta} \right)}\left( \frac{T^2}{\sin^2\left( \frac{ \varphi_x-\varphi_y }{2} \right)} \right)^\Delta + \left( \Delta \leftrightarrow \bar{\Delta} \right)\bm{1} \right) \right. \\ 
     &\left. -i \left( \frac{\Gamma(\bar{\Delta}-\Delta) \Gamma\left( \frac12 + \Delta \right)}{\Gamma \left( \frac12 - \Delta \right)}\left( \frac{T^2}{\sin^2\left( \frac{ \varphi_x-\varphi_y }{2} \right)} \right)^\Delta +\left( \Delta \leftrightarrow \bar{\Delta} \right) \right){\ell}  \slashed{n} \right] \Lambda(\mathbf{x},\mathbf{y})  \, ,  
\end{split}
\end{equation}
where now $\Delta = \frac12 + i \ell m$ and we can see that at late times, the principal series decays and thus its contribution washes away in the deep IR regime of the theory. 

\section{Realising the Discrete Series: A proposal}
\label{GTFermions}
So far we have shown that certain free field theories realise the fermionic discrete series in their classical mode solution space. In particular, the mode solutions of these theories furnish a direct sum of two discrete series UIRs of $SL(2, \mathbb{R})$. This comes at a high price, as such theories are endowed with a purely imaginary mass term which makes the unitarity of the field theory not evident. Canonical quantisation is  not straightforward as the naive action functional is non-hermitian and there is no clear gauge fixing procedure to deal with the shift symmetry these discrete series fermions possess. Despite this, an Euclidean analysis allowed us to put forward a proposal for the two-point function of these fields. Thus, if a sensible quantisation procedure is possible, then the putative theory should alleviate the aforementioned tensions and reproduce the corresponding correlation functions. 

Similar tensions are also found in the case of the bosonic discrete series, where they correspond to tachyonic instabilities \eqref{def:tachyon_mass}. While a general procedure to deal with this is not known, it has been advocated that such fields should be part of a bigger gauge theory. Examples of this involve the $\Delta = 1$ discrete series which can be realised as a massless scalar field with its shift symmetry gauged or, the general $\Delta = n$ case which is contained in the operator content of a $SL(N,\mathbb{R})$ BF-theory \citep{Anninos:2023exn}. Another example involves the Schwinger model in a fixed de Sitter background \citep{Anninos:2024fty}, which in the weak coupling regime contains a $\Delta = 1$ field which exactly reproduces the discrete two-point function put forward in \citep{Anninos:2023exn}.  

Yet, there is still a lack of proposals for gauge theories that realise the fermionic discrete series UIRs as a sector of a bigger gauge theory. The only known example, to the best of our knowledge, was recently given in \citep{Anninos:2023exn} where a supersymmetric timelike Liouville theory coupled to superconformal matter was analysed. In two spacetime dimensions the supergravity multiplet does not encode any propagating degree of freedom, yet the fluctuations around the de Sitter saddle point encode a $\Delta = \frac32$ fermion. 

In this section, we will put forward two proposals for theories that contain fermionic discrete series UIRs in their operator content, whether such operators endure after imposing the corresponding gauge constraints is beyond the scope of this paper and is left for future work. Building on \citep{Pethybridge:2024qci}, we will first analyse a supersymmetric extension of the $q = 2$ SYK and show that it has an infinite tower of both bosonic and fermionic discrete UIRs in its spectrum. Armed with this and the known relation between SYK and JT-gravity, we propose a $\mathcal{N}=1$ supersymetric version  of de Sitter JT-gravity and show that the fluctuations around the saddle point correspond to a $\Delta = 2$ and $\Delta = \frac{3}{2}$ fields. While the full analysis of both these theories and their generalisations are left for future work, we present evidence that they contain the fermionic discrete UIRs in their spectrum. 

\subsection{\texorpdfstring{$q = 2$ Super-SYK}{q = 2 Super-SYK}}

The bosonic discrete UIRs of $SO(1,2)$ can be realised at the level of the pre-Hilbert space of the field operators of a BF gauge theory on a fixed dS$_2$ \citep{Anninos:2023lin, Pethybridge:2024qci}. For example, the $\Delta = 2$ UIR was shown to be realised in the linearised field equations of $SL(2,\mathbb{R})$ BF-theory which is semi-classically equivalent to a JT gravity. Upon quantisation, when all of the constraints of the gauge choice have been taken into account, the Hilbert space states collapse onto a finite dimensional subspace. However, the discrete series operators remain, and are precisely the gauge invariant operators of the model. This construction can be generalised to a $SL(N,\mathbb{R})$ BF gauge theory which in turn, corresponds to a higher-spin generalisation of JT gravity. This topological theory has an operator algebra, and pre-Hilbert space that includes states corresponding to the bosonic UIR scaling dimensions
\begin{equation} \label{res:weights_bosons}
\Delta = 2, 3,..., N  ~. 
\end{equation}
In two dimensions, AdS-JT is microscopically described by the SYK model with a $q= 4,6,...$-body interaction \citep{Almheiri:2014cka,Maldacena:2016upp,Cotler:2016fpe,Jensen:2016pah}. One natural question is whether a similar quantum mechanical description of dS$_2$ can be found. This has lead to the suggestion \citep{Anninos:2023lin} that the bulk $SL(N,\mathbb{R})$ BF theory, in the $N \rightarrow\infty$ limit might be described microscopically by the complex $q = 2$ SYK model. This model was found to include operators with precisely the integer  scaling dimensions \eqref{res:weights_bosons} required to provide this potential one-dimensional boundary dual in \citep{Pethybridge:2024qci}. In this paper we have generalised the discussion to fermionic discrete UIRs, and we would like to discuss if such UIRs can also appear in a generalised SYK model. It is thus suggestive to think in the possibility of a supersymmetrisation of the complex $q = 2$ model. 

From the quantum mechanical boundary, there have been many supersymmetric extensions of the SYK model for generic $q$ described previously \citep{Fu:2016vas,Marcus:2018tsr,Benini:2024cpf,Anninos:2016szt,Anninos:2017cnw, Biggs:2023mfn}. In particular, a possible model with dynamical bosons and $\mathcal{N} = 2$ supersymmetry is the version of the $q=2$ SYK studied in \citep{Anninos:2017cnw,Biggs:2023mfn} with a superpotential 
\begin{equation}
    W(\phi_i)  = \Gamma_{ij}\Phi_i\Phi_j~,
\end{equation}
where $i = 1,...N$ and $\Gamma_{ij}$ are drawn from a zero mean Gaussian distribution with variance set to $1/N$ for the purposes of this outlook. The superfield $\Phi_i$ is defined as 
\begin{equation}
    \Phi_i = \phi_i + i \theta_\alpha \psi_i^\alpha +i \epsilon_{\alpha\beta}\theta^\alpha \theta^\beta F_i
\end{equation}
where $\phi_i$ and $F_i$ are complex scalars and $\psi_i^\alpha$ is a fundamental $SU(2)$ spinor with two complex components.  The action in terms of the components of the supermultiplet is 
\begin{equation} \label{superSYK}
    S_E = \int d\tau \left[\psi_i^\alpha\Dot{\psi}_i^\alpha + \Dot{\phi}_i\Dot{\phi}_i - F_i F_i + \Gamma_{ij} \left(\phi_i F_j - \psi_i \psi_j \epsilon_{\alpha \beta} + \text{c.c.}\right)\right]~,
\end{equation}
from which we identify a decoupled fermionic $q=2$ SYK sector with the interaction being $-\Gamma_{ij}\psi_i \psi_j \epsilon_{\alpha \beta}$ as well as bosonic sector. Performing the disorder average over $\Gamma_{ij}$ as detailed in \citep{Anninos:2017cnw} reveals that the theory in the large $N$ limit has an effective description at low energy in terms of the two propagators 
\begin{equation}
    Q (\tau)  \equiv \frac{1}{N}\braket{\bar{\phi}_i(\tau)\phi_i(0)}  \xrightarrow[IR]{}\frac{\sqrt{\pi}}{2}\left(\log|\tau|+ \gamma\right)~, \hspace{1cm} S(\tau) \equiv\frac{1}{N}\braket{\bar{\psi}^\alpha_i\psi_i^\alpha}  \xrightarrow[IR]{}\frac{1}{\tau}~,
\end{equation}
where $\gamma$ is the Euler-Mascheroni constant.

In the original SYK model the bosonic operators corresponding to the discrete scaling dimensions \eqref{res:weights_bosons} can be defined in terms of the  fermions $\psi_i$ as  \citep{Pethybridge:2024qci,Gross:2017hcz}
\begin{equation} \label{Bosonic_OSYK}
    \mathcal{O}_n(\vartheta) = \frac{i}{\sqrt{N}} \sum_{k=0}^n d_{nk} \partial_{\vartheta}^k \bar{\psi}_i(\vartheta) \partial_\vartheta^{n-k} \psi_i(\vartheta) \, ,
\end{equation}
where $d_{nk}$ ensures that the operator is a primary, 
\begin{equation}
    d_{nk} = \frac{\pi(-1)^k(-n)_k^2}{\sqrt{\Gamma(2n+1)}\Gamma(k+1)^2} \, .
\end{equation}
Repeating the analysis for the supersymmetrised model \eqref{superSYK} describes an operator algebra with bosonic operators of the same form as \cref{Bosonic_OSYK}, along with fermionic operators. For example, the following two-point function 
\begin{equation}
    \frac{1}{N^2}\braket{(\bar{\psi}_i^\alpha \dot{\phi}_i)(\tau)(\psi_j^\alpha \dot{\bar{\phi}}_j)(0)} \sim \frac{1}{N|\tau|^3} + \mathcal{O}(N^{-2})~,
\end{equation}
defined in analogy to the two-point function of \eqref{Bosonic_OSYK}, suggests that the theory has states with
\begin{equation}
    \Delta = \frac{3}{2}\, , \frac{5}{2} \, , \cdots \, 
\end{equation}
The fermionic operators have the form 
\begin{equation} \label{SYKferDS}
    \mathcal{O}_{\Delta}^\alpha = \frac{1}{\sqrt{2}N^{\frac32}} \sum_{k=0}^{r} \tilde{d}_{rk} \left(\partial^k \bar{\psi}^\alpha_i(\tau) \partial^{r-k} \phi_i(\tau) + c.c.\right)~, 
\end{equation}
where $\Delta =r+\frac12$ and 
\begin{equation}
    \tilde{d}_{rk} = \frac{2 i^{r+1} (-1)^{k} \Gamma (r) \Gamma (r+1) (1-\delta
   _{r,k})}{\pi^{\frac14} \sqrt{\Gamma (2 r)} \Gamma (k+1)^2 \Gamma
   (r-k) \Gamma (r-k+1)}~.
\end{equation}
The existence of these operators supports the claim that a supersymmetric extension of the SYK could microscopically describe a  theory of discrete series fermions in dS$_2$. Furthermore, it also suggests that it might be possible to define a bulk higher spin theory containing both bosonic and fermionic UIRs. 

\subsection{$\mathcal{N}=1$ de Sitter JT gravity}
Recent years have seen an increasing interest in lower dimensional theories of gravity as an arena in which theoretically tractable models allow one to access quantum features of spacetimes. Notably, gravitational quantum effects are simplest to formulate and interpret in asymptotically AdS spacetimes, see \citep{Mertens:2022irh} and references therein. On the other hand, for cosmological spacetimes the scope of models at our disposal is scarce at best. Attempts to realise a quantum dS$_2$ in non-supersymmetric JT gravity face certain obstructions such as the appearence of a non-normalisable Hartle-Hawking wavefunction and a divergent sphere partition function \citep{Maldacena:2019cbz, Cotler:2019nbi, Nanda:2023wne}, see \citep{Held:2024rmg, Alonso-Monsalve:2024oii} for recent attempts at addressing these issues. 

Motivated by the spectrum of the supersymmetric $q=2$ SYK discussed in the previous section we will analyse a supersymmetric extension of JT gravity admitting an Euclidean de Sitter saddle, the round $S^2$,  whose Gaussian fluctuations carry a discrete series multiplet containing a $\Delta = 2$ and a $\Delta = \frac32$ field. It has been long known that supersymmetry and de Sitter are hard to reconcile \citep{Lukierski:1984it, Pilch:1984aw} and that, at least in two dimensions, a two-dimensional supergravity with a positive cosmological constant is possible \citep{Polyakov:1981re}, see \citep{Anninos:2023exn} for a recent discussion. 

In Lorentzian signature, the two dimensional off-shell $\mathcal{N} = (1,1)$ supergravity multiplet contains the zweibein $e_\mu^a$, a spin-$3/2$ Majorana gravitino $\psi_\mu$ and a real scalar field $A$. For the following discussion, we will consider Euclidean signature. It is well known that in two dimensions there are no Majorana-Weyl spinors in Euclidean signature, and thus the content of the Lorentzian multiplet has to be complexified, doubling the degrees of freedom \citep{vanNieuwenhuizen:1996tv}. {We note in passing that, since the final objective in Euclidean signature is the computation of the partition function (which we will not analyse here),  a half-dimensional contour must be picked in the complexified field space in order to perform the integration. In other words, while performing the path integration, the doubled degrees of freedom will be again reduced to half, to match the degrees of freedom of the original (uncomplexified) Lorentzian theory.}  We will follow the conventions of \citep{Howe:1978ia},\footnote{See \citep{Fan:2021wsb,Cribiori:2024jwq} for a modern exposition.} and consider the following (Euclidean) super-JT action
\begin{equation} \label{dSJT}
    S = \int_{\Sigma_h}  \dd^2 x~ e \left[ \phi \left( R + 2 i \mu A + \frac{i \mu}{2 } \bar{\psi}_\mu \gamma^{\mu \nu} \psi_\nu \right) - 2F \left( A-i \mu\right) + \bar{\lambda} \left( 2 \gamma^{\mu \nu} \nabla_\mu \psi_\nu - i \mu \gamma^\mu \psi_\mu \right) \right],
\end{equation}
where we introduced the Majorana conjugate of a spinor 
\begin{equation}
    \bar{\psi}_\mu = \psi^T_\mu \mathcal{C} \, ,
    \label{EuclideanMajo}
\end{equation}
defined using the charge conjugation matrix $\mathcal{C}$ satisfying 
\begin{equation}
    \gamma_{\alpha}^{T} = - \mathcal{C} \gamma_{\alpha}\mathcal{C}^{-1} \, ,
\end{equation}
see Appendix A of \citep{Anninos:2023exn} for details on the conventions. Note that in this Section the `bar' above the spinors does not denote the same operation as in the previous sections - in (\ref{def:MassiveFermAction}) it stands for Dirac conjugation in Lorentzian signature, while in \eqref{def:MassiveFermActionEuclidean}, \eqref{EuclideanDirac} it stands for hermitian conjugation in Euclidean signature. The parameter $\mu$ is real, while the theory is defined on a compact manifold $\Sigma_h$ of genus $h$. Now, our supermultiplet in (\ref{dSJT}) consists of a Dirac gravitino $\psi_{\nu}$, a complex zweibein $e^{a}_{\mu}$, and a complex scalar field $A$. The covariant derivative acting on  $\psi_{\nu}$ in (\ref{dSJT}) is defined
as 
\begin{equation}
\nabla_{\mu} \psi_{\nu} = (\partial_{\mu} + \frac{1}{4}\omega_{\mu bc} \gamma^{bc} )  \psi_{\nu} - \Gamma_{\mu \nu}^{\lambda}  \psi_{\lambda} \, ,    
\end{equation}
while the Christoffel symbol term drops out due to the anti-symmetry between $\mu$ and $\nu$: $\gamma^{\mu \nu}  \nabla_{\mu} \psi_{\nu} =\gamma^{\mu \nu}  \nabla_{[\mu} \psi_{\nu]} $. Furthermore, there is  a topological contribution given by 
\begin{equation} \label{dsJTtopo}
    S_h =  \frac{\vartheta}{4\pi} \int_{\Sigma_h} \dd^2 x e R = \vartheta \chi_h\, , 
\end{equation}
where $\chi_{h} = 2 - 2h$ is the Euler number of $\Sigma_h$ and $\vartheta$ is a real parameter that upon fixing a $S^2$ saddle point captures the de Sitter entropy.\footnote{In the context of bosonic JT-gravity the parameter $\vartheta$ is normally referred as $\phi_0$.} All in all, the $\mathcal{N}=(1,1)$ super-JT gravity with positive cosmological constant is defined using both contributions \eqref{dSJT} and \eqref{dsJTtopo}, as 
\begin{equation}
    S_{\textnormal{sJT}} = S[e_\mu^a, \psi_\mu, A, \phi, {\lambda}, F ] + S_h[\vartheta] \, ,  
\end{equation}

The action is invariant under the corresponding $\mathcal{N}=(1,1)$ supergravity transformations\footnote{Both actions (\ref{dSJT}) can be obtained from a superfield construction by considering a super-Berezinian $\mathbb{E}$, a super-curvature $\mathbb{R}$ and a super Dilaton field $\Phi$. Thus the topological contribution can be seen to arise from the superfield generalisation of $eR$, $e R \to \mathbb{E} \mathcal{R}$.  }  
\begin{equation}
    \delta_{\epsilon} e_\mu^a = \frac12 \bar{\epsilon} \gamma^a \psi_\mu \,, \qquad \delta_{\epsilon} \psi_\mu = \nabla_\mu \epsilon + \frac{1}{2} A \gamma_\mu \epsilon \,, \qquad \delta_{\epsilon} A = \frac12 \bar{\epsilon} \gamma^{\mu \nu} \left( \nabla_\mu \psi_\nu - \frac12 A \gamma_\nu \psi_\mu \right) \,.
\end{equation} 
While for the super-dilaton field we have 
\begin{equation}
    \delta_\epsilon \phi = \frac{1}{2} \bar{\epsilon} \lambda \, , \quad \delta_\epsilon \lambda = \frac{1}{2} \gamma^\mu \epsilon \hat{D}_\mu \phi - \frac12 F \epsilon \, , \quad \delta_\epsilon F = \frac{1}{2} \bar{\epsilon} \gamma^\mu \hat{D}_\mu\lambda \, , 
\end{equation}
where 
\begin{equation}
    \hat{D}_\mu \phi = \partial_\mu \phi -\frac12 \bar{\psi}_\mu \lambda \, , \quad \hat{D}_\mu \lambda = \nabla_\mu \lambda -\frac12 \gamma^\nu \psi_\mu \hat{D}_\nu \phi - \frac12 F \psi_\mu \, ,
\end{equation}
For the following analysis, it is sufficient to note that the super Dilaton multiplet $(\phi,\bar{\lambda},F)$ serves the role of a Lagrange multiplier enforcing the following equations of motion 
\begin{equation} 
\begin{split}
    \frac{\delta S}{\delta \phi} &: \left( R + 2 i \mu A + \frac{i \mu}{2 } \bar{\psi}_\mu  \gamma^{\mu \nu} \psi_\nu \right) = 0 \, , \\ 
    \frac{\delta S}{\delta F} &: A - i \mu  = 0 \, , \\ 
    \frac{\delta S}{\delta \bar{\lambda}} &: \left( 2 \gamma^{\mu \nu} \nabla_\mu \psi_\nu - i \mu \gamma^\mu \psi_\mu \right) = 0 \, .
\end{split}
\end{equation}
Classical solutions of the theory have vanishing fermionic fields, implying $R = 2 \mu^2$ which corresponds to an Euclidean nearly de Sitter solution \citep{Maldacena:2016upp,Anninos:2018svg}.

A path integral analysis of the theory requires a suitable definition of the path integration measure, which is not defined prior to gauge fixing. We can exploit the diffeomorphism invariance of the theory to impose a super-Weyl gauge \citep{Howe:1978ia}, which is given by 
\begin{equation}
    e^a_\mu = e^{b \varphi} \tilde{e}^{~a}_\mu \,, \qquad \psi_\mu = e^{\frac{b}{2}\varphi} \tilde{\gamma}_\mu \psi \, , \qquad A = e^{-b \varphi} G, 
\end{equation} 
where the zweibein $\tilde{e}^{~a}_{\mu}$ refers to a fixed fiduciary metric $\tilde{g}_{\mu \nu} = \tilde{e}^{~a}_{\mu} \delta_{a b}\tilde{e}^{~b}_{\nu}$. The curved-space gamma matrices $\tilde{\gamma}_{\mu}$ and the constant flat-space gamma matrices are related as usual by $\tilde{\gamma}_{\mu} = \tilde{e}^{~a}_{\mu} \gamma_{a}$. After fixing the gauge, the degrees of freedom of the theory are the complex scalar and Dirac spinor fields, $\left( \varphi, \psi \right)$, which capture the dynamics of the metric and the gravitino, respectively. We find that the gauge-fixed action is
\begin{align} \label{dSJT_gauge fixed}
    S_{\textnormal{gf}} = \int_{\tilde{M}} \dd^2 x \tilde{e} \Bigg[ &\phi \left( \tilde{R} - 2 b \tilde{\nabla}^2 \varphi + 2 \mu i e^{b \varphi} G - i \mu  e^{b \varphi} \bar{\psi} \psi  \right) - 2 F e^{b\varphi} \left( G - i \mu e^{b \varphi} \right) \nonumber \\
    &+ 2 \bar{\lambda} e^{\frac{b}{2}\varphi} \left( \tilde{\slashed{\nabla}} \psi - i \mu  e^{b\varphi} \psi\right) \Bigg],
\end{align}
where $\tilde{\nabla}^{2} = \tilde{g}^{\mu \nu} \tilde{\nabla}_{\mu}  \tilde{\nabla}_{\nu}$ is the Laplace-Beltrami operator on $\tilde{M}$ and $\tilde{\slashed{\nabla}} = \tilde{\gamma}^{\mu} \tilde{\nabla}_{\mu}$ is the  Dirac operator; the spinor covariant derivative is defined in terms of the spin connection on $\tilde{M}$ as $\tilde{\nabla}_{\mu} = \partial_{\mu} + \frac{1}{4} \tilde{\omega}_{\mu b c } \gamma^{bc}$.
{To arrive at the expression (\ref{dSJT_gauge fixed}) for the gauge-fixed action, we have used that in the super-Weyl gauge the scalar curvature and spin connection become:
\begin{align}
    &   e R  =\tilde{e} \left( \tilde{R}-2b \tilde{\nabla}^{2}\varphi \right) ,  \\
     & \omega_{\mu  d   c}  = \tilde{\omega}_{\mu  d   c}+ {2}\,b \,   \tilde{e}_{\mu  [d}~ \tilde{e}^{\rho}_{c]}    \, \partial_{\rho}\varphi,
\end{align}
where the last equation implies $\nabla_{[\mu}  \psi_{\nu]}  = \tilde{\nabla}_{[\mu}  \psi_{\nu]}   + \frac{b}{2} \left(\partial_{\rho}\varphi \right) \tilde{\gamma}_{[\mu}\,^{\rho}  ~ \psi_{\nu]}$.  } 

We can now integrate out the auxiliary field $G$ and arrive at the equations of motion for the (complex) Weyl mode and the Dirac spinor, as 
\begin{equation} \label{EOMWeyl}
\begin{split}
     - \tilde{\nabla}^2 \varphi &= - \frac{1}{2b} \tilde{R} +  \frac{\mu^2}{b} e^{2 b \varphi} + i \frac{\mu}{2b} e^{b\varphi} {\psi}^{T}\mathcal{C} \psi  \\ 
   \tilde{\slashed{\nabla}} \psi &= i \mu e^{b \varphi} \psi \, .  
\end{split}
\end{equation}
The equations now admit a saddle point with a constant value for the Weyl mode $\varphi_*$ and vanishing fermionic fields. Recall that after all, we are looking for a de Sitter saddle and thus we want our fiducidary metric to be an $S^2$, namely $\tilde{R} = \frac{2}{\ell^2}$. At this point it is convenient to bring the equations to a canonical form by a field and parameter redefinition given by $\bar{\mu} = \frac{{\mu}}{\beta}$, $\Psi = \frac{{i^{1/2}}}{\beta} \psi$, $b = \beta$ we then have 
\begin{equation}
    - \tilde{\nabla}^2 \varphi =  \bar{\mu}^2 \beta e^{2 \beta \varphi} -\frac{1}{\beta \ell^2}   + \frac{1}{2} \bar{\mu} \beta^2 e^{\beta \varphi} \bar{\Psi} \Psi \, , \qquad \tilde{\slashed{\nabla}} \Psi = i \bar{\mu} \beta e^{\beta \varphi} \Psi \, ,
\end{equation}
which we identify as the equations of motion of $\mathcal{N} = 1$ super Liouville on Euclidean de Sitter of \citep{Anninos:2023exn} in the limit $\beta \to 0^+$. The de Sitter saddle corresponds to the following configurations 
\begin{equation} \label{DSSaddle}
    \Psi_* = 0 \, , \qquad \varphi_* = \frac{1}{2 \beta} \log \left( \frac{1}{\ell^2 \bar{\mu}^2 \beta^2} \right) \, ,  
\end{equation}
where $\varphi_*$ is real for $\bar{\mu}^2 \beta^2 > 0$. The physical metric is given by $g_{\mu \nu} = e^{2 \beta \varphi_* } \tilde{g}_{\mu \nu}$ and corresponds to the real metric of the round 2-sphere, i.e. Euclidean $dS_{2}$. Furthermore, in the limit $\beta \to 0^+$ the radius of the sphere grows as $\frac{1}{\mu^2 \beta^2}$ and consequently $\vartheta$ which captures the classical de Sitter entropy is now given by $\vartheta \gg 1$, this makes the higher topological contributions to the path integral exponentially suppressed. 

After fixing the super-Weyl gauge and making the saddle point approximation, the theory is invariant in the $\beta^+ \to 0$ limit under the subgroup of (gauge-preserving) diffeomorphisms that are well defined on $S^2$ namely $OSp(1\lvert2, \mathbb{C})$. The corresponding SUSY variations on the fields are \citep{Benini:2012ui,Doroud:2012xw}  
\begin{equation}
    \delta_{\epsilon} \varphi = \bar{\epsilon} \Psi \, , \qquad \delta_{\epsilon} \Psi = i \tilde{\slashed{\partial}} \varphi ~\epsilon + \frac{i}{\beta } \tilde{\slashed{\nabla}} \epsilon - i G \epsilon \, , \qquad \delta_{\epsilon} G = - \bar{\epsilon} \tilde{\slashed{\nabla}}\Psi \, ,
\end{equation}
{where $\epsilon$ are Dirac conformal Killing spinors on $S^{2}$.}
We can now define the Gaussian theory by the fluctuations around the saddle point \eqref{DSSaddle} namely 
\begin{equation}
    \Psi = \Psi_* + \delta \Psi \, , \qquad \varphi = \varphi_* + \delta \varphi \, ,
\end{equation}
and expanding up to first order in $\beta$, eqs. \eqref{EOMWeyl} give
\begin{equation} \label{DSfluctuations}
    \left( -\tilde{\nabla}^2 - \frac{2}{\ell^2} \right) \delta \varphi = 0 \, , \qquad \left( \tilde{\slashed{\nabla}} - \frac{i}{\ell} \right) {\delta \Psi} = 0 \, ,
\end{equation}
where $\delta \varphi$ and $\delta \Psi$ are complex fluctuations, the imaginary-mass Dirac equation in (\ref{DSfluctuations})  does not admit Majorana solutions in Euclidean signature, but in Lorentzian it does.
We note that the bosonic fluctuation has a negative mass squared, which in terms of the UIRs of $SO(1,2)$ \eqref{def:tachyon_mass} corresponds to a $\Delta = 2$ discrete series field. Similarly, the non-standard imaginary mass term for the fermionic fluctuation corresponds to the $\Delta=3/2$ discrete series discussed in the previous sections. As discussed before, these non-standard mass values characterise the fluctuations around the saddle point and are pieces of an underlying gauge theory. Recall that in $2d$ the supergravity multiplet does not encode any propagating degree of freedom. We thus expect that the gauge constraints, once imposed, should remove the propagating degrees of freedom.

\section{Summary \& Outlook}
In this note, we have studied the discrete series representation of $SL(2,\mathbb{R})$ and its free field theory realisation as fermions propagating in a fixed two-dimensional de Sitter background. We have shown that the corresponding UIR forces us to consider spinor fields with \emph{purely imaginary} mass \eqref{eq: imaginary mass Dirac eqn}.  These imaginary-mass spinors enjoy a gauge symmetry corresponding to a (spinorial) shift symmetry \eqref{eq:shift_symFerm r=2}. These features suggest that canonical quantisation is non-standard as now the hermiticity of the action functional is not evident and a gauge-fixing procedure is required. 

We analysed the free two-point function of such discrete series spinors and showed that it is not possible to find one with the right singularity structure \eqref{singstructureDS}. An Euclidean analysis shows that there are now zero-modes that render this correlation function ill-defined at the corresponding discrete series mass values. We proposed a modification of the Euclidean two-point function  \eqref{discrete2ptdif} to address this issue and showed how to explicitly obtain a correlation function \eqref{DS2ptcorrect} with the desired singularity structure for the case of $\Delta = \frac32$. The differential equations that have to be satisfied by the fermionic discrete series two-point functions with the zero-modes removed were found for arbitrary $\Delta$. {We stress that such equations can be solved for any given value of $\Delta$}. We expect that any sensible theory on dS$_2$ that has a discrete fermion in its spectrum should yield such correlation functions. 

Finally, we have presented two theories that contain both fermionic and bosonic discrete series fields in their spectrum, and thus, could serve as a playground to understand the features of such representations on de Sitter spacetimes. The first one corresponds to a supersymmetric $q=2$ SYK theory which has a higher-spin tower of $\Delta = \frac32, 2, \frac52, 3 , \cdots$ fields \eqref{Bosonic_OSYK} and \eqref{SYKferDS}. This microscopic theory could be relevant to formulate a holographic description of such discrete series fields in dS$_2$. The second one, corresponds to a super-JT gravity theory that admits a de Sitter saddle. The fluctuations around this saddle were shown to correspond to a discrete series multiplet with $\Delta = 2$ and  $\Delta= \frac32$ \eqref{DSfluctuations}. 

We would like to conclude this note with an outlook and future directions regarding the fate of these representation and their role in de Sitter QFT. 

\subsection*{$\mathcal{N} = (1,1) \, \& \,  \mathcal{N} = (2,2)$  dS super-JT gravity}
We have presented a $\mathcal{N} = 1$ supersymmetric extension of JT-gravity which admits an Euclidean de Sitter saddle point and whose corresponding fluctuations encode a multiplet of $SL(2, \mathbb{R})$ containing a $\Delta = 2$ and $\Delta = \frac32$ field. After gauge fixing the theory has a residual gauge symmetry that corresponds to $OSp(1\lvert2,\mathbb{C})$. This analysis serves as the starting point to a proper path integral computation of the theory and determine whether the problems found in the non-supersymmetric case, such as the presence of a non-normalizable Hartle-Hawking state, get alleviated in the $\mathcal{N}=(1,1)$ theory. Furthermore, if the sphere path integral is finite, we expect the one-loop partition function around the fluctuations \eqref{DSfluctuations} to encode the corresponding Harish-Chandra characters of the discrete series UIR. This will signal towards the unitarity of the theory despite the apparent non-reality of the action in Lorentzian signature. Finally, it should also be possible to generalise the theory to contain now a $\mathcal{N}=(2,2)$ symmetry, where now the fluctuations around the de Sitter saddle point should correspond to the $\mathcal{N} = (2,2)$ super-Liouville equations of \citep{Anninos:2023exn}. 

\subsection*{A Lorentzian view for de Sitter super-JT}
The analysis carried out in this note for the $\mathcal{N}=(1,1)$ super-JT gravity was carried out  in the Euclidean formulation of the theory. Recently, there has also been efforts to understand the Lorentzian Hilbert space of two-dimensional theories of quantum gravity with $\Lambda > 0$ through an analysis of the Wheeler-DeWitt equation  \citep{Anninos:2024iwf}. It would be interesting to carry out such analysis in the case of the supersymmetric de Sitter JT theory. 

\subsection*{Super Higher-Spin dS$_2$ gravity}
The spectrum of the $\mathcal{N} = 2$, $q = 2$ super-SYK \eqref{superSYK} suggests the presence of a higher-spin tower of (bosonic and fermionic) discrete series fields in a potential dual theory living on dS$_{2}$. In other words,  it should be possible to construct a $\mathcal{N} = 2$ super-SL$(N,\mathbb{R})$  BF theory which would be an analogue of a lower-dimensional super higher-spin gravity theory in dS$_2$. Interestingly enough{, in the context of dS holography,} examples of higher-spin gravity theories in dS$_4$ have been studied before \citep{Anninos:2017eib}. Developing the tools to address the lower-dimensional case might pave the way for a 4-dimensional analogue, extending the higher-spin theories of \citep{Anninos:2017eib} to also contain fermions.  

\subsection*{Late time for Discrete Fermions}
Contrary to the principal series case, the two-point function for discrete series fermions shows a growing behaviour in the deep IR or late-time regime of de Sitter. One way out of this problem is to require that such fields are not present in any sensible QFT on de Sitter as they will disrupt the stability of the solution. Yet, the corresponding UIRs seem to be an important part of recent formulations of de Sitter quantum gravity \citep{Anninos:2023exn}. It was recently shown \citep{Anninos:2024fty} that such logarithmic growths in the late time can actually be resummed to yield a de Sitter invariant correlation function that actually decays in the late-time. It would be interesting to develop an analogue set-up to address where the loop corrections to the fermionic discrete two-point functions can also be resummed.

\section*{Acknowledgements}

We would like to thank Dionysios Anninos, Tarek Anous, Guillermo Silva, Volodia Schaub, Gizem \c{S}eng\"or, Atsushi Higuchi, Pietro Benetti Genolini and F.F. John for interesting disussions, suggestions and, support along the way. V.A.L. is grateful to the Eleni Gagon Survivor’s Trust for supporting his work. ARF is funded by the Royal
Society under the grant “Concrete Calculables in Quantum de Sitter”.
\appendix

\section{Matrix elements of \texorpdfstring{$SL(2,\mathbb{R})$}{SL(2,R)} and details on the \texorpdfstring{$\Delta=3/2$}{Δ=3/2} character}
 \label{Appendix_matrix elements}
Here we present the details on how the anti-hermitian generators $\hat{J}_{21}, \hat{J}_{01}, \hat{J}_{02}$ of the algebra $\mathfrak{so}(1,2)$ act on an abstract representation space. Then, we find an explicit expression for a (finite) boost matrix element and use this expression to compute the $\Delta = 3/2$ discrete series character~(\ref{Delta=3/2 character}) - see also Appendix F.1. of \citep{Sun:AdS} for an alternative computation of the characters for any $\Delta$.

We denote states in our Hilbert space (representation space) as $\ket{\Delta,n}$ for fixed $\Delta$ and we consider the basis on which $\hat{J}_{21}$ acts diagonally
\begin{equation}\label{J21 on states abstract}
     \hat{J}_{21} \ket{\Delta,n} = i n \ket{\Delta,n}.
\end{equation}
The Hilbert space is equipped with a positive definite scalar product such that 
\begin{equation}
    \braket{\Delta, n | \Delta,n'} = \delta_{nn'} \, .
\end{equation}
The value of $\Delta$, and the range of $n$, depend on the $\mathfrak{so}(1,2)$ UIR under consideration - see Subsection~\ref{Subsec_classification of UIRs}. The way in which the rest of the $\mathfrak{so}(1,2)$ generators act on the states is well-known~\citep{schwarz}: 
\begin{align}\label{J02 on states abstract}
    \hat{J}_{02}  \ket{\Delta, n}  =-\frac{i}{2} \sqrt{n(n+1)-\Delta(\Delta-1)}  \, \ket{\Delta, n+1}- \frac{i}{2} \sqrt{(n-1)n-\Delta(\Delta-1) } \, \ket{\Delta, n-1} ,
\end{align}
\begin{align}\label{J01 on states abstract}
    \hat{J}_{01}  \ket{\Delta, n} &=   [\hat{J}_{02}, \hat{J}_{21} ] \ket{\Delta, n} \nonumber\\
    &=-\frac{1}{2} \sqrt{n(n+1)-\Delta(\Delta-1)}  \, \ket{\Delta, n+1}+ \frac{1}{2} \sqrt{(n-1)n-\Delta(\Delta-1) } \, \ket{\Delta, n-1} .
\end{align}
Clearly, the raising operator $\hat{J}^{(+)} \equiv \hat{J}_{02} + i \hat{J}_{01}$ defined in (\ref{raising/lowering ops J basis}) acts on states as $$\hat{J}^{(+)}\ket{\Delta,n} = - i \sqrt{n(n+1)-\Delta(\Delta-1)}  \, \ket{\Delta, n+1},$$
while the lowering operator, $\hat{J}^{(-)} \equiv \hat{J}_{02} - i \hat{J}_{01},$ defined in (\ref{raising/lowering ops J basis}), acts as $$\hat{J}^{(-)}\ket{\Delta,n} = - i \sqrt{n(n-1)-\Delta(\Delta-1)}  \, \ket{\Delta, n-1}.$$

The expressions~(\ref{J21 on states abstract})-(\ref{J01 on states abstract}) are consistent with the anti-hermiticity of the generators, i.e.:
$$ \bra{\Delta,n}\hat{J}_{AB}\ket{\Delta,n\pm1}+ \left( \hat{J}_{AB}\ket{\Delta,n}
  \right)^{\dagger} \ket{\Delta,n\pm1}=0 , $$
  for $A,B \in \{ 0,1,2 \}$. Also, using~(\ref{J21 on states abstract}), (\ref{J02 on states abstract}) and (\ref{J01 on states abstract}), it is straightforward to verify the eigenvalue~(\ref{def:Cas_groupEigen}) of the quadratic Casimir as
  \begin{align}
 \Big(   \left(\hat{J}_{01}\right)^{2} +\left(\hat{J}_{02}\right)^{2}-   \left(\hat{J}_{21}\right)^{2} \Big) \ket{\Delta,n} = \Delta (\Delta-1) \, \ket{\Delta,n} .
\end{align}
Note that unitarity requires the eigenvalue of the Casimir to be real. The reason is that the Casimir is a hermitian operator.

 Knowing the action of the generators on arbitrary states allows us to obtain relations between group matrix elements defined by 
\begin{equation}\label{def:Melem}
    M^\Delta_{n',n}(t) \equiv \bra{\Delta,n'} e^{-t \hat{J}_{01}} \ket{\Delta,n} \, .
\end{equation}
To be specific, we will obtain relations among the matrix elements $ M^\Delta_{n',n}(t),  M^\Delta_{n',n-1}(t)$ and $ M^\Delta_{n',n+1}(t)$, as originally presented in~\citep{Wilson}. Using these relations, we will arrive at a second-order differential equation for $M^\Delta_{n',n}(t)$ and, then, we will find an explicit expression for $M^\Delta_{n',n}(t)$ as in \citep{Wilson}. The contribution of this part of the present paper is that we use the explicit expression of $M^\Delta_{n',n}(t)$ to compute the $\Delta = 3/2$ character directly. Without further ado, let us start by reviewing the steps followed in \citep{Wilson}. First, let us re-write the group matrix elements as 
\begin{equation}
    M^\Delta_{n',n}(t) = - \frac{i}{n'} \bra{\Delta,n'} \hat{J}_{21} e^{-t \hat{J}_{01}} \ket{\Delta,n} \, .
\end{equation}
Now, using\footnote{A quick way to verify~\eqref{noice BCH formula} is to consider the following representation for the $\mathfrak{so}(1,2)$ generators: 
\begin{equation*}
    j_{01} = \frac12 \sigma^1 \, , \quad j_{02} = - \frac{1}{2} \sigma^2 \, , \quad j_{21} = \frac{i}{2} \sigma^3\, ,
\end{equation*} 
where $\sigma^i$ are the Pauli matrices. Therefore
\begin{equation*}
    \exp(t j_{01}) = \bm{1} \cosh \frac{t}{2} + 2 j_{01} \sinh \frac{t}{2}
\end{equation*}
from where it is straightforward to verify \eqref{noice BCH formula}. }
\begin{equation} \label{noice BCH formula}
    \hat{J}_{21} e^{-t \hat{J}_{01}} = e^{-t \hat{J}_{01}} \left( \hat{J}_{21} \cosh t - \hat{J}_{02} \sinh t \right) \, ,
\end{equation}
we obtain 
\begin{align}\label{rec 2}
    \frac{n'-n \cosh{t}}{\sinh{t}   } M^\Delta_{n',n}(t) =& ~ \frac{1}{2} \sqrt{ n(n+1)-\Delta(\Delta-1)} M^\Delta_{n',n+1}(t) \nonumber\\
    &+\frac{1}{2} \sqrt{n(n-1)-\Delta(\Delta-1) } M^\Delta_{n',n-1}(t) ,
\end{align}
where we have made use of (\ref{J21 on states abstract}) and (\ref{J02 on states abstract}). Now we can obtain one more such relation.
Differentiating the group matrix element (\ref{def:Melem}) with respect to $t$, and using Eq.~(\ref{J01 on states abstract}), we find
\begin{equation}\label{rec 1}
\begin{split}
        \partial_t M^\Delta_{n',n}(t) = &\frac12 \sqrt{n(n+1)-\Delta(\Delta-1)} M^\Delta_{n',n+1}(t) - \frac12 \sqrt{n(n-1)-\Delta(\Delta-1)} M^\Delta_{n',n-1}(t) \, .
\end{split}
\end{equation}
Combining (\ref{rec 1}) and (\ref{rec 2}), and following the steps described in \citep{Wilson}, we obtain the following differential equation for the matrix elements $M^\Delta_{n',n}(t)$: 
\begin{align}\label{diff eqn matrix elements}
  \left(  \frac{d^{2}}{  d   t^{2}}   + \coth{t}  \frac{d}{d  t}    - \frac{1}{\sinh^{2}{t}}  (n^{2} + n^{'2} - 2 n\, n' \cosh{t})  - \Delta (\Delta-1) \right)   M^\Delta_{n',n}(t) = 0.
\end{align}
For $n'=n$ the solution that is regular at $\cosh{t} = 1$ is given by~(\ref{solution for SL(2,R) matrix element}). Let us now use this solution to compute the $\Delta=3/2$ character.

\noindent \textbf{Comment for $D_\Delta^+$ and $D_\Delta^-$ characters.} Before computing the $\Delta=3/2$ discrete series character~(\ref{Delta=3/2 character}), let us explain why the $D_\Delta^+$ character is equal to the $D_\Delta^-$ character [this is true for any $\Delta$ in the discrete series]. Recall that the UIRs $D_\Delta^\pm$ are characterised by the following $\mathfrak{so}(2)$ quantum numbers:
\begin{align*}
   & n = \Delta, \Delta + 1, \cdots ~\text{for}\hspace{5mm}D_\Delta^+ ,\\
    &n = \cdots, - \Delta - 1, - \Delta ~\text{for}\hspace{5mm}D_\Delta^- \, .
\end{align*}
For convenience, let us denote these quantum numbers as $n_{\pm}$ for $D_\Delta^\pm$, respectively. In other words, we have  
\begin{equation}
    n_+ = \Delta, \Delta + 1, \cdots \, , \qquad n_- = \cdots, - \Delta - 1, - \Delta \, .
\end{equation}
Let us first show that $M^\Delta_{n_{+}, n_{+}}(t) = M^\Delta_{n_{-},n_{-}}(t)$ for any $\Delta$ in the discrete series. 
For convenience, let us express $n_{\pm}$ in a more compact form as
\begin{equation} \label{convenient quntum number so(2)}
    n_\pm = \pm \left( \Delta + \tilde{n} \right) \, , \qquad \tilde{n}  = 0, 1, \cdots ~.
\end{equation}
Then, using the explicit expression~(\ref{solution for SL(2,R) matrix element}) for the matrix element, as well as the standard property \citep{bateman1953higher}
\begin{equation}
    F(A,B;C;z)=(1-z)^{C-A-B}\,F(C-A, C-B;C;z),
\end{equation} 
it is straightforward to show that 
\begin{equation}\label{matrix el D- = matrix el D+}
    M^\Delta_{\Delta+ \tilde{n}, \Delta + \tilde{n}}(t) = M^\Delta_{-(\Delta + \tilde{n}),-(\Delta+\tilde{n})} \, , 
\end{equation}
i.e. the group matrix elements for $D^{+}_{\Delta}$ and $D^{-}_{\Delta}$ are equal. The $D^{\pm}_{\Delta}$ characters are given by 
\begin{equation} \label{start point character}
  \chi_{D^+_\Delta}  = \chi_{D^-_\Delta} = \sum_{\tilde{n}=0}^{\infty}  M^\Delta_{\pm \left(\Delta+\tilde{n}\right), \pm \left( \Delta + \tilde{n}\right)}(t).
\end{equation}
Note that, although the corresponding characters are the same, $D^{+}_{\Delta}$ and $D^{-}_{\Delta}$ are in-equivalent UIRs. 

\noindent   \textbf{$\Delta = 3/2$ character computation}

We now proceed to the computation of the $\Delta = \frac{3}{2}$ character. Substituting $\Delta = \frac{3}{2}$ in (\ref{solution for SL(2,R) matrix element}), we find that the matrix elements are expressed as
\begin{equation} \label{Mcharacter}
  M^{\Delta= \frac{3}{2}}_{n_-,n_-}(t) =   M^{\Delta= \frac{3}{2}}_{n_+,n_+}(t) = \left(1-x(t)\right)^{-\tilde{n}-3/2} ~_{2}F_{1}\left(-\tilde{n}-2  ,    -\tilde{n} ; 1; x(t)  \right),
\end{equation}
where we have expressed $n_{\pm}$ in terms of $\tilde{n}$ as in~(\ref{convenient quntum number so(2)}), and we have defined
$$ x(t) \equiv  \frac{1-\cosh{t}}{2}.$$
The matrix elements (\ref{Mcharacter}) can be rewritten in a more convenient form as follows. First, let us exploit the following identity (see Appendix B of \citep{Letsios:2020twa})
\begin{equation}
    \begin{split}
        \left[ a(b-c) -a(b-a-1)z - (a+1-b)z(1-z) \frac{d}{dz}  \right] {_2}F_1(a,b;c;z) = a(b-c)~_{2}F_{1}(a+1,b-1;c;z)  ,
    \end{split}
\end{equation}
which we take with $a=b=-\tilde{n}-1$ and $c=1$.
Then, we re-write (\ref{Mcharacter})  as
\begin{align}\label{final raise a lower b}
  \frac{(1-x)^{-\tilde{n}-3/2}}{(\tilde{n}+1)(\tilde{n}+2)}  \Big( (\tilde{n}+1)(\tilde{n}+2)-(\tilde{n}+1)x-x(1-x)\frac{d}{dx}\Big)&  \,_{2}F_{1}(-\tilde{n}-1,-\tilde{n}-1;1;x)~  \nonumber \\
  =&~ M^{\Delta= \frac{3}{2}}_{n_-,n_-}(t) = ~  M^{\Delta= \frac{3}{2}}_{n_+,n_+}(t).
\end{align}
Let us now use \citep{website}
\begin{equation}
    {_2}F_1\left(a,b;a-b+1;z \right) = \Gamma(a-b+1) \left( -z \right)^{\frac{b-a}{2}}(1-z)^{-b} P_{-b}^{b-a}\left( \frac{1+z}{1-z} \right) \, ,~~ z \notin (1, \infty )
\end{equation}
where $P_\nu^\mu$ are the associated Legendre polynomials (with $\mu=0$ corresponding to the standard Legendre polynomials), and we take $a = b = - \tilde{n}-1$. Substituting this in (\ref{final raise a lower b}), we can express (\ref{final raise a lower b}) (and thus (\ref{Mcharacter})) as 
\begin{equation}\label{kontevoume}
  M^{\Delta =\frac32}_{n_+, n_+}(t) =   M^{\Delta =\frac32}_{n_-, n_-}(t) = -(1-x)^{-\frac12} \left[ \frac{x(1-x)}{(\tilde{n}+1)(\tilde{n}+2)} \frac{d}{dx} -1 \right] P_{\tilde{n}+1}\left( \frac{1+x}{1-x}\right) \, ,
\end{equation}

Having followed all these steps, let us, for convenience, write here again the sum we want to compute:
\begin{equation} \label{start point character}
    \chi_{D^\pm_\Delta} = -\sum_{\tilde{n}=0}^{\infty}  (1-x)^{-\frac12} \left[ \frac{x(1-x)}{(\tilde{n}+1)(\tilde{n}+2)} \frac{d}{dx} -1 \right] P_{\tilde{n}+1}\left( \frac{1+x}{1-x}\right)
\end{equation}
(recall that $x = x(t) \equiv  \frac{1-\cosh{t}}{2}$).
Finally, we introduce the variable
\begin{equation}
    w = w(x) = \frac{1+x}{1-x} \, ,
\end{equation} 
(where $|w| < 1$) and find 
\begin{equation}\label{ending point character}
    \chi_{D^\pm_\Delta} = \sum_{\tilde{n}=0}^\infty \frac{(1-x)^{\frac12}}{2} \left[ (1-w^2) \frac{d}{dw} \frac{P_{\tilde{n}+1}(w)}{\tilde{n}+1} -(1-w^2) \frac{d}{dw}\frac{P_{\tilde{n}+1}(w)}{\tilde{n}+2} + (w+1) P_{\tilde{n}+1}(w)  \right] \, ,
\end{equation}
This can be now computed to obtain 
\begin{equation}
  \chi_{D^+_{\Delta=\frac32}} (t) =  \chi_{D^-_{\Delta=\frac32}} (t) = \frac{e^{-\frac32 \lvert t \lvert}}{1-e^{-\lvert t \lvert}} \, ,
\end{equation}
in agreement with the known results \citep{sugiura}. This is our final result for the $\Delta=3/2$ character. Some details are in order that help the computation.

$\bullet$ In order to calculate:
$$ \sum^{\infty}_{ \tilde{n}=0 }\frac{P_{\tilde{n}+1}(w) }{\tilde{n}+1} = \sum^{\infty}_{ k = 1 }\frac{P_{k}(w) }{k} $$
that appears in (\ref{ending point character}), we go to page 700 of \citep{PrudnikovVol2}, chapter 5, Sec. 10.1 ``Series of the form $\sum a_{k} P_{nk+m}(x)$'', equation 4, which reads
$$\sum^{\infty}_{ k = 1 } \frac{\tau^{k}}{k} {P_{k}(y) } = ln \left( \frac{2}{1-\tau y +\sqrt{1-2\tau y   +  \tau^{2}}}   \right), ~~ |y|  \leq 1 .$$
Obviously, we have to let $\tau =1$ and $y=   w =(1+x)/(1-x)$ for our case (\ref{ending point character}).

$\bullet$  In order to calculate (recall $P_{0}(w)=1$):
$$ \sum^{\infty}_{ \tilde{n}=0 }\frac{P_{\tilde{n}+1}(w) }{\tilde{n}+2} =-P_{0}(w) +\sum^{\infty}_{ k = 0 }\frac{P_{k}(w) }{k+1} = -1 +\sum^{\infty}_{ k = 0 }\frac{P_{k}(w) }{k+1}, $$
which appears in (\ref{ending point character}), we go to page 700 of \citep{PrudnikovVol2}, chapter 5, Sec. 10.1, equation 5, which reads
$$\sum^{\infty}_{ k = 0 } \frac{1}{k+1} {P_{k}(y) } = ln \left( 1+\sqrt{\frac{2}{1-y}}   \right) .$$
We have to let $y=   w =(1+x)/(1-x)$ to apply this to our case (\ref{ending point character}).

$\bullet$ In order to calculate:
$$ \sum^{\infty}_{ \tilde{n}=0 }{P_{\tilde{n}+1}(w) } = -1+\sum^{\infty}_{ k = 0 }{P_{k}(w) }, $$
which appears in (\ref{ending point character}), we go to page 700 of \citep{PrudnikovVol2}, chapter 5, Sec. 10.1, equation 1, which reads
$$\sum^{\infty}_{ k = 0 } {(\pm 1)^{k}} {P_{k}(y) } = \frac{1}{\sqrt{2 (1 \mp y)}}, ~~~~ |y|<1.$$
Obviously, we have to choose the plus sign and let $y=   w =(1+x)/(1-x)$ for our case (\ref{ending point character}).


\section{Conventions}
\subsection{Lorentzian fermions} \label{App:fermions}

We denote the curved (i.e. coordinate basis) indices as $\mu, \nu, \cdots$ and flat local indices as $a,b,\cdots$. The Minkowskian two-dimensional gamma matrices $\gamma^a$ ($a = 0, 1$) satisfy the anti-commutation relations

\begin{equation}\label{anticommutation_relations_gamma}
   \{\gamma^{a}, \gamma^{b}\}  = 2 \eta^{ab} {\bm{1}} \, , \qquad  a,b=0,1\, , 
\end{equation}
where $\bm{1}$ is the spinorial identity matrix and $\eta^{a b}= \textnormal{diag}(-1,1)$. The zweibein fields $\bm{e}_{a}=e^{\mu}{\hspace{0.2mm}}_{a}\partial_{\mu}$, determining an orthonormal frame, satisfy
\begin{align}
    e_{\mu}{\hspace{0.2mm}}^{a} \, e_{\nu}{\hspace{0.2mm}}^{b}\eta_{ab}=g_{\mu \nu}, \hspace{4mm}e^{\mu}{\hspace{0.2mm}}_{a}\,e_{\mu}{\hspace{0.2mm}}^{b}=\delta^{b}_{a},
\end{align}
where the co-zweibein fields $\bm{e}^{a}=e_{\mu}{\hspace{0.2mm}}^{a}\,dx^{\mu}$ define the dual coframe. The gamma matrices with coordinate basis indices are defined using the zweibein fields as $\gamma^{\mu}(x) \equiv e^{\mu}{\hspace{0.2mm}}_{a}(x) \gamma^{a}$ which in turn obey 
\begin{equation}
    \{ \gamma^\mu, \gamma^\nu \} = 2 g^{\mu \nu} \bm{1}\, .
\end{equation}
In two dimensions, the frame gamma matrices can be expressed in terms of Pauli matrices. We choose the following conventions:
\begin{equation} \label{even_gammas}
    \gamma^0 = i \sigma^1 = \begin{pmatrix}
         0 & i \\
         i & 0
    \end{pmatrix} \, , \qquad \gamma^1 = - \sigma^2  = \begin{pmatrix}
         0 & i \\
         -i & 0
    \end{pmatrix}\, , \qquad \gamma_* = \sigma^3  = \begin{pmatrix}
        1 & 0 \\
        0 & -1
    \end{pmatrix}\, ,
\end{equation}
where $\gamma_*$ anti-commutes with $\gamma^{1}$ and with $\gamma^{0}$.

The covariant derivative for a spinor field (in both Euclidean and Lorentzian signature) is  
  \begin{equation}\label{def:covDspin}
      \nabla_{\nu} \Psi = \partial_{\nu}  \Psi  + \frac{1}{4} \omega_{\nu bc} \gamma^{bc}  \Psi,
  \end{equation}
where $\omega_{\nu b c  }=\omega_{\nu [b c]  } =e_{\nu}{\hspace{0.2mm}}^{a}\omega_{a b c  }$ is the spin connection and $\gamma^{bc} = \gamma^{[b}  \gamma^{c]}$.  According to our sign convention, we have
 \begin{equation}\label{zweibein postulate}
      \partial_{\mu} e^{\rho}\hspace{0.1mm}_{b} + {\Gamma}^{\rho}_{\mu \sigma}e^{\sigma}\hspace{0.1mm}_{b} - \omega_{\mu}\hspace{0.1mm}^{c}\hspace{0.1mm}_{b}  \,e^{\rho}\hspace{0.1mm}_{c}=0,
\end{equation}
where ${\Gamma}^{\rho}_{\mu \sigma}$ are the Christoffel symbols. 

For the global coordinate chart (\ref{global}) the non-zero Christoffel symbols are
\begin{align}\label{Christoffels_dS}
    &\Gamma^{\tau}_{\hspace{0.2mm}\vartheta \vartheta}=\cosh{\tau} \sinh{\tau} , \hspace{2mm} \Gamma^{\vartheta}_{\hspace{0.2mm}\vartheta \tau} =\tanh{\tau}.
\end{align}
We choose the following expressions for the zweibein fields on $dS_{2}$:
\begin{equation}\label{vielbeins}
    e^{\tau}{\hspace{0.2mm}}_{0}={1}, \hspace{5mm}  e^{\vartheta}{\hspace{0.2mm}}_{1}=\frac{1}{\cosh{\tau}}. 
\end{equation} The non-zero components of the spin connection on $dS_{2}$ are given by
\begin{equation}\label{spin_connection_dS}
     \omega_{101}= -\omega_{110}=  -\tanh{\tau}.
\end{equation}

\subsection{Embedding space formalism}

Here we collect the relevant information for utilising the embedding space formalism while referring the reader to \citep{Pethybridge:2021rwf} for further details. We consider dS$_{d+2}$ as a submanifold of $\mathbb{R}^{1,d+1}$ defined by the embedding 
\begin{equation}
    -X_0^2 + X^i X_i = \ell^2 \, .
\end{equation}
Tensor fields in de Sitter space dS$_{d+1}$ can be obtained through constrained tensor fields defined in the embedding space $\mathbb{R}^{1,d+1}$. For example, a symmetric traceless tensor (STT) with components $T_{A_1, A_2, \cdots , A_J}(X)$ defines a STT on dS$_{d+1}$ given that 
\begin{equation}
    X^{A_1} T_{A_1, A_2, \cdots, A_J}(X) = 0 \, ,
\end{equation}
the dS tensor is simply obtained via
\begin{equation}
    t_{\mu_1,\cdots , \mu_J} = \frac{\partial X^{A_1}}{\partial x^{\mu_1}}\cdots \frac{\partial X^{A_J}}{\partial x^{\mu_J}} T_{A_1, \cdots A_J}(X) \, .
\end{equation}
In a similar way, one can define ambient spinor fields and constrain them in order to obtain irreducible spinors on the dS slice \citep{Pethybridge:2021rwf}. Let $\Psi^A(X)$ be an ambient space spinor satisfying 
\begin{equation}
    \Gamma_M X^M \Psi^A = \Psi^A \, ,
    \label{spinor_project}
\end{equation}
then $\Psi^A$ defines a spinor in the dS slice. Instead of working with constrained spinors it is more convenient to introduce a polarisation spinor $\bar{S}$ which incorporates the projection \eqref{spinor_project}, allowing us to work with 
\begin{equation}
    \tilde{\Psi}^A(X,\bar{S}) \equiv \bar{S} \Psi^A(X) \, , 
\end{equation}
where the constrain now simply reads 
\begin{equation}
    X^M \Gamma_M S = - S \, , \qquad \bar{S}X^M \Gamma_M = \bar{S} \, .
\end{equation}
Finally, as in the tensor case a dS spinor is a homogeneous scalar polynomial in the (commuting) polarisation variable 
\begin{equation}
    \tilde{\Psi}^A(X,\lambda \bar{S}) = \lambda \tilde{\Psi}^A(X,\bar{S}) \, .
\end{equation}
Note that, with these definitions, writing all the possible spinor-structures that obey the constraint is simply equivalent to writing all non-vanishing scalar objects satisfying the constraint on $S$ and the homogeneity requirement. 

\section{Expansions on \texorpdfstring{$S^2$}{S2} in terms of spherical harmonics}\label{appenidx:expansions}
\noindent \textbf{Scalars.} Scalar functions $\Phi(\mathbf{x})$ defined on $S^2$ can be expanded in a complete basis of eigenfunctions of the two-sphere Laplacian $\nabla^2_{{S}^2} $ (Laplace-Beltrami operator) as 
\begin{equation}\label{eq:bosonexpansion}
    \Phi(\mathbf{x}) = \sum_{l=0}^\infty \sum_{k=-l}^l c_{lk} Y_{lk}(\mathbf{x}) \, ,
\end{equation}
where the $Y_{lk}(\mathbf{x})$ are a basis of $\emph{real}$ spherical harmonics that satisfy
\begin{equation}\label{eq:YLM}
    -\nabla^2_{{S}^2} Y_{l k} = \frac{l(l+1)}{\ell^2} Y_{lk} \, , \qquad \int_{S^2} \textnormal{d}^2 x \sqrt{g}\, Y_{lk} Y_{l'k'} = \ell^2 \delta_{ll'} \delta_{kk'} \, ,
\end{equation}
as well as the completeness relation 
\begin{equation}
    \sum_{l=0}^\infty \sum_{k=-l}^l Y_{lk}(\mathbf{x}) Y_{lk}(\mathbf{y}) = \frac{\delta(\mathbf{x} - \mathbf{y})}{\sqrt{g}} \, .
\end{equation}
Furthermore, we have the addition theorem 
\begin{equation}\label{app:scalar_addition}
    \sum_{k=-l}^l Y_{lk}(\mathbf{x}) Y_{lk}(\mathbf{y}) = \frac{2l+1}{4\pi} P_l\left( \cos \Theta_{xy} \right) \, ,
\end{equation}
where $P_{l}$ are the Legendre polynomials and $\cos \Theta_{xy}$ is given in \eqref{def: geodesic distance S2}. 

\noindent  \textbf{Spinors.} Fermionic fields $\Psi(\mathbf{x})$ on $S^2$ can be expanded in a complete basis of spinor eigenmodes with Grassmann-valued coefficients 
\begin{equation}
    \Psi(\mathbf{x}) = \frac{1}{\sqrt{\ell}}\sum_n c_n \psi_n(\mathbf{x}) \, , \qquad \bar{\Psi}(\mathbf{x}) = \frac{1}{\sqrt{\ell}} \sum_n \psi_n^\dagger(\mathbf{x}) \bar{c}_n \, ,
\end{equation}
where $n$ denotes collectively quantum numbers that will be discussed below.
When computing fermionic path integrals, one needs to keep in mind how to deal with Grassmann-valued integrals. Some useful expressions concerning integration are
\begin{equation}
\begin{split}
    \int \mathrm{d}c_n &= 0 \,, \qquad \int \mathrm{d}c_n\, c_n = 1 \, , \\ 
    \int \mathrm{d}\bar{c}_n &= 0 \,, \qquad \int \mathrm{d}\bar{c}_n\, \bar{c}_n = 1 \, ,
\end{split}
\end{equation}
from which we see that un-saturated coefficients will give a vanishing result.

The spinor eigenfunctions of the Dirac operator $\slashed{\nabla} = \gamma^{\mu} \nabla_{\mu}$ on $S^{2}$, $\psi_n(\mathbf{x})$, are known \citep{Camporesi:1995fb} to satisfy: 
\begin{equation}\label{eq:spheredirac}
    \slashed{\nabla} \psi_n(\mathbf{x}) \equiv  \slashed{\nabla} \psi^{(\sigma)}_{\pm, N L } (\mathbf{x}) = \pm i \frac{\left( N + 1 \right)}{\ell} \psi^{(\sigma)}_{\pm,N L } (\mathbf{x}) \, , .
\end{equation}
where $N \in \{ 0,1, ...\}$, and $N \geq L \geq 0$, while the label $\sigma$ takes the values $\pm$. For each fixed value of $N$, the set of eigenfunctions $\{ \psi^{(\sigma)}_{+, N L } \}$ (with all allowed values of $\sigma$ and $L$) forms the representation space for a $\mathfrak{so}(3)$ representation with weight $N + 1/2$. The two sets $\{ \psi^{(\sigma)}_{+, N L } \}$ and $\{ \psi^{(\sigma)}_{-, N L } \}$ form equivalent representations. Under the decomposition $\mathfrak{so}(3)  \supset \mathfrak{so}(2)$, the $\mathfrak{so}(2)$ content of the $\mathfrak{so}(3)$ representation with weight $N+1/2$ is labelled by the $\mathfrak{so}(2)$ weights $M \in \mathbb{Z} + \tfrac{1}{2}$. We parameterise $M$ as $M =\sigma (L+1/2$). The $\mathfrak{so}(3)  \supset \mathfrak{so}(2)$ branching rules are $N+1/2 \geq M \geq - (N+1/2)$, which lead to $N \geq L \geq 0$. The spinor covariant derivative is given by the expression in (\ref{def:covDspin}) and the `zweiben postulate' by (\ref{zweibein postulate}). Working in the coordinate system \eqref{eq:metricS2}, and with the following representation of gamma matrices and zweibein
\begin{align}
  &  \gamma^{2} = \begin{pmatrix}
         0 & 1 \\
         1 & 0
    \end{pmatrix}, ~~~~~~~~~~~~~\gamma^{1} = \begin{pmatrix}
         0 & i \\
         -i & 0
    \end{pmatrix},  \nonumber \\
    & \gamma^{\varphi} = e^{\varphi}\,_{2} ~ \gamma^{2} = \gamma^{2},~ ~~~~~\gamma^{\vartheta} = e^{\vartheta} \,_{1}  \gamma^{1} = \frac{1}{ \sin{\varphi}} \gamma^{1},
\end{align}
the eigenfunctions take the form \citep{Camporesi:1995fb} 
\begin{equation}\label{eq: spinor eigenmodes S2}
    \psi^{(+)}_{\pm,N L}(\varphi,\vartheta)  =\frac{c_{N L}}{\sqrt{2}} e^{i(L+\frac12) \vartheta} \begin{pmatrix} i \Psi_{N L}(\varphi) \\ \pm  \Phi_{N L}(\varphi) 
    \end{pmatrix} \, , \qquad  \psi^{(-)}_{\pm,N L}(\varphi,\vartheta) = \frac{c_{N L}}{\sqrt{2}} e^{-i(L+\frac12) \vartheta} \begin{pmatrix}   \Phi_{N L}(\varphi) \\ \pm   i \Psi_{N L}(\varphi)  \end{pmatrix} \, ,
\end{equation}
where $0 \leq \vartheta < 2 \pi$, $0 \leq \varphi \leq \pi  $ and 
\begin{align}
      \Phi_{N L }(\varphi) &= \cos^{L+1} \left(\frac{\varphi}{2} \right) \sin^L\left(\frac{\varphi}{2}  \right) P_{N-L}^{(L,L+1)}(\cos \varphi) \, ,  \label{eq: spinor eigenmodes S2 Phi} \\
        \Psi_{N L }(\varphi) &= \cos^L \left(\frac{\varphi}{2}  \right) \sin^{L+1} \left(\frac{\varphi}{2}  \right) P_{N-L}^{(L+1,L)}(\cos\varphi) = (-1)^{N-L} \Phi_{N L }\left(\pi - \varphi\right) \label{eq: spinor eigenmodes S2 Psi}\, ,
\end{align}
where $P_{n}^{(a,b)}(x)$ are the real-valued Jacobi polynomials \citep{NIST:DLMF}
\begin{align} \label{def: Jacobi in terms of 2F1}
  P_{n}^{(a,b)}(x) =   \frac{\Gamma(n + a +1)}{\Gamma(n+1)   \Gamma(a+1)}~ _{2}F_{1} \left(-n, n+a+b+1, a+1 ; \frac{1-x}{2}  \right).
\end{align}
  The exponentials in \eqref{eq: spinor eigenmodes S2} are  the spinor harmonics on $S^1$ \citep{Camporesi:1995fb}, and they correspond to eigenfunctions of the \emph{Dirac operator} on $S^1$ 
 \begin{align}
     \partial_{\vartheta}   e^{i \, M \, \vartheta} = i M\, e^{i \, M \, \vartheta}, \hspace{5mm} M \in \mathbb{Z}+\frac{1}{2}.
 \end{align}
In terms of the angular momentum quantum number $L=0,1,...$ we have
 \begin{align} \label{def: convenient quantum number}
   |M|  \equiv L+\frac{1}{2}  \, , 
 \end{align}
and then, the (normalised) spinor harmonics are conveniently expressed as 
\begin{align}
    \frac{1}{\sqrt{ 2 \pi}}e^{\pm \, i \, (L+1/2 )\, \vartheta},\hspace{5mm} L = 0 ,1,2,...  ~.
\end{align}

Let us note that for $\varphi =0 $ the functions  \eqref{eq: spinor eigenmodes S2 Phi} and \eqref{eq: spinor eigenmodes S2 Psi} satisfy
\begin{align}
      \Phi_{N L }(0) &= \delta_{L0},  \label{eq: spinor eigenmodes S2 Phi varphi=0} \\
        \Psi_{N L }(0) &= 0 \label{eq: spinor eigenmodes S2 Psi varphi=0}.
\end{align}
 The normalisation factors in (\ref{eq: spinor eigenmodes S2}) are given by 
\begin{equation} \label{eq: normalistn const S2}
    c_{N L }^2 = \frac{(N+L+1)!(N-L)!}{2\pi ~ \left(N! \right)^2} \, ,
\end{equation}
and ensure the following normalisation conditions 
\begin{equation} \label{fermionorm}
   \int \dd \Omega \sqrt{g}  ~ \psi^{(\sigma) \dagger}_{P, N L }(\varphi , \vartheta)~ \psi^{(\sigma')}_{P', N' L'} (\varphi, \vartheta) = \ell^2 \delta_{N N'} \delta_{L L'} \delta_{P P'} \delta_{\sigma \sigma'} \, , \qquad P,P' = \pm  .
\end{equation}

Note that there is a spinorial analogue of the scalar addition theorem \eqref{app:scalar_addition}:
\begin{equation}\label{app:spinor_addition_pm}
\begin{split}
    \sum_{L=0}^{N}\sum_{\sigma = \pm} \psi^{(\sigma)}_{\pm,N L} \left( \mathbf{x} \right) \otimes \left( \psi_{\pm,N L}^{(\sigma)} \left( \mathbf{y} \right) \right)^\dagger = \frac{N+1}{4\pi} &\left[ \cos\left( {\frac{\Theta_{xy}}{2}} \right)  P_{N}^{{(0,1)}}\left( \cos\left( \Theta_{xy} \right) \right) \bm{1} \right. \\ 
    &\left. \pm i \sin\left( {\frac{\Theta_{xy}}{2}} \right) P_{N}^{(1,0)}\left( \cos \left( {\Theta_{xy}} \right) \right) {\ell} {n}_{\mu}\gamma^{\mu} \right] \Lambda \left( \mathbf{x}, \mathbf{y} \right),
\end{split}
\end{equation}
where  $n_{\mu}$ is the tangent vector at $\mathbf{x}$ to the (shortest) geodesic connecting the points $\mathbf{x}$, $\mathbf{y}$ {and in our conventions it has units of $\ell^{-1}$}, while $\Lambda(\mathbf{x}, \mathbf{y})$ is the spinor parallel propagator - for more details on the tangent vectors and the parallel propagator see Appendix \ref{App:2ptdetails}. The addition theorem can be re-expressed in the form of the following two relations:
\begin{align}\label{app:spinor_addition_first}
    \sum_{L=0}^{N}\sum_{\sigma = \pm}  \Bigg( &\psi^{(\sigma)}_{+,N L} \left( \mathbf{x} \right) \otimes \left( \psi_{+,N L}^{(\sigma)} \left( \mathbf{y} \right) \right)^\dagger+ \psi^{(\sigma)}_{-,N L} \left( \mathbf{x} \right) \otimes \left( \psi_{-,N L}^{(\sigma)} \left( \mathbf{y} \right) \right)^\dagger\Bigg)  \nonumber  \\
    = & \frac{N+1}{2\pi} \cos{\left(\frac{\Theta_{xy}}{2}\right)}   P_{N}^{{(0,1)}}\left(\cos{\left( \Theta_{xy}\right)}\right)~ \Lambda \left( \mathbf{x}, \mathbf{y} \right)
\end{align}
\begin{align}\label{app:spinor_addition_second}
    \sum_{L=0}^{N}\sum_{\sigma = \pm}  \Bigg( &\psi^{(\sigma)}_{+,N L} \left( \mathbf{x} \right) \otimes \left( \psi_{+,N L}^{(\sigma)} \left( \mathbf{y} \right) \right)^\dagger- \psi^{(\sigma)}_{-,N L} \left( \mathbf{x} \right) \otimes \left( \psi_{-,N L}^{(\sigma)} \left( \mathbf{y} \right) \right)^\dagger\Bigg)  \nonumber  \\
    = & ~i\frac{N+1}{2\pi} \sin{\left(\frac{\Theta_{xy}}{2}\right)}   P_{N}^{{(1,0)}}\left(\cos{\left( \Theta_{xy}\right)}\right)~ {\ell}\gamma^{\mu}n_{\mu}~\Lambda \left( \mathbf{x}, \mathbf{y} \right).
\end{align}
Eq. (\ref{app:spinor_addition_first}) can be proved following steps that are very similar to the ones followed in \citep{Camporesi:1995fb} to obtain the spinor heat kernel as a mode sum on spheres. Eq. (\ref{app:spinor_addition_second}) can be obtained by acting with the Dirac operator on both sides of (\ref{app:spinor_addition_first}).

Note also the completeness relation
\begin{equation}\label{completeness spinors}
\begin{split}
  \sum_{N=0}^{\infty}   \sum_{L=0}^{N}\sum_{\sigma = \pm} \left(\psi^{(\sigma)}_{+,N L} \left( \mathbf{x} \right) \otimes \left( \psi_{+,N L}^{(\sigma)} \left( \mathbf{y} \right) \right)^\dagger+ \psi^{(\sigma)}_{-,N L} \left( \mathbf{x} \right) \otimes \left( \psi_{-,N L}^{(\sigma)} \left( \mathbf{y} \right) \right)^\dagger \right)= \frac{\delta(\mathbf{x}-\mathbf{y})}{\sqrt{g}} \, \bm{1}.
\end{split}
\end{equation}

\section{Details for the principal series two-point function} \label{App:2ptdetails}
In this Appendix, we present the details of the Euclidean computation of the two-point function for a principal series (i.e. real mass) spinor. By definition, the correlation function is given by 
\begin{align}
S_{f}\left(\mathbf{x},   \mathbf{y}   \right) &=   \frac{\int \mathcal{D}\overline{\Psi}\mathcal{D}\Psi\, {\Psi}(\Omega) \overline{\Psi}(\Omega') e^{- S[\overline{\Psi},\Psi]}}{\int \mathcal{D}\overline{\Psi} \mathcal{D}\Psi \, e^{-S[\bar{\Psi},\Psi]}} \nonumber \\
&=\frac{1}{\ell} \sum_{N=0}^\infty \sum_{L=0}^N \sum_{\sigma = \pm} \left( \frac{ \psi_{+,NL}^{(\sigma)}\left(\varphi_{x}, \vartheta_{x}  \right)  \otimes \left(\psi_{+,NL}^{(\sigma)} \left(\varphi_{y},   \vartheta_{y} \right)\right)^\dagger   }{m \ell + i (N+1)} + \frac{ \psi_{-,NL}^{(\sigma)}\left(\varphi_{x} , \vartheta_{x} \right)  \otimes \left(\psi_{-,NL}^{(\sigma)}\left(\varphi_{y} ,    \vartheta_{y} \right) \right)^\dagger  }{m \ell - i (N+1)}\right) \, ,
\end{align}
where we have expanded in terms of the spinor spherical harmonics on $S^{2}$ - see Appendix \ref{appenidx:expansions}.
 The expression for the spinor two-point function on spheres in terms of intrinsic geometric objects was given in \citep{Mueck:1999efk}, and here we will reproduce this expression from the viewpoint of the Euclidean path integral - note that there is a misprint in the expression for the two-point function in \citep{Mueck:1999efk} which has been corrected in \citep{Letsios:2020twa}.
 Substituting the explicit expressions for the spinor spherical harmonics (\ref{eq: spinor eigenmodes S2}) into the two-point function, we can express it as a $2 \times 2$ matrix: 
 \begin{align}
    & {\ell}~ S_{f}\left( (\varphi_{x}, \vartheta_{x}  )  , (\varphi_{y} , \vartheta_{y}) \right)   \nonumber   \\
     =&\sum_{N=0}^\infty \sum_{L=0}^N  \, \frac{ c^{2}_{NL}  \,  e^{-i(L+1/2)(\vartheta_{x}-\vartheta_{y})}  }{\ell^{2}m^{2}+(N+1)^{2}}    
 \begin{pmatrix}
     \ell m \, \Phi_{NL}(\varphi_{x})  \Phi_{NL}(\varphi_{y}) &  & -(N+1)\, \Phi_{NL}(\varphi_{x})  \Psi_{NL}(\varphi_{y})\\ \\
    (N+1) \Psi_{NL}(\varphi_{x})  \Phi_{NL}(\varphi_{y}) &   &  \ell m\, \Psi_{NL}(\varphi_{x})  \Psi_{NL}(\varphi_{y})
 \end{pmatrix}   \nonumber\\  \nonumber \\
+&\sum_{N=0}^\infty \sum_{L=0}^N  \, \frac{ c^{2}_{NL}  \,  e^{+i(L+1/2)(\vartheta_{x}-\vartheta_{y})}  }{\ell^{2}m^{2}+(N+1)^{2}}    
 \begin{pmatrix}
     \ell m \, \Psi_{NL}(\varphi_{x})  \Psi_{NL}(\varphi_{y}) &  & (N+1)\, \Psi_{NL}(\varphi_{x})  \Phi_{NL}(\varphi_{y})\\  \\
   - (N+1) \Phi_{NL}(\varphi_{x})  \Psi_{NL}(\varphi_{y}) &   &  \ell m\, \Phi_{NL}(\varphi_{x})  \Phi_{NL}(\varphi_{y})
 \end{pmatrix} .
     \end{align}
 We will take advantage of the invariance of the Euclidean spinor two-point function under $\mathfrak{so}(3)$. For simplicity, we first let $\vartheta_{x}  = \vartheta_{y}\equiv \vartheta$ without loss of generality, and we see that the two-point function becomes independent of $\vartheta$. Also, the spinor parallel propagator for the points $\mathbf{x}=(\varphi_{x} , \vartheta)$ and $\mathbf{y} = (\varphi_{y} , \vartheta)$ is equal to the spinorial identity matrix $\bm{1}$ \citep{Camporesi:1995fb} (this will simplify the calculation). We have
\begin{align}
    & {\ell}~S_{f}\left( (\varphi_{x}, \vartheta  )  , (\varphi_{y} , \vartheta) \right)   \\
     =&\sum_{N=0}^\infty \sum_{L=0}^N  \, \frac{ c^{2}_{NL}  \,    }{\ell^{2}m^{2}+(N+1)^{2}}    
\ell m   \left( \Phi_{NL}(\varphi_{x})  \Phi_{NL}(\varphi_{y}) + \Psi_{NL}(\varphi_{x})  \Psi_{NL}(\varphi_{y})  \right) \, \bm{1} \nonumber\\
-&\sum_{N=0}^\infty \sum_{L=0}^N  \, \frac{ c^{2}_{NL} ~(N+1) \,    }{\ell^{2}m^{2}+(N+1)^{2}}    \left( \Phi_{NL}(\varphi_{x})  \Psi_{NL}(\varphi_{y}) - \Psi_{NL}(\varphi_{x})  \Phi_{NL}(\varphi_{y})  \right) \, \gamma^{2}.
     \end{align}
Let us recall that the geodesic distance (\ref{def: geodesic distance S2}) [for $\vartheta_{x} = \vartheta_{y}$] between the two points is $\Theta_{xy} = {\varphi_{x} - \varphi_{y}}$. Moreover, let us recall that the tangent vector $n^{\mu}$ at $\mathbf{x}$ to the (shortest) geodesic connecting the points $\mathbf{x} ,   \mathbf{y}$ is defined as \citep{Mueck:1999efk}
\begin{align}
    n_{\mu}(\mathbf{x} ,   \mathbf{y} ) = \partial_{\mu} \Theta_{xy}, ~~~~~~~~~~~~ \mu \in \{ \varphi_{x} , \vartheta_{x}   \},
\end{align}
where $\Theta_{xy}$ is defined in (\ref{def: geodesic distance S2}) ($\Theta_{xy}$ is dimensionless in our conventions).
 This definition holds for any two points on $S^{2}$, but here we specialise to our case $\Theta_{xy} = {\varphi_{x} - \varphi_{y}}$, and we have
 \begin{align}
     n_{\mu} = (n_{\varphi}, n_{\vartheta}) = {\frac{1}{\ell}}(  1, 0 ).
 \end{align}
 Now we can express $S_{f}\left( (\varphi_{x}, \vartheta  )  , (\varphi_{y} , \vartheta) \right)$ as
 \begin{align}
    & {\ell}~ S_{f}\left( (\varphi_{x}, \vartheta  )  , (\varphi_{y} , \vartheta) \right) = \left(  \mathcal{A}_{m}(\varphi_{x} , \varphi_{y})   +  \mathcal{B}_{m}(\varphi_{x} , \varphi_{y}) \, {\ell}\slashed{n}   \right) \bm{1},
     \end{align}
where
     \begin{align}
         \mathcal{A}_{m}(\varphi_{x} , \varphi_{y}) =\ell m~  \sum_{N=0}^\infty \sum_{L=0}^N  \, \frac{ c^{2}_{NL}  \,    }{\ell^{2}m^{2}+(N+1)^{2}}    
   \left( \Phi_{NL}(\varphi_{x})  \Phi_{NL}(\varphi_{y}) + \Psi_{NL}(\varphi_{x})  \Psi_{NL}(\varphi_{y})  \right) ,
     \end{align}
     \begin{align}
         \mathcal{B}_{m}(\varphi_{x} , \varphi_{y})=  -\sum_{N=0}^\infty \sum_{L=0}^N  \, \frac{ c^{2}_{NL}~ (N+1) \,    }{\ell^{2}m^{2}+(N+1)^{2}}    \left( \Phi_{NL}(\varphi_{x})  \Psi_{NL}(\varphi_{y}) - \Psi_{NL}(\varphi_{x})  \Phi_{NL}(\varphi_{y})  \right).
     \end{align}
     To simplify the computation further, we let the point $\mathbf{y} =  (\varphi_{y} , \vartheta) $ coincide with the North Pole, i.e. we set $\varphi_{y} = 0$. This simplifies $S_{f}$ because 
     \begin{align}
       \Phi_{N L }(0) = \delta_{0 L},~~~~  \Psi_{N L }(0) = 0,
     \end{align}
     and thus only the $L=0$ term will survive in the sum.
     Also, $c^{2}_{N0} = (N+1) / 2  \pi$ - see eq. (\ref{eq: normalistn const S2}).
     Thus, we now have 
     \begin{align} 
    {\ell}~  S_{f}\left( (\varphi_{x}, \vartheta  )  , (0 , \vartheta) \right) =& ~   \frac{\ell m}{2\pi} \sum_{N=0}^{\infty} \frac{N+1}{\ell^{2}m^{2}+(N+1)^{2}}  \Phi_{N0}(\varphi_{x}) ~ \bm{1}\nonumber \\
    &+  \frac{1}{ 2\pi}\sum_{N=0}^{\infty} \frac{ (N+1)^{2} }{\ell^{2}m^{2}+(N+1)^{2}}  \Psi_{N0}(\varphi_{x})\, ~{\ell}\slashed{n} ~  \bm{1}.
     \end{align}
     Expressing $\Phi_{N0}(\varphi_{x})$ and $\Psi_{N0}(\varphi_{x})$ in terms of the Jacobi polynomials (see \cref{appenidx:expansions}), we find
     \begin{align}\label{eq: almost final S2 principal 2pt}
   {\ell}~  S_{f}\left( (\varphi_{x}, \vartheta  )  , (0 , \vartheta) \right) =& ~   \frac{\ell m}{2\pi}\cos{\frac{\varphi_{x}}{2}} ~\sum_{N=0}^{\infty} \frac{N+1}{\ell^{2}m^{2}+(N+1)^{2}}  P_{N}^{(0,1)}(\cos{\varphi_{x}}) ~ \bm{1}\nonumber \\
    &+ \frac{1}{2\pi}\,\sin{\frac{\varphi_{x}}{2}}~ \sum_{N=0}^{\infty} \frac{( N+1)^{2} }{\ell^{2}m^{2}+(N+1)^{2}}  P^{(1,0)}_{N}(\cos{\varphi_{x}})\, ~{\ell}\slashed{n} ~  \bm{1}.
     \end{align}
     
     Keeping in mind that for our current choice of points on $S^{2}$ we have $\Theta_{xy} =  \varphi_{x} - \varphi_{y} = \varphi_{x}$, and that the spinor parallel propagator is $\Lambda(\mathbf{x} , \mathbf{y}) = \bm{1}$, we re-express the two-point function (\ref{eq: almost final S2 principal 2pt}) in a manifestly $\mathfrak{so}(3)$ invariant form:
      \begin{align} 
   {\ell}~  S_{f}\left( \mathbf{x} , \mathbf{y} \right)    =& ~   \frac{\ell m}{2 \pi} \cos{\frac{\Theta_{xy}}{2}} ~\Bigg(\sum_{N=0}^{\infty} \frac{N+1}{\ell^{2}m^{2}+(N+1)^{2}}  P_{N}^{(0,1)}(\cos{\Theta_{xy}})\Bigg) ~ \Lambda(\mathbf{x} , \mathbf{y})\nonumber \\
    &+ \frac{1}{2 \pi}\,\sin{\frac{\Theta_{xy}}{2}}~ \Big(\sum_{N=0}^{\infty} \frac{( N+1)^{2} }{\ell^{2}m^{2}+(N+1)^{2}}  P^{(1,0)}_{N}(\cos{\Theta_{xy}})\Big)\, ~{\ell}\slashed{n} ~  \Lambda(\mathbf{x} , \mathbf{y}).
     \end{align}
     This form is in agreement with the expression given in \citep{Mueck:1999efk, Letsios:2020twa}. The only remaining step is to compute the series of Jacobi polynomials and show that they match the corresponding functions of the geodesic distance $\Theta_{xy}$ in \citep{Mueck:1999efk, Letsios:2020twa} - we stress again that there is a misprint in \citep{Mueck:1999efk} that has been corrected in \citep{Letsios:2020twa}. The Jacobi polynomials $P_{N}^{(0,1)}$ and $P_{N}^{(1,0)}$ are expressed in terms of the Gauss hypergeometric function as follows:
     \begin{align}
        P_{N}^{(0,1)}(\cos{\Theta_{xy}}) &=~ _{2}F_{1}\left( -N, N+2;1;\frac{1 -\cos{\Theta_{xy}} } {2} \right) , \nonumber \\
         P_{N}^{(1,0)}(\cos{\Theta_{xy}}) &=~(N+1)~ _{2}F_{1}\left( -N, N+2;2;\frac{1 -\cos{\Theta_{xy}} } {2} \right) .
     \end{align}
     Comparing with the expression of the two-point function in \citep{Mueck:1999efk, Letsios:2020twa}, we conclude that:
     \begin{align}\label{Conjecture 1}
        \sum_{N=0}^{\infty} & \frac{2(N+1)}{\ell^{2}m^{2}+(N+1)^{2}}  ~P_{N}^{(0,1)}\left( \cos{\Theta_{xy}} \right)    =  \sum_{N=0}^{\infty}  \frac{2(N+1)}{\ell^{2}m^{2}+(N+1)^{2}}  ~_{2}F_{1}\left( -N, N+2;1;\frac{1 -\cos{\Theta_{xy}} } {2} \right)  \nonumber \\
             &\overset{!}{=}  \left| \Gamma(1 + i \ell m)  \right|^{2} \, ~_{2}F_{1}\left( 1+ i \ell m, 1- i \ell m;2;z \right) ,~~~z= \frac{1 + \cos{\Theta_{xy}}}{2}
     \end{align}
  \begin{align}\label{Conjecture 2}
        \sum_{N=0}^{\infty} & \frac{2(N+1)^{2}}{\ell^{2}m^{2}+(N+1)^{2}}  ~P_{N}^{(1,0)}\left( \cos{\Theta_{xy}} \right)    =  \sum_{N=0}^{\infty}  \frac{2(N+1)^{3}}{\ell^{2}m^{2}+(N+1)^{2}}  ~_{2}F_{1}\left( -N, N+2;2;\frac{1 -\cos{\Theta_{xy}} } {2} \right)  \nonumber \\
             &\overset{!}{=}  \left| \Gamma(1 + i \ell m)  \right|^{2} \, ~_{2}F_{1}\left( 1+ i \ell m, 1- i \ell m;1;z \right) .
     \end{align}
  While we were not able to find an analytic proof of these relationships (and hence the use of the exclamation mark above the equality sign), it is straightforward to check them numerically.  Thus, eqs. (\ref{Conjecture 1})  and (\ref{Conjecture 2}) serve as conjectures for series involving hypergeometric functions.

\bibliographystyle{utphys2}
\bibliography{refs.bib}

\end{document}